\documentclass[10pt,journal,compsoc]{IEEEtran}
\usepackage{amsmath,amsfonts}
\usepackage[export]{adjustbox}
\usepackage{url}
\usepackage{graphicx}
\usepackage{cite}
\usepackage{multirow}
\usepackage{booktabs}
\usepackage{amssymb}
\usepackage{stmaryrd}
\usepackage[justification=centering,caption=false]{subfig}
\usepackage{hyperref}
\graphicspath{{./fig_com/}}
\DeclareGraphicsExtensions{.pdf,.png,.jpg} 
\usepackage[final]{changes}
\hypersetup{hidelinks} 
\usepackage{array}
\usepackage{textcomp}
\usepackage{stfloats}
\usepackage{epsfig}
\usepackage{float}
\usepackage{color}
\usepackage{tabularx}
\usepackage{longtable}
\usepackage{tabu}
\usepackage{bbding}
\newcommand{\etal}{et al.}
\newcommand{\ie}{i.e.}
\newcommand{\eg}{e.g.}

\long\def\comment#1{}
\begin{document}

\title{A Perceptually Optimized and Self-Calibrated Tone Mapping Operator}

\author{Peibei Cao,
        Chenyang Le,
        Yuming Fang,~\IEEEmembership{Senior Member,~IEEE},
        and Kede Ma,~\IEEEmembership{Senior Member,~IEEE}

\IEEEcompsocitemizethanks{\IEEEcompsocthanksitem Peibei Cao and Kede Ma are with the Department of Computer Science, City University of Hong Kong, Kowloon, Hong Kong (e-mail: peibeicao2-c@my.cityu.edu.hk, kede.ma@cityu.edu.hk).
\IEEEcompsocthanksitem Chenyang Le and Yuming Fang are with the School of Information Management, Jiangxi University of Finance and Economics, Nanchang, China (e-mail:leshier12@gmail.com, fa0001ng@e.ntu.edu.sg).}
\thanks{Corresponding author: Kede Ma.}
}

\IEEEtitleabstractindextext{
\begin{abstract}
With the increasing popularity and accessibility of high dynamic range (HDR) photography, tone mapping operators (TMOs) for dynamic range compression are practically demanding. In this paper, we develop a two-stage neural network-based TMO that is self-calibrated and perceptually optimized. In Stage one, motivated by the physiology of the early stages of the human visual system, we first decompose an HDR image into a normalized Laplacian pyramid. We then use two lightweight deep neural networks (DNNs), taking the normalized representation as input and estimating the Laplacian pyramid of the corresponding LDR image. We optimize the tone mapping network by minimizing the normalized Laplacian pyramid distance (NLPD), a perceptual metric aligning with human judgments of tone-mapped image quality. In Stage two, the input HDR image is self-calibrated to compute the final LDR image. We feed the same HDR image but rescaled with different maximum luminances to the learned tone mapping network, and generate a pseudo-multi-exposure image stack with different detail visibility and color saturation. We then train another lightweight DNN to fuse the LDR image stack into a desired LDR image by maximizing a variant of the structural similarity index for multi-exposure image fusion (MEF-SSIM), which has been proven perceptually relevant to fused image quality. The proposed self-calibration mechanism through MEF enables our TMO to accept uncalibrated HDR images, while being physiology-driven. Extensive experiments show that our method produces images with consistently better visual quality. Additionally, since our method builds upon three lightweight DNNs, it is among the fastest local TMOs
\end{abstract}

}

\maketitle

\IEEEdisplaynontitleabstractindextext

\IEEEpeerreviewmaketitle

\section{Introduction}
\begin{figure}[t]
  \centering
    \subfloat[Linear rescaling]{\includegraphics[width=0.23\textwidth]{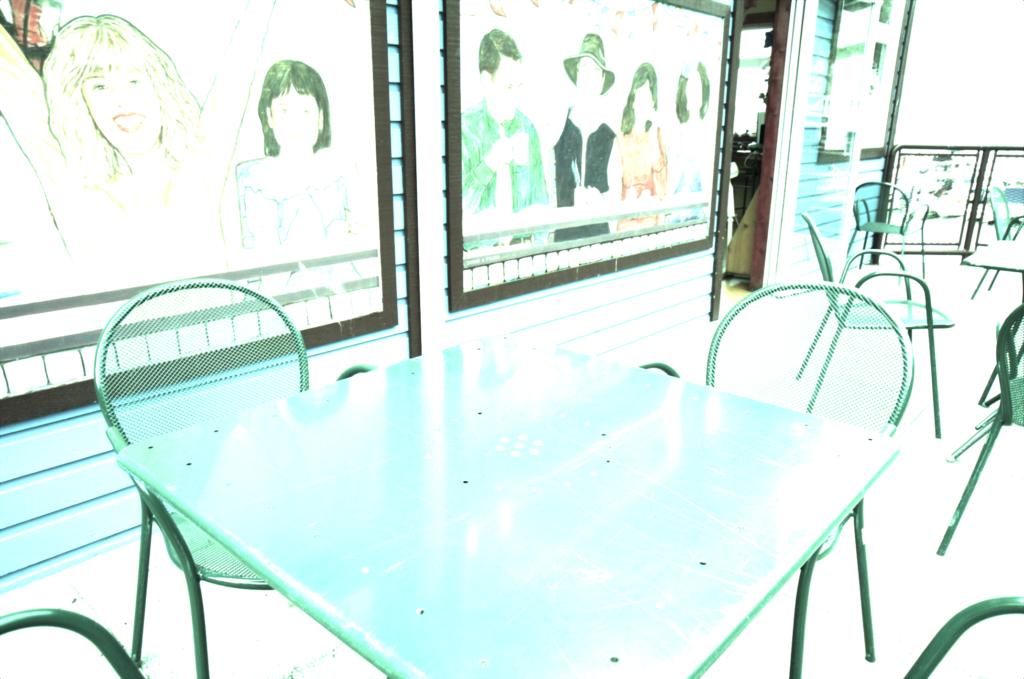}} \hskip0.3em
    \subfloat[Drago03]{\includegraphics[width=0.23\textwidth]{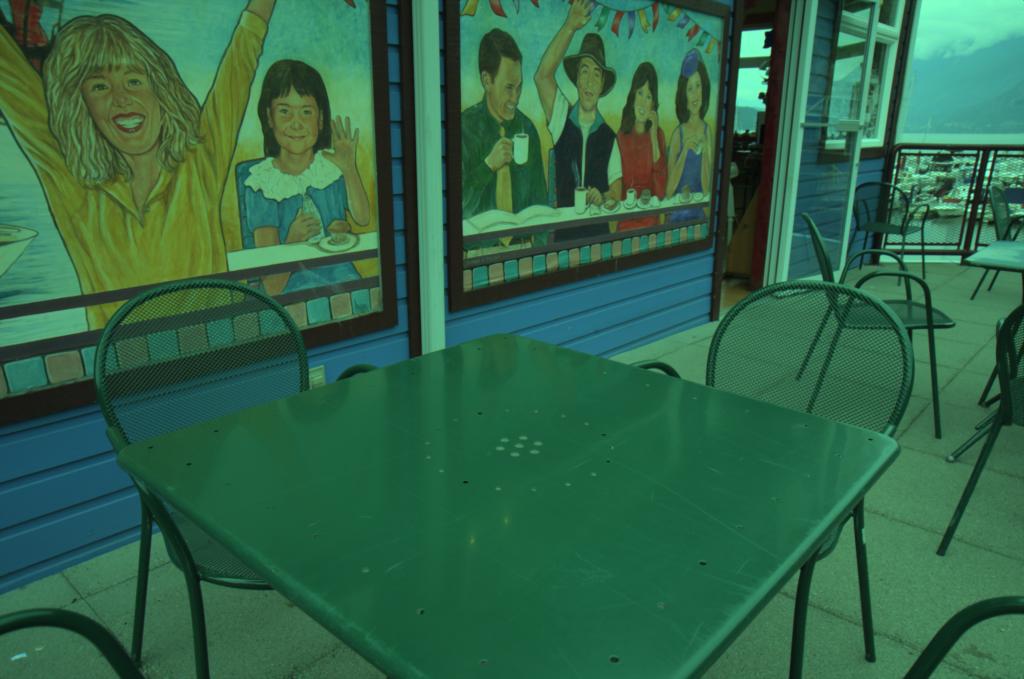}} \\ \vspace{-0.5em}
    \subfloat[Liang18]{\includegraphics[width=0.23\textwidth]{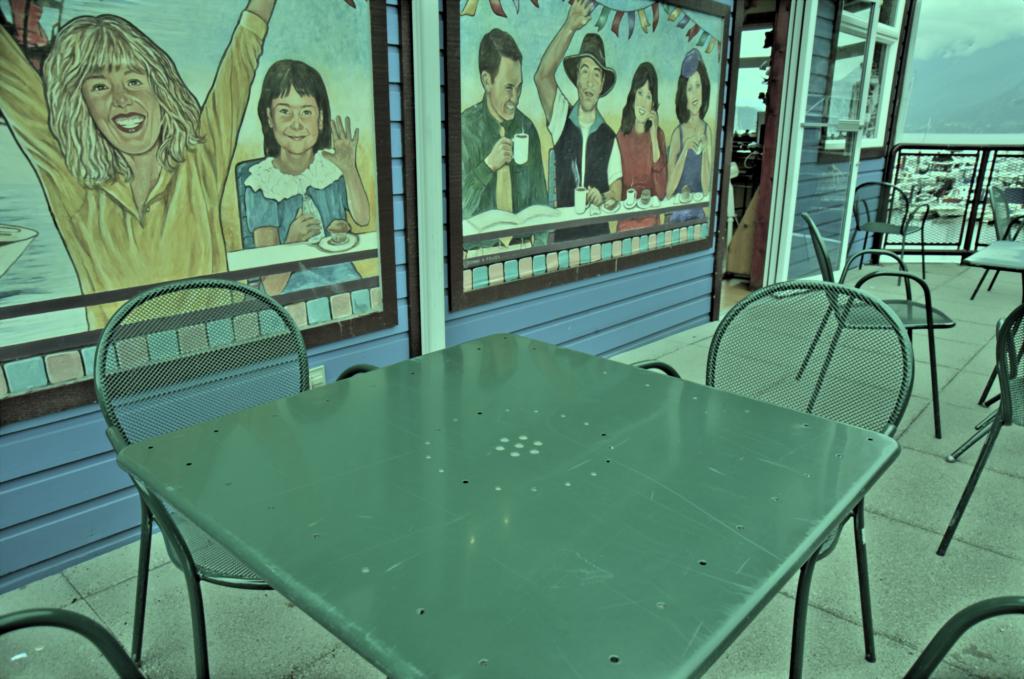}} \hskip0.3em
    \subfloat[Vinker21]{\includegraphics[width=0.23\textwidth]{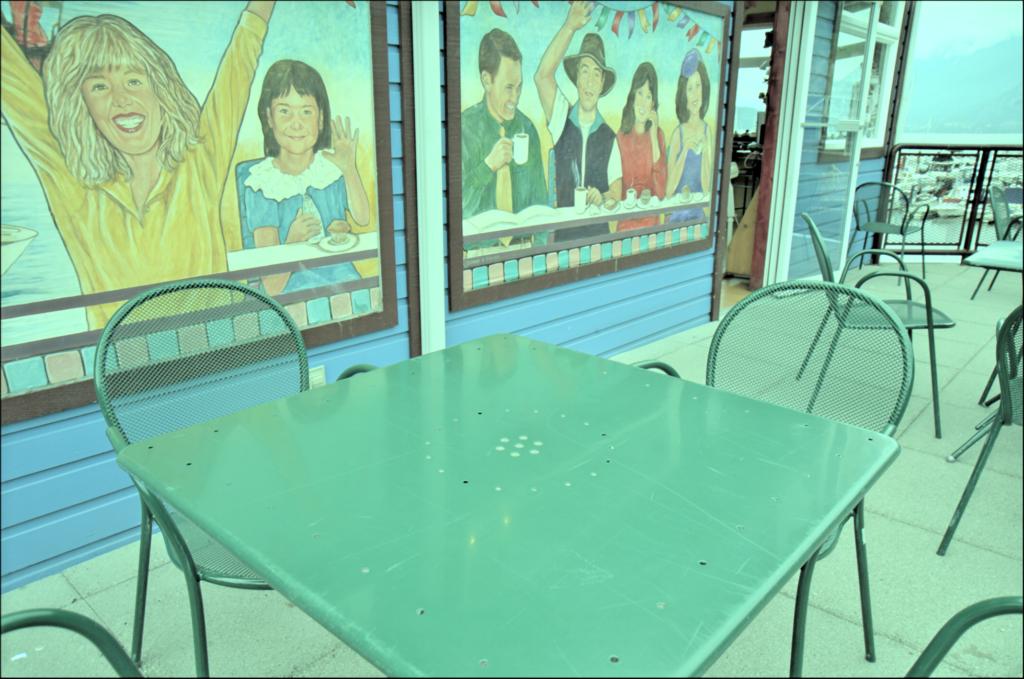}} \\  \vspace{-0.5em}
    \subfloat[Zhang21]{\includegraphics[width=0.23\textwidth]{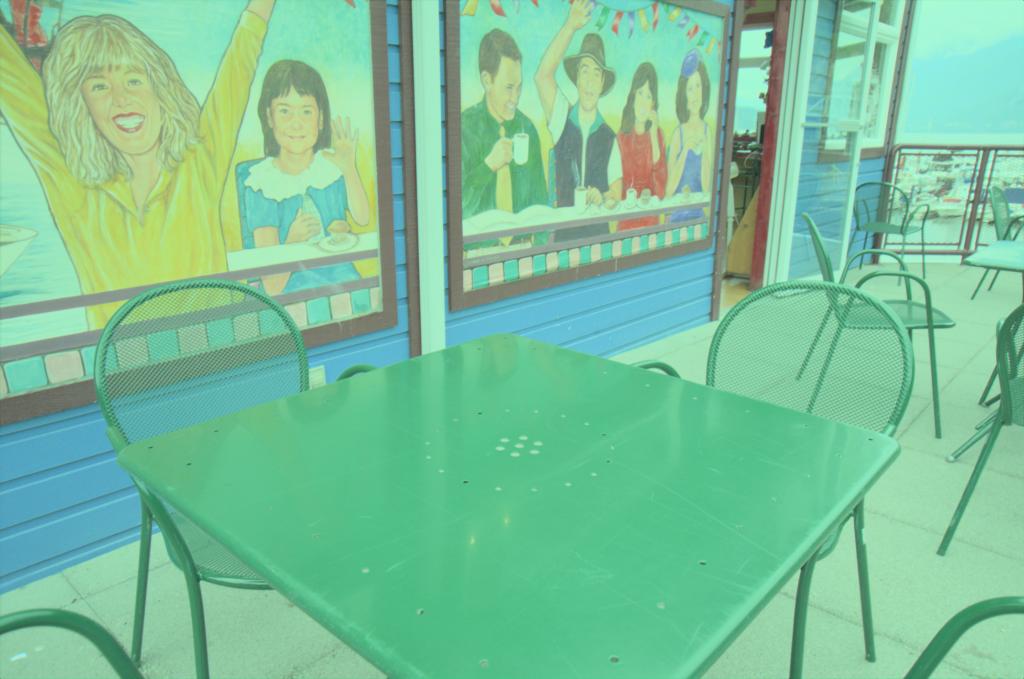}} \hskip0.3em
    \subfloat[PS-TMO]{\includegraphics[width=0.23\textwidth]{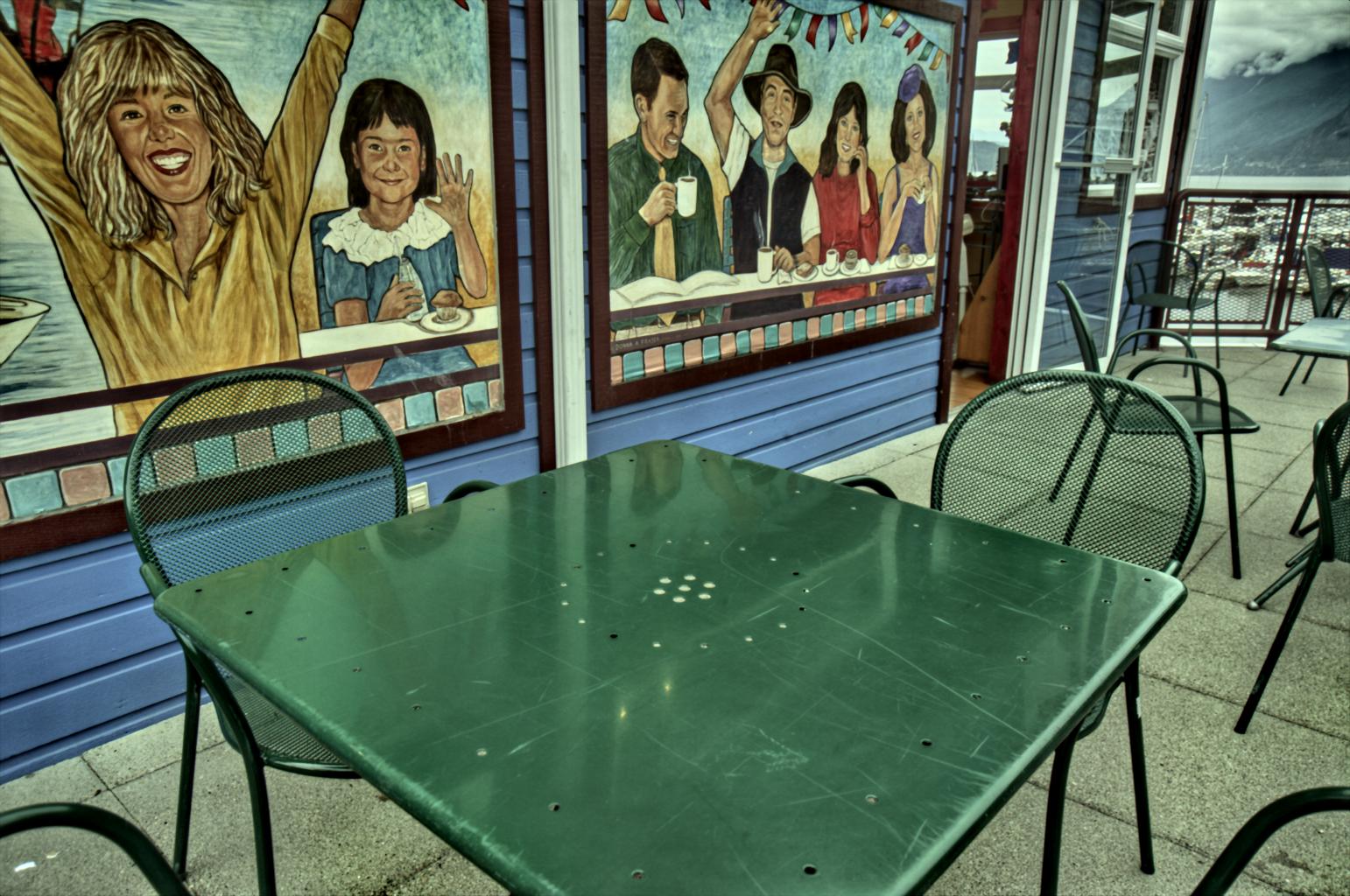}} \\ 
  \caption{Tone mapping results of the ``Outdoor Table'' HDR scene. Compared to existing TMOs, the proposed PS-TMO produces a more natural and engaging appearance with rich details.} 
  \label{fig:table}
  \vspace{-0.6em} 
\end{figure}

\IEEEPARstart{W}{ith} the steady improvements in photography technologies, current image sensors (often powered by computational imaging methods~\cite{mertens2009exposure}) are able to capture pictures with a high dynamic range up to eight orders of magnitude, closely approximating the sensitivity of human vision in the photopic regime~\cite{hoefflinger2007high}. However, existing monitors, projectors, and print-outs, are limited to a lower dynamic range than that can be captured by current sensors~\cite{reinhard2010high}, and thus are inadequate to reproduce the full spectrum of luminance values presented in natural scenes. When rendering high dynamic range (HDR) images on low dynamic range (LDR) display devices, tone mapping operators (TMOs) are a prerequisite for dynamic range compression, preserving visual features that are important to describe the original scenes and perceptually noticeable to the human eye. An example is given in Fig.~\ref{fig:table}, in which we tone map the ``Outdoor Table'' HDR scene using six different TMOs.

The na\"{i}ve way of HDR image tone mapping is to \textit{linearly} rescale the luminances of the HDR image to the range that the display can reproduce. However, images produced this way are often severely under- or over-exposed, due to the existence of local regions with high luminances (see Fig.~\ref{fig:table} (a)). In the past twenty years, extensive effort has been dedicated to developing TMOs for \textit{non-linear} dynamic range compression with faithful tone reproduction and detail preservation. These can be broadly categorized into global and local methods. Global TMOs perform the same computation to all pixels (\ie, translation-invariant), which are more computationally efficient at the cost of contrast decrease and detail loss~\cite{ward1994radiance, larson1997visibility, tumblin1993tone, drago2003adaptive, reinhard2005dynamic, kim2008consistent}. Local TMOs~\cite{durand2002fast,farbman2008edge,paris2011local,BRUCE201412,shibata2016gradient,liang2018hybrid}, on the other hand, aim to preserve and enhance local contrast often within a two-layer decomposition framework~\cite{durand2002fast}. Although these methods can produce images with better visual quality, it remains difficult to balance global and local contrast, and to prevent edge-related artifacts. Moreover, these TMOs rely on pre-defined computational graphs with few justifications for the perceptual optimality of such structures. Besides, manual hyper-parameter adjustment (\eg, setting maximum luminances for uncalibrated HDR images) is often needed to produce reasonable results, which are, however, no better than conventional photographs on challenging HDR scenes~\cite{Ledda2005, Eilertsen2013}.

Recently, deep neural networks (DNNs) began to show their potential in HDR image tone mapping~\cite{ zhang2019hsv, Montulet2019DeepLF, rana2020deep, Panetta2021}. However, unlike traditional image processing tasks such as Gaussian image denoising and image compression, there are no easy-to-obtain ground-truth images available for supervised training \textit{in the LDR domain}.
One popular strategy is to choose the best tone-mapped image from a candidate set  produced by multiple existing TMOs with the help of objective quality metrics~\cite{Montulet2019DeepLF,rana2020deep} or subjective experiments~\cite{yang2021deep}. Although with the goal of creating a ``super-method'', the resulting TMO may be biased by the common failures of base TMOs. Another approach is to ask photography experts to manually  compress the dynamic range of HDR images~\cite{zhang2019hsv,Zhang2021}, which is prohibitively slow and suffers from subjective biases. To alleviate this, semi-supervised~\cite{Zhang2021} and adversarial learning~\cite{vinker2021unpaired} techniques have been explored for tone mapping, which turn out to be less accurate and less robust (see Fig.~\ref{fig:table}).

Tumblin~\etal~\cite{tumblin1993tone} pioneered perceptual optimization of HDR image tone mapping \textit{in a cross-dynamic-range} setting. They first advocated optimizing TMOs capable of producing tone-mapped images that perceptually match the appearance of the original scenes. Yeganeh and Wang searched over the space of all feasible tone-mapped images for the closest one with respect to the original scene, measured by a structural fidelity index~\cite{yeganeh2012objective}. Ma~\etal~\cite{ma2015high} improved this method by incorporating a statistical naturalness measure. Laparra~\etal~\cite{laparra2017perceptually} formulated HDR image tone mapping as a more general image rendering problem, with the objective function defined as the normalized Laplacian pyramid distance (NLPD). The above methods are computationally expensive iterative TMOs, which limit their wide adoption in real-world time-sensitive applications.

In this paper, we describe a two-stage DNN-based TMO for rendering HDR images, which is 1) perceptually optimized, 2) self-calibrated, and 3) computationally efficient. Specifically, being physiology-driven,  we explicitly model how the early stages of the human visual system (HVS) respond to different light levels by decomposing the input HDR image into a normalized Laplacian pyramid~\cite{laparra2017perceptually}, a multi-scale non-linear representation derived from Laplacian pyramid~\cite{burt1983laplacian}. This allows us to artificially manipulate the maximum luminance of the original scene, giving rise to different detail visibility and color saturation~\cite{laparra2017perceptually}. As will be clear shortly, we will take advantage of this nice property to self-calibrate the input HDR image with respect to perceptual quality (not physical plausibility).

 Instead of iteratively optimizing over the space of all feasible tone-mapped images, we train two feed-forward DNNs (collectively referred to as the \textit{tone mapping network}) in Stage one. One network accepts all bandpass channels and the highpass channel, while the other network processes the lowpass channel of the normalized Laplacian pyramid~\cite{laparra2017perceptually} of an (randomly photometrically calibrated) HDR image. Together, they predict the Laplacian pyramid of the corresponding LDR image. 
Unlike most TMOs, the tone mapping network is optimized in the \textit{cross-dynamic-range} setting by minimizing a \textit{perceptual} image quality metric - the normalized Laplacian pyramid distance (NLPD)~\cite{laparra2017perceptually} between the input HDR scenes and the estimated LDR images.

 After training of the tone mapping network in Stage one, we are able to perform \textit{self-calibration} of the input HDR image for final LDR image generation in Stage two. Specifically, we first generate a pseudo-multi-exposure LDR image stack  using the learned tone mapping network by varying the maximum luminance of the input HDR image. As a result, the image stack shares the same visual content but with different structural and color appearances. We then train another DNN (referred to as the \textit{fusion network}) to fuse the LDR image stack into a desired LDR image by maximizing a variant of another \textit{perceptual} image quality metric~\cite{ma2015perceptual} - the structural similarity index for multi-exposure image fusion (MEF-SSIM). Both NLPD and MEF-SSIM, used in Stages one and two, respectively, have been subject-verified on databases of human perceptual scores~\cite{laparra2017perceptually,ma2015perceptual} and proven effective in optimizing image rendering algorithms~\cite{laparra2017perceptually,ma2020deep}. Moreover, the tone mapping and fusion networks are designed to be highly lightweight with a total of $100,839$ model parameters, making the entire method \textit{computationally efficient}.

We have conducted extensive experiments to demonstrate the superiority of  
the proposed method, which we name Perceptually optimized and Self-calibrated TMO (PS-TMO) against fifteen existing TMOs. We find that PS-TMO performs consistently better than the competing TMOs both qualitatively (via a formal debiased subjective experiment~\cite{cao2021debiased}) and quantitatively (in terms of objective metrics, TMQI~\cite{yeganeh2012objective} and NLPD \cite{laparra2017perceptually}). Meanwhile, the proposed self-calibration mechanism through MEF makes PS-TMO fully automatic to work with uncalibrated HDR images (with unknown maximum luminances), while being physiology-driven. Besides, PS-TMO is among the fastest local TMOs, and runs in real-time on standard-grade GPUs.
 
The preliminary results of this paper were published in its six-page conference version~\cite{chenyang2021}. The current journal article provides a complete design and a more comprehensive analysis of the proposed PS-TMO, \ie, the self-calibration via the LDR stack fusion (in Sec. \ref{subsec:fusion}), the debiased subjective experiment (in Sec. \ref{sec:subjective}), and the design choice and hyper-parameter analysis (in Sec. \ref{sec:ablation experiment}).

\section{Related Work}\label{sec:relatedwork}
In this section, we provide a brief review of existing TMOs, with emphasis on DNN-based methods. As the proposed PS-TMO involves fusing a pseudo-multi-exposure image stack, we also review MEF, an alternative approach to HDR image tone mapping. 

\subsection{Existing TMOs}
\subsubsection{Conventional TMOs}
TMOs can be classified into several categories under different sets of criteria~\cite{dufaux2016high}, among which we adopt the taxonomy of global and local operators. Global TMOs~\cite{ward1994radiance, larson1997visibility, tumblin1993tone, drago2003adaptive, reinhard2005dynamic, kim2008consistent} rely on a family of parametric functions, specified by some global image statistics, and applied to
all pixels in an HDR image. These include histogram equalization, homography (\eg, $\frac{S}{S+1}$), gamma mapping (\eg, $S^{\gamma}$), logarithmic function~\cite{drago2003adaptive}, and sigmoid non-linearity~\cite{reinhard2005dynamic}. Glocal methods remain the fastest TMOs as each pixel in $S$ undergoes the same simple non-linear
 transformation. They preserve overall contrast well but may lose some detailed information. Local TMOs~\cite{durand2002fast, farbman2008edge, paris2011local, gu2012local, BRUCE201412, shibata2016gradient, liang2018hybrid} are a set of sophisticated methods, which preserve relative contrast between neighboring pixels (\eg, in the form of local gradients) that the human eye is more sensitive to. A common design principle is the layer decomposition originated from the retinex theory~\cite{land1971lightness}. Among many variants~\cite{pattanaik1998multiscale, tumblin1999boundary}, the two-layer decomposition by Durand~\etal~\cite{durand2002fast} was the most widely accepted, in which tone compression is applied to the base layer, while detail reproduction or enhancement is applied to the detail layer. Many subsequent methods~\cite{farbman2008edge, gu2012local, shibata2016gradient,liang2018hybrid} have been proposed based on this design principle, differing mainly in how the two-layer image decomposition is performed in a more effective and perceptual way. On many HDR scenes, local TMOs lead to excellent improvements in local contrast preservation. However, this often comes at the cost of increased computational complexity and manual hyper-parameter tuning~\cite{paris2011local}. Besides, global contrast may be compromised, and local artifacts such as halo-like glows may appear, resulting in unnatural and unrealistic tone-mapped images.

The design of the above-mentioned TMOs is mostly based on empirical rules, with little validity of perceptual optimality of such rules. Perceptual optimization of tone mapping in a cross-dynamic-range setting has been investigated by Tumblin~\etal~in~\cite{tumblin1993tone} and later by Tumblin~\etal~in~\cite{tumblin1999two} and Mantiuk~\etal~in~\cite{mantiuk2008display}, who employed simple parametric functions with limited expressiveness. More generally, HDR image tone mapping can be formulated as a constrained optimization problem~\cite{ma2015high,laparra2017perceptually}:
\begin{align} \label{eq:optimization}
I^{\star} = \mathop{\arg\min}_{I}\ell(S,I), \quad {\rm{s.t.}} \quad I \in \mathcal{C},
\end{align} 
 where $S$ denotes a photometrically calibrated HDR image, and $\mathcal{C}$ is the set of feasible tone-mapped images given physical constraints (\eg, the minimum and maximum luminances of a given display device). $\ell(\cdot,\cdot)$ denotes an objective metric that is capable of measuring the perceptual distance between two images of different dynamic ranges. $I^{\star}$ is the optimal tone-mapped image under the criterion $\ell(S,\cdot)$. Note that traditional objective quality metrics such as the mean squared error (MSE), the structural similarity (SSIM) index~\cite{wang2004image}, and HDR visible difference predictor (HDR-VDP)~\cite{Mantiuk04Visible, Mantiuk2011VDP2} are not suitable here because they assume that the two images being compared have the same dynamic range (see Sec. \ref{sec:ablation experiment} and Fig. \ref{fig:optimization}). Common choices for $\ell(\cdot,\cdot)$ include TMQI~\cite{yeganeh2012objective} and NLPD~\cite{laparra2017perceptually}. Due to the non-convexity of TMQI and NLPD and the high-dimensionality of the constrained optimization problem, gradient-based iterative solvers were originally proposed, which are computationally prohibitive.

\begin{figure*}[t]
\centering
\includegraphics[width=0.98\linewidth]{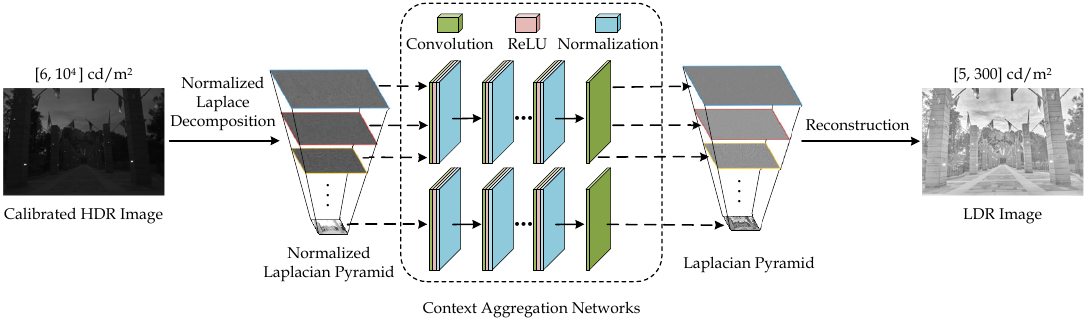}
\vspace{-0.5em} 
   \caption{The schematic diagram of the proposed tone mapping network. The input HDR image is first decomposed into a normalized Laplacian pyramid. All bandpass channels and the highpass channel share the same DNN, whereas the lowpass channel has its own. The outputs of the two DNNs constitute the Laplacian pyramid of the corresponding LDR image, which has a displayable luminance range of $[5, 300]$ $ \rm cd/m^{2}$. The HDR image is represented by simple linear rescaling.}
\label{fig:tmo}
\vspace{-0.6em} 
\end{figure*}

\subsubsection{DNN-based TMOs}
The primary effort of many DNN-based TMOs~\cite{rana2020deep,yang2021deep} is to create a number of ground-truth LDR images for paired training in the LDR domain. Montulet~\etal~\cite{Montulet2019DeepLF} and Zhang~\etal~\cite{zhang2019hsv} applied a list of existing TMOs to each HDR image, and selected the best tone-mapped one in terms of TMQI as the ground-truth. Many subsequent studies have followed this path~\cite{rana2020deep}, except for Yang~\etal~\cite{yang2021deep}, who resorted to formal subjective experiments for the best image selection. Panetta~\etal~\cite{Panetta2021} trained the tone mapping network over a combination of low-light datasets, which contain the ground-truth normal-light images. Despite the effort, the created ground-truths may be biased by the adopted objective metrics or human annotators. For instance, although TMQI performs well in \textit{quality assessment} of tone-mapped images, it has its own ``blind spots,'' especially when used as a \textit{perceptual optimization} objective (see Sec. \ref{sec:ablation experiment} and Fig. \ref{fig:optimization}). As a consequence, different combinations of loss functions have been proposed to encourage the creation of better-quality images in a rather ad hoc way. Candidate losses for combination include mean absolute error (MAE), MSE, gradient profile loss~\cite{Panetta2021}, and VGG content loss~\cite{Montulet2019DeepLF,rana2020deep,yang2021deep}. Zhang~\etal~\cite{Zhang2021} proposed a semi-supervised learning scheme, employing the adversarial loss and the cycle-consistency loss to match the distribution of high-quality LDR images. Vinker~\etal~\cite{vinker2021unpaired} achieved tone mapping with a deep generative adversarial network, where the structural similarity is enforced by patch-wise Pearson correlation. Instead of working in the LDR domain with difficult-to-obtain ground-truths, we perform perceptual optimization of HDR image tone mapping in a cross-dynamic-range setting, where we treat the available HDR image containing richer information of the captured natural scene as the ground-truth. This is made possible by cross-dynamic-range quality metrics such as NLPD~\cite{laparra2017perceptually}.

\subsection{MEF Methods}
MEF refers to a class of techniques that fuse a sequence of LDR images with different exposures into a single high-quality LDR image with a better overall appearance~\cite{mertens2009exposure}. The prevailing scheme for MEF follows a weighted summation framework, where each exposure image is associated with a weight map of the same size. Burt and Adelson proposed the Laplacian pyramid in 1983~\cite{burt1983laplacian}, which has a profound impact on MEF~\cite{burt1993enhanced, mertens2009exposure}. To reproduce or enhance the local details, various edge-preserving filters, including bilateral filter~\cite{tomasi1998bilateral} and guided filter~\cite{He2013Guided}, have been used for weight map computation. Entering the era of deep learning, a similar trend in the MEF field has been observed that researches tried to specify ground-truth fused images~\cite{Cai18learning} and to combine various loss functions~\cite{MA2020Infrared, zhang2020IFCNN, Yang2021GANFuse} so as to enable end-to-end optimization of MEF networks. 

Switching to quality assessment, Ma~\etal~\cite{ma2015perceptual} developed one of the first quality metrics - MEF-SSIM, and successfully applied it to perceptual optimization of MEF methods in the space of raw pixels~\cite{MA2018} and DNN parameters~\cite{ma2020deep}, respectively. In Stage two and as part of the self-calibration procedure, PS-TMO generates a sequence of LDR images, which can be treated as a pseudo-multi-exposure image stack because they correspond to the same HDR scene with different (simulated) maximum luminances. Similar techniques for pseudo-multi-exposure generation have been widely practiced in the related field of inverse tone mapping~\cite{endoSA2017}.

\section{Proposed PS-TMO}\label{sec:method}
In this section, we describe the proposed PS-TMO for HDR image tone mapping, which is perceptually optimized, self-calibrated, and computationally efficient. Fig.~\ref{fig:tmo} and Fig.~\ref{fig:fusionnetwork} together show the schematic diagrams. In Stage one, after preprocessing, we decompose the color-space-transformed and randomly photometrically calibrated HDR image into a normalized Laplacian pyramid, and input it to the \textit{tone mapping network}, consisting of two DNNs for Laplacian pyramid estimation. In Stage two, we use the trained tone mapping network to self-calibrate the input HDR image by producing a pseudo-multi-exposure image stack out of it  with different maximum luminances. We then train a \textit{fusion network} for weight map estimation. The final high-quality LDR image is computed by a weighted fusion.

\subsection{Stage One: Tone Mapping Network}
\subsubsection{Preprocessing}
It is pivotal for PS-TMO to work with \textit{photometrically calibrated} HDR images, meaning that all pixels record the true luminance values (in the unit of candela per square meter, $\rm cd/m^{2}$). This is because the responses of the HVS to different light levels are highly non-linear~\cite{carandini2012}. HDR image calibration (also known as the photometric calibration) allows TMOs to make correct distinctions between bright and dim scenes. Otherwise, a day-lit HDR image in arbitrary luminance units may be tone-mapped to a night scene with loss of structural details. However, in the real world, the majority of HDR images circulated on the Internet are acquired without calibration, in which the recorded measurements $R$ are linearly proportional to the true luminances $S$ with an unknown scaling factor. To apply HVS-based TMOs to an uncalibrated HDR image, educated guesses about the minimum and maximum luminances of the original scene~\cite{laparra2017perceptually}, denoted by $S_{\mathrm{min}}$ and $S_{\mathrm{max}}$, respectively, need to be made. Nevertheless, this is by itself a very challenging computer vision task. One significant advantage of PS-TMO is that during training the tone mapping network, the HDR scenes can be calibrated with arbitrary minimum and maximum luminances as a form of data augmentation, followed by self-calibration through MEF in Stage two to generate the final HDR image. After specifying $S_{\mathrm{min}}$ and $S_{\mathrm{max}}$, we convert the HDR image from the RGB to HVS color space~\cite{fattal2002gradient}, and linearly rescale the luminance measurements:
\begin{align} \label{eq:calibration}
&\bar{R}=\frac{R-R_{\mathrm{min}}}{R_{\mathrm{max}}-R_{\mathrm{min}}} \in [0,1],\\
&S=(S_{\mathrm{max}}-S_{\mathrm{min}})\cdot \bar{R}+S_{\mathrm{min}}.
\end{align}
We then decompose the ``calibrated'' luminance channel into the normalized Laplacian pyramid~\cite{laparra2017perceptually}. 

\subsubsection{Network Architecture}
The core of our tone mapping network are two DNNs to estimate the Laplacian pyramid of the LDR image from the normalized Laplacian pyramid of the input HDR image. One DNN is shared to process all bandpass channels and the highpass channel, while the other is reserved for the lowpass channel. From a number of alternative networks, we employ the context aggregation network (CAN)~\cite{yu2015multi,chen2017fast}  as our default architecture, which has been used to approximate and accelerate a wide range of image processing applications, including $\ell_{0}$ smoothing, style transfer, and pencil drawing. It allows receptive field expansion without compromising spatial resolution, which effectively aggregates global context information. The two CANs share the same architecture with six convolution layers, whose outputs have the same resolution as the inputs. The details are specified in Table~\ref{tab:CAN_conf}, which are manually optimized to be highly lightweight. Convolutions, except the last one, are followed by the adaptive normalization (AN):
\begin{align} 
\label{eq:AN1}
    \mathrm{AN}(Z)=\lambda_{1}Z+\lambda_{2}\mathrm{BN}(Z),
\end{align}
where $\lambda_{1}$ and $\lambda_{2} \in \mathbb{R}$ are two learnable parameters, and $Z$ denotes  intermediate representation. The weight sharing across all bandpass channels and the highpass channel allows PS-TMO to process a normalized Laplacian pyramid of arbitrary levels. We employ the leaky rectified linear unit (LReLU) as the non-linear activation function (also known as the half-wave rectification in the signal processing field):
\begin{align} 
\label{eq:relu}
    \mathrm{LReLU}(Z)=\max(\lambda_{3}Z,Z),
\end{align}
where the parameter $0\leq\lambda_{3}<1$ is made fixed during training. The lowpass channel is compressed by the other CAN with the same architecture. The output LDR image, constrained to have a luminance range of $[5, 300]$ $\rm cd/m^{2}$, is reconstructed by collapsing the estimated Laplacian pyramid from the tone mapping network. In other words, we assume a fixed display device with the minimum and maximum luminances of $I_\mathrm{min} = 5$ and $I_\mathrm{max} = 300$, respectively, which are typical specifications of consumer-grade displays of standard dynamic ranges.

\begin{table}[t]
	\centering
	\caption{Specification of the two CANs in PS-TMO for tone mapping in Stage one. Exclusion of the bias terms makes PS-TMO scaling-invariant, which improves generalization to unseen luminance levels}
	\begin{tabular}{l|c c c c c c}
		\toprule
		Layer & 1 & 2 & 3 & 4 & 5  & 6 \\
		\midrule
        Convolution & 3 & 3 & 3 & 3 & 3  & 3  \\
		Dilation & 1 & 2 & 4 & 8 & 1  &1 \\
		Width & 32 & 32 & 32 & 32 & 32  & 1 \\
		Bias & \XSolidBrush & \XSolidBrush & \XSolidBrush & \XSolidBrush & \XSolidBrush & \XSolidBrush\\
		Adaptive Normalization & \Checkmark & \Checkmark & \Checkmark & \Checkmark & \Checkmark  & \XSolidBrush \\
		LReLU Non-linearity & \Checkmark & \Checkmark & \Checkmark & \Checkmark & \Checkmark & \XSolidBrush\\
		\bottomrule
	\end{tabular}
	\label{tab:CAN_conf}
 \vspace{-1em} 
\end{table}

\begin{figure}[t]
\centering
\addtocounter{subfigure}{0}
\subfloat[With bias terms]{\includegraphics[height=0.39\linewidth]{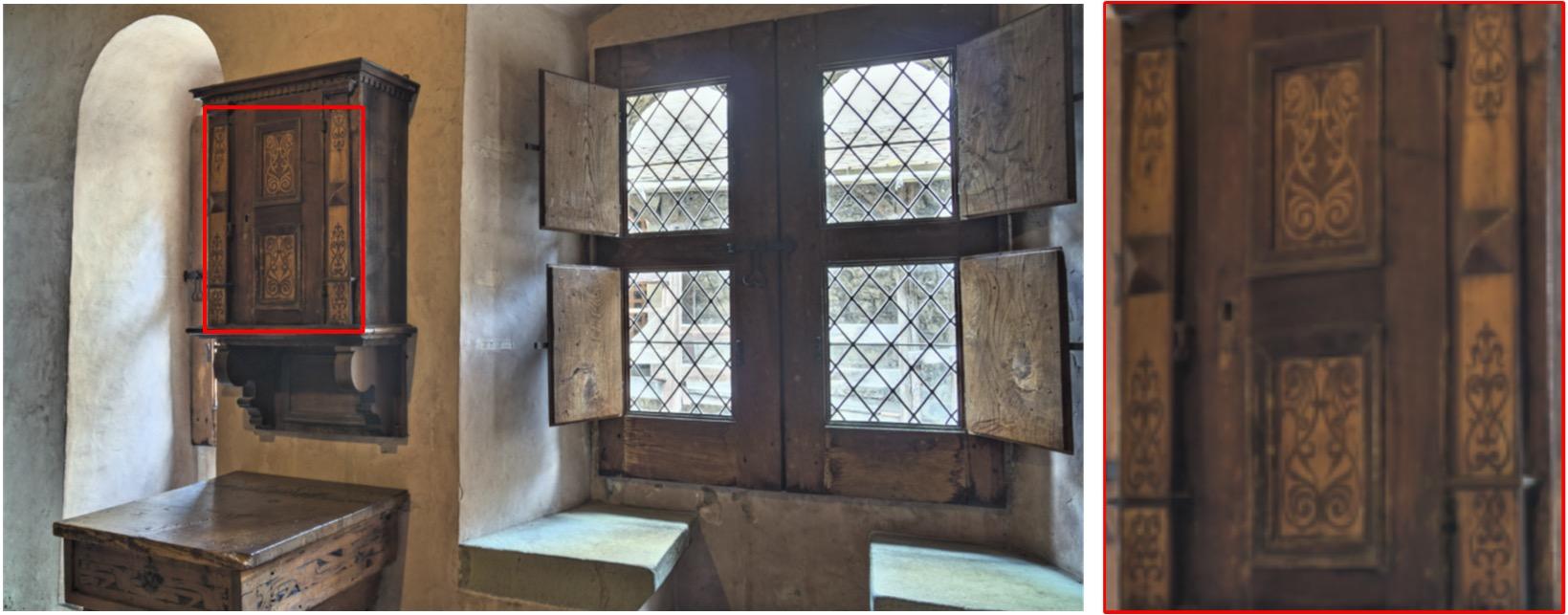}}\\ \vspace{-0.5em} 
\subfloat[Without bias terms]{\includegraphics[height=0.39\linewidth]{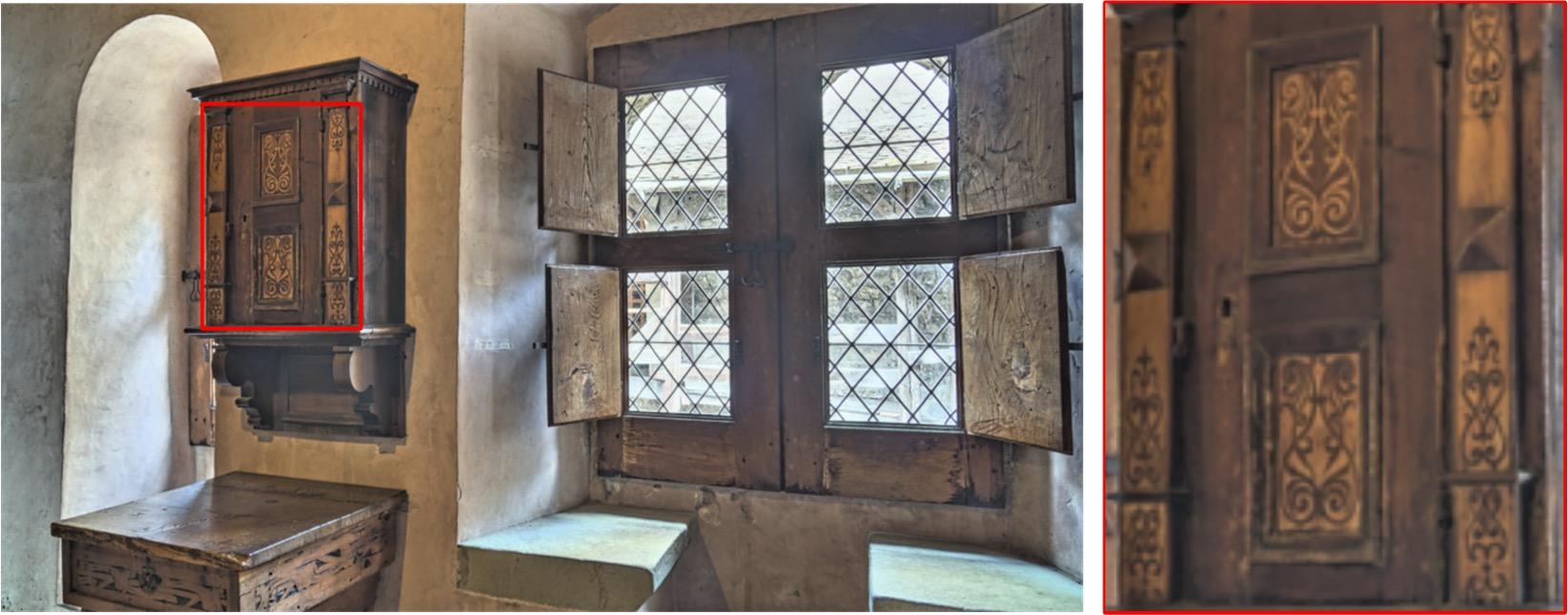}}\\ 
   \caption{Visual comparison of PS-TMO with and without bias terms. The bias-free PS-TMO is robust to the ``Old House'' scene with a higher dynamic range, which is not seen during training, and produces the tone-mapped image with more faithful local structures.}
\label{fig:bias}
\vspace{-0.6em} 
\end{figure}

A worth-mentioning difference of our tone mapping network compared to the original CAN~\cite{chen2017fast} is that all bias terms, including those in adaptive normalization, are removed. As proved
 in~\cite{mohan2019robust}, a bias-free DNN with piece-wise linear activation function (\eg, LReLU) is \textit{scaling-invariant}. That is, if the input is rescaled by a constant value, the output will be rescaled by the same amount:
\begin{align} 
\label{eq:rescale}
    g(\alpha Z_{j})=\alpha g(Z_{j}), 
\end{align}
where $j$ indexes the coefficients of the intermediate representation $Z$. Empirically, scaling-invariance renders the tone mapping network more robust to maximum luminance variations during training and testing (see Fig.~\ref{fig:bias}).

\subsubsection{Perceptual NLPD as the Loss Function}
The NLPD metric, proposed in~\cite{laparra2017perceptually} and adopted as the objective function for our tone mapping network, is inspired by the physiology of the early visual system. Specifically, the luminances of the calibrated HDR image $S$ are firstly preprocessed by a power function, approximating the transformation of light to the response of retinal photoreceptors~\cite{laparra2017perceptually}: 
\begin{align} \label{eq:gammaS}
    S^{(1)} = S^{\gamma}.
\end{align}
After that, $S^{(1)}$ is partitioned recursively into frequency subbands via luminance subtraction, which mimics the center-surround receptive fields in the retina and the lateral geniculate nucleus~\cite{laparra2017perceptually}:
\begin{align}\label{eq:pyramid}
    &X^{(i+1)} = DLX^{(i)}, \quad i \in \left \{1,\ldots,M-1 \right\}, \\
    &Z^{(i)} = X^{(i)}-LUX^{(i+1)}, \\
    &Z^{(M)} = X^{(M)},
\end{align}
where $D$ and $U$ represent linear down-/up-sampling operations, respectively, and $L$ denotes the lowpass filter, which is inherited from the Laplacian pyramid~\cite{burt1983laplacian}. $M$ is the number of pyramid levels. The normalized Laplacian pyramid can be computed by dividing each coefficient with a weighted summation of neighboring coefficients (plus a constant) within each subband:
\begin{align} \label{eq:norm}
    Y^{(i)} = Z^{(i)} \oslash (P\vert Z^{(i)}\vert + C_0),
\end{align}
where $\oslash$ represents the Hadamard division, and $P$ is a convolution filter optimized to eliminate the statistical redundancies~\cite{laparra2017perceptually}. $C_0$ is a small positive constant to avoid potential division by zero.  
The normalized Laplacian pyramid representations of the HDR and tone-mapped images can be expressed as
\begin{align} \label{eq:tone-mapped}
    f(S) = \left\{Y^{(i)}\right\}_{i=1}^{M} \mbox{ and } f(I) = \left\{\tilde{Y}^{(i)}\right\}_{i=1}^{M},
\end{align}
based on which we compute the NLPD metric:
\begin{align} \label{eq:nlpd}
\ell(S,I) = \left[\frac{1}{M} \sum_{i=1}^{M}\left(\frac{1}{n^{(i)}}\sum_{j=1}^{n^{(i)}}\vert Y_{j}^{(i)}-\tilde{Y}_{j}^{(i)}\vert^{\alpha}\right)^{\frac{\beta}{\alpha}}\right]^{\frac{1}{\beta}},
\end{align}
where $n^{(i)}$ denotes the number of coefficients in the $i$-th subband. The two exponents $\alpha$ and $\beta$ are applied to each frequency subband and for all subbands, respectively, which are optimized to match the human perception of image quality on a subject-rated image quality database~\cite{ponomarenko2009tid2008}. 

\begin{figure}[t]
\centering
\includegraphics[width=\linewidth]{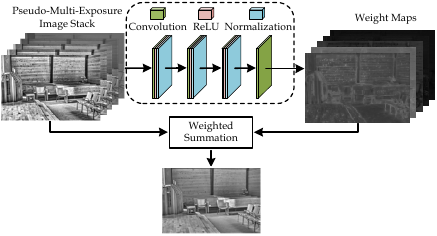}
\vspace{-1em} 
   \caption{The schematic diagram of the fusion network for self-calibrating the HDR image and producing final LDR image.  It accepts the pseudo-multi-exposure image stack corresponding to the same HDR image calibrated with different maximum luminances, and estimates a set of weight maps that highlight perceptually important local regions. The final LDR image is obtained by a weighted summation of the LDR image stack.}
\label{fig:fusionnetwork}
\vspace{-0.6em} 
\end{figure}

\begin{figure*}[t]
\centering
\subfloat{\includegraphics[width=0.975\linewidth]{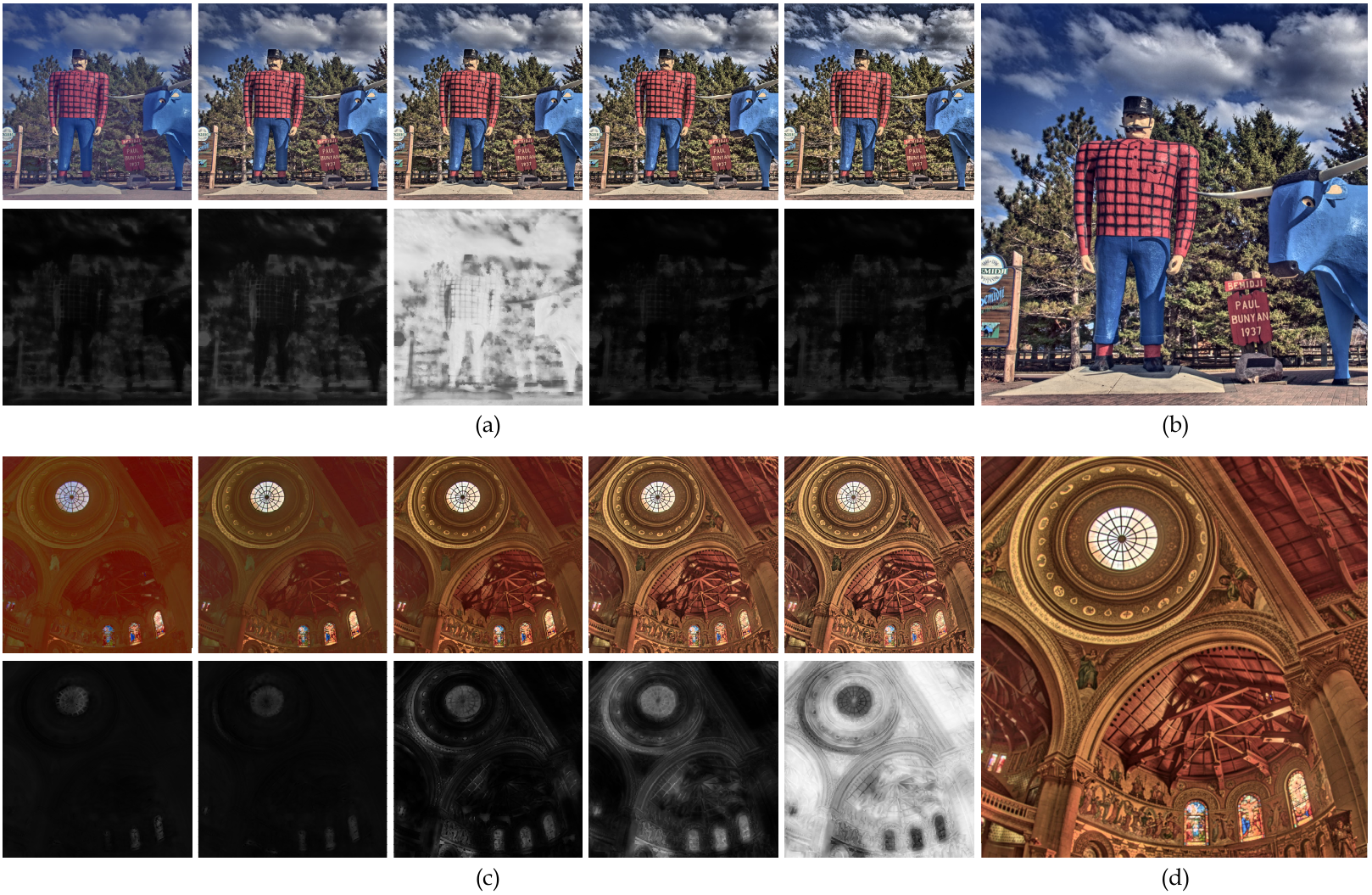}}\\
\vspace{-0.5em} 
   \caption{Pseudo-multi-exposure image stacks produced by the tone mapping network and the corresponding weight maps, together with the final LDR images by the fusion network. A brighter pixel in the weight map indicates that the corresponding LDR image pixel contributes more to the final image. It is clear that images with higher maximum luminances are tone-mapped with richer structural details but also severer degrees of noise, while images with lower maximum luminances are more color-saturated and less detailed. The fusion network optimized by a variant of MEF-SSIM is able to generate reasonable weight maps that show a strong preference for clean, high-contrast, well-exposed, and well-saturated patches. As a result, the output images combine the best perceptual aspects of the LDR image stacks.  \textbf{(a)} and \textbf{(c)}: LDR image stacks and the learned weight maps  of ``Paul Bunyan 
  and Babe the Blue Ox Statues" and ``Stanford Memorial Church'' HDR scenes, corresponding to  maximum luminances of $10^{3}$, $10^{4}$, $10^{5}$, $10^{6}$, and $10^{7}$ $ \rm cd/m^{2}$, respectively.  \textbf{(b)} and \textbf{(d)}: Output LDR images.}
\label{fig:fusion}
\vspace{-0.6em} 
\end{figure*}

\subsection{Stage Two: Fusion Network}\label{subsec:fusion}
\subsubsection{Generation of Pseudo-Multi-Exposure Image Stack}\label{subsubsec:pg}
The forward propagation of the tone mapping network and the computation of the NLPD metric in Stage one require the exact specification of the minimum and maximum luminances for uncalibrated HDR scenes. While most TMOs are fairly robust to the minimum luminance $S_\mathrm{min}$, making its setting straightforward\footnote{Throughout this paper, we set $S_\mathrm{min} = I_\mathrm{min}=5$ $\rm cd/m^2$.}, this is not the case for the maximum luminance $S_\mathrm{max}$, which involves extensive human expertise and is thus time-consuming. Empirically, a higher estimated $S_\mathrm{max}$ means that more simulated light is cast into the original scene~\cite{laparra2017perceptually}, leading to better visibility of local structures, especially in dark regions (see Fig.~\ref{fig:fusion} (a) and (c)). As there is no free lunch in computational photography, the measurement noise is also likely to be amplified, when $S_\mathrm{max}$ is set extremely high (\eg, $S_\mathrm{max} = 10^7$ $\rm cd/m^2$). Here, instead of manually picking a scene-dependent $S_\mathrm{max}$ as done previously~\cite{laparra2017perceptually,chenyang2021}, we make PS-TMO fully automatic, that is, the HDR image can be \textit{self-calibrated} by means of MEF. Specifically, we first linearly sample $K$ maximum luminances from the range of $[10^3, 10^7]$ $\rm cd/m^{2}$ in the logarithmic scale, and calibrate the HDR image with each of the $K$ values. We then feed the calibrated images to the trained tone mapping network to create a sequence of $K$ candidate LDR images, which we call the pseudo-multi-exposure image stack, and will be fused to produce the final LDR image. We use the prefix ``pseudo'' because the image stack does not contain under- and over-exposure distortions, but instead may suffer from color saturation and noise artifacts. We will take advantage of these distortion characteristics to make a slight modification of the MEF-SSIM metric~\cite{ma2015perceptual}.

\subsubsection{Network Architecture}
As shown in Fig.~\ref{fig:fusionnetwork}, our fusion network is also implemented by a CAN that predicts the weight maps $\left\{W^{(k)}\right\}$ with the same resolution of the input  pseudo-multiple-exposure image stack $\left\{I^{(k)}\right\}$. The network specification is given in Table~\ref{tab:CAN_fusion}, which is shallower than the tone mapping network. The parameters are shared by all pseudo-exposure images, allowing an arbitrary-length stack to be handled. The last layer predicts the weight maps, which are used to compute the final output image by a weighted summation:
\begin{align} 
\label{eq:ws}
    F=\sum_{k=1}^{K}W^{(k)} \odot I^{(k)},
\end{align}
where $\odot$ denotes the Hadamard product.

\subsubsection{Perceptual MEF-SSIM Variant as the Loss Function}
The MEF-SSIM metric proposed in~\cite{ma2015perceptual} provides an accurate quality characterization of multi-exposure fused images. It first decomposes an image patch $x^{(k)}$ into three conceptually decorrelated components - mean intensity, signal contrast, and signal structure:
\begin{align} \label{eq:spd}
    x^{(k)} & = \Vert x^{(k)}-\mu_{{k}}\Vert_2 \cdot \frac{x^{(k)}-\mu_{k}}{\Vert x^{(k)}-\mu_{k}\Vert_2} + \mu_{k}\nonumber \\
    & = \Vert\tilde{x}^{(k)}\Vert_2 \cdot \frac{\tilde{x}^{(k)}}{\Vert\tilde{x}^{(k)}\Vert_2} +\mu_{k}\nonumber \\
    & = c_{k} \cdot s_{k}+l_{k},
\end{align}
where $\Vert \cdot \Vert_2$ denotes the $\ell_{2}$-norm. $l_k = \mu_{k}$, $c_{k}=\Vert \tilde{x}^{(k)}\Vert_2$, and $s_{k}=\frac{\tilde{x}^{(k)}}{\Vert \tilde{x}^{(k)}\Vert_2}$  represent the mean intensity, the signal contrast, and the signal structure, respectively.
MEF-SSIM computes the intensity of the desired patch by
\begin{align} \label{eq:di}
    \hat{l}=\frac{\sum_{k=1}^{K}w_{l}(g_{k},l_{k})\mu_{k}}{\sum_{k=1}^{K}w_{l}(g_{k},l_{k})},
\end{align}
where $w_{l}(\cdot)$ is specified by a two-dimensional Gaussian to measure the well-exposedness:
\begin{align} \label{eq:gaussian}
    w_{l}(g_{k},l_{k})=\mathrm{{exp}}\left(-\frac{(g_{k}-\tau)^{2}}{2\sigma^{2}_{g}}-\frac{(l_{k}-\tau)^{2}}{2\sigma^{2}_{l}}\right).
\end{align}
$\sigma_{g}$ and $\sigma_{l}$ are the variances as a measure of the spread, and $\tau = 0.5$ stands for the mid-intensity value in the range of $[0,1]$. The desired contrast is defined as the highest one across all exposures: 
\begin{align} \label{eq:contrast}
    \hat{c}=\max\limits_{1\leq k\leq K}c_{k}.
\end{align}

\begin{table}[t]
	\centering
	\caption{Specification of the CAN in PS-TMO for HDR image self-calibration and final LDR image generation in Stage two}
	\begin{tabular}{l|c c c c c c c}
		\toprule
		Layer & 1 & 2 & 3 & 4 \\
		\midrule
        Convolution & 3 & 3 & 3 & 1  \\
		Dilation & 1 & 2 & 4 & 1 \\
		Width & 24 & 24 & 24 & 1 \\
		Bias & \XSolidBrush & \XSolidBrush & \XSolidBrush & \Checkmark \\
		Adaptive Normalization & \Checkmark & \Checkmark & \Checkmark & \XSolidBrush \\
		LReLU Non-linearity & \Checkmark & \Checkmark & \Checkmark & \XSolidBrush\\
		\bottomrule
	\end{tabular}
	\label{tab:CAN_fusion}
 \vspace{-1em} 
\end{table}

The desired structure is calculated by a weighted summation followed by $\ell_2$-normalization:
\begin{align} \label{eq:structure}
    \hat{s} = \frac{\bar{s}}{\Vert \bar{s} \Vert_2}, \quad \mathrm{{where}} \quad \bar{s}=\frac{\sum_{k=1}^{K}w_{s}(\tilde{x}^{(k)})s_{k}}{\sum_{k=k}^{K}w_s(\tilde{x}^{(k)})}.
\end{align}
In the original MEF-SSIM for perceptual optimization~\cite{MA2018,ma2020deep},
$w_{s}(\cdot)$ is a Kronecker delta function that identifies the structure vector corresponding to the maximum contrast (\ie, $\hat{c}$). This is primarily motivated to avoid selecting under- or over-exposed patches with essentially no structural information. However, such choice is not wise for the pseudo-multi-exposure image stack created in Sec. \ref{subsubsec:pg}, where the structure vector with the maximum contrast is highly likely to contain (amplified) measurement noise (see Fig.~\ref{fig:fusion} (a)), while under- or over-exposure distortions are in fact not present. Here, we propose a simple yet effective remedy for MEF-SSIM: change the Kronecker delta function $w_{s}(\cdot)$ to select the structure vector with the \textit{median} signal contrast instead. As shown in Fig.~\ref{fig:fusion}, the fusion network optimized by the MEF-SSIM variant is able to generate weight maps that show a strong preference for clean, high-contrast, well-exposed, and well-saturated patches. 

The remaining construction of MEE-SSIM is left intact, where we first compute the desired image patch by fusing the three components:
\begin{align}\label{eq:fused}
    \hat{x} =  \hat{c} \cdot \hat{s}+\hat{l},
\end{align}
and make SSIM-like local quality measurements:
\begin{align} \label{eq:ssim}
    S\left(\left\{x^{(k)}\right\},f\right) = \frac{(2\mu_{\hat{x}}\mu_{f}+C_{1})(2\sigma_{\hat{x}f}+C_{2})}{(\mu_{\hat{x}}^{2}+\mu_{f}^{2}+C_{1})(\sigma_{\hat{x}}^{2}+\sigma_{f}^{2}+C_{2})},
\end{align}
where $\mu_{\hat{x}}$ and $\mu_{f}$ represent the mean intensities of the desired patch $\hat{x}$ and a given fused patch $f$, respectively. $\sigma_{\hat{x}}$, $\sigma_{f}$,  and  $\sigma_{\hat{x}f}$ indicate the local variances of $\hat{x}$ and $f$, and their covariance, respectively. $C_{1}$ and $C_{2}$ are two small positive constants for numerical stability. The local quality measurements are averaged to produce an overall quality estimate 
for the fused image.

We conclude this subsection by further discussing the proposed self-calibration mechanism implemented by MEF. First, the final LDR image corresponds non-linearly to the input HDR image with a particular maximum luminance $S^\star_\mathrm{max}\in [10^3, 10^7]$ $\rm cd/m^{2}$, optimized for perceptual quality (as approximated by MEF-SSIM) rather than physical plausibility (\ie, the true maximum luminance as measured by a photometer). Second, our self-calibration mechanism can be made linear by selecting (in a post hoc way) a maximum luminance for photometric calibration corresponding to  the most important weight map, \ie, the map with the largest aggregated weight value, $\max_k \sum_j W^{(k)}_j$. As expected, we observe that the linearized self-calibration would slightly sacrifice the perceptual quality of the final LDR image.

\subsection{Color Reproduction}
As discussed before, both the tone mapping network and the fusion network work with the luminance channel due primarily to the fact that the two perceptual metrics NLPD~\cite{laparra2017perceptually} and MEF-SSIM~\cite{ma2015perceptual} accept grayscale images only.  To recover the color appearance of the final LDR image, we adopt the method in~\cite{tumblin1999boundary, fattal2002gradient}, which is also widely adopted by other recent methods~\cite{liang2018hybrid,chenyang2021}:
\begin{align} 
\label{eq:color1}
    F^{(c)} = \left(\frac{S^{(c)}}{S}\right)^{\rho}F,
\end{align}
where $c\in \{R, G, B\}$ indexes the RGB channels, and $\rho$ controls the color saturation.  $S$ and $F$ represent the luminance channel before and after tone mapping, respectively.

\subsection{Model Training and Testing}\label{sec:setting}
We collect $714$ HDR images mainly from~\cite{tumblin1993tone, Debevec97, fairchild2007hdr,froehlich2014, korshunov2014, Kundu2017, Nima2017, eilertsen2021cheat, Hanji2022}, among which $634$ are utilized for training while $80$ for testing. Random cropping and flipping have been employed to augment the training data.
We choose to train PS-TMO sequentially for fast convergence. We first optimize the tone mapping network by minimizing the NLPD metric with the default setting~\cite{laparra2017perceptually}. Specifically, the non-linearity parameter $\gamma$ in Eq.~\eqref{eq:gammaS} is set to $\frac{1}{2.6}$. For bandpass and highpass channels, the convolutional filter $P$ is set to $[0.05,0.25,0.4,0.25,0.05]$, and the constant $C_0$ in Eq.~\eqref{eq:norm} is set to $0.17$. For the lowpass channel, the filter $P$ is set to the identity matrix and $C_0$ is set to $4.86$. The two optimized exponents $\alpha$ and $\beta$ in Eq.~\eqref{eq:nlpd} are set to $2.0$ and $0.6$, respectively. We set the negative slope $\lambda_{3} = 0.2$ in  LReLU. During the training of the tone mapping network, each HDR image is arbitrarily calibrated by maximum luminances sampled randomly from $\{10^{3}, 10^{4}, 10^{5}, 10^{6}, 10^{7}\}$ $\rm cd/m^{2}$, and each calibrated image is decomposed into a normalized Laplacian pyramid with five levels.  

 Adam~\cite{Kingma2014adam} is employed as the stochastic optimizer with an initial learning rate of  $10^{-3}$ and a mini-batch size of $4$. We decay the learning rate every $1,000$ epochs by a factor of $10$, and we train the tone mapping network for a total of $2,000$ epochs. After Stage one optimization, we use the trained tone mapping network to create the pseudo-multi-exposure image stack for each HDR 
scene by setting the number $K=5$. That is, we sample the same five discrete luminance values $\{10^{3}, 10^{4}, 10^{5}, 10^{6}, 10^{7}\}$ $\rm cd/m^{2}$ for stack generation. It is noteworthy that the fusion network is designed to accept an LDR image stack of arbitrary length and resolution. The two Gaussian spread parameters in Eq.~\eqref{eq:gaussian} of MEF-SSIM are inherited from previous publications~\cite{ma2020deep, MA2018}: $\sigma_{g}=0.2$ and $\sigma_{l}=0.5$. The training of the fusion network is nearly identical to that of the tone mapping network, except that we allocate the LDR image stack along the batch dimension as an efficient implementation of parameter sharing. This makes the mini-batch size to be one.

During testing, we  keep the original size of the input HDR image, and calibrate it with five maximum luminance values $\{10^{3}, 10^{4}, 10^{5}, 10^{6}, 10^{7}\}$ $\rm cd/m^{2}$. We feed the calibrated HDR images of the same content to the tone mapping network to generate the pseudo-multi-exposure image stack, which will be subsequently fed into the fusion network to produce the final high-quality LDR image (followed by color reproduction with $\rho=0.6$ in Eq.~\eqref{eq:color1}).

\begin{figure*}[t]
  \centering
    \subfloat[Kim08]{\includegraphics[width=0.24\linewidth]{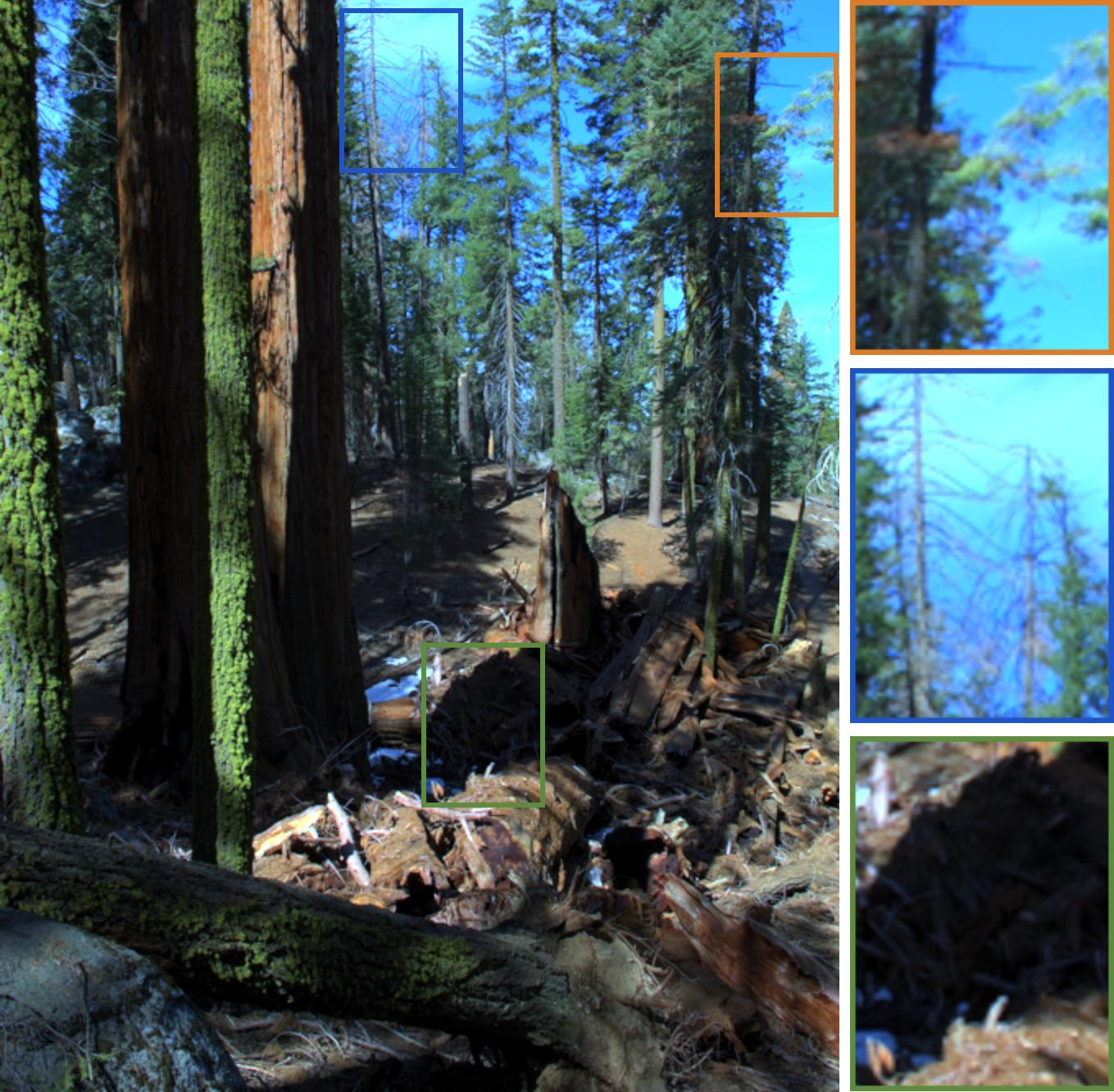}} \hskip0.3em
    \subfloat[WLS]{\includegraphics[width=0.24\linewidth]{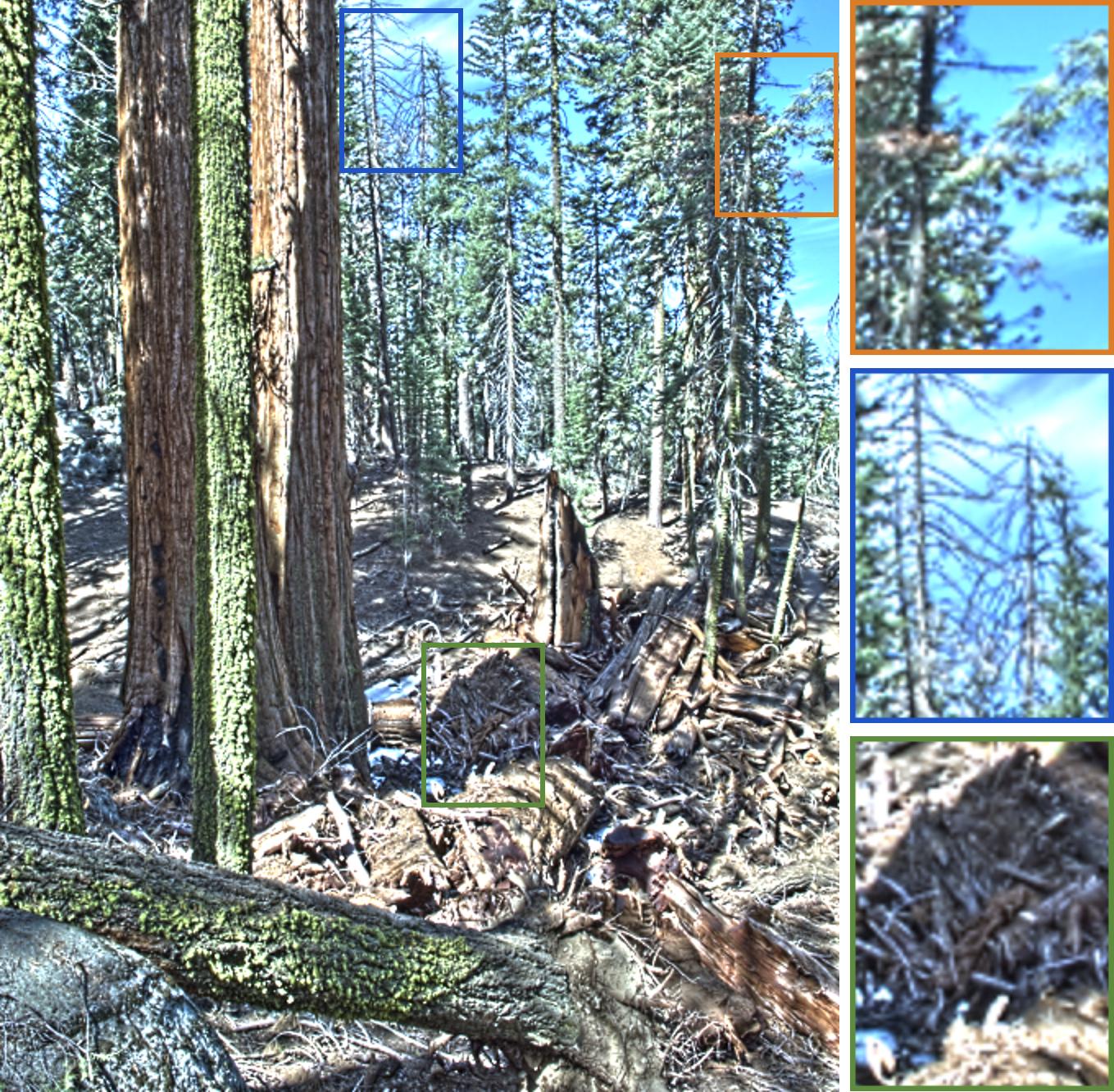}} \hskip0.3em
    \subfloat[GR]{\includegraphics[width=0.24\linewidth]{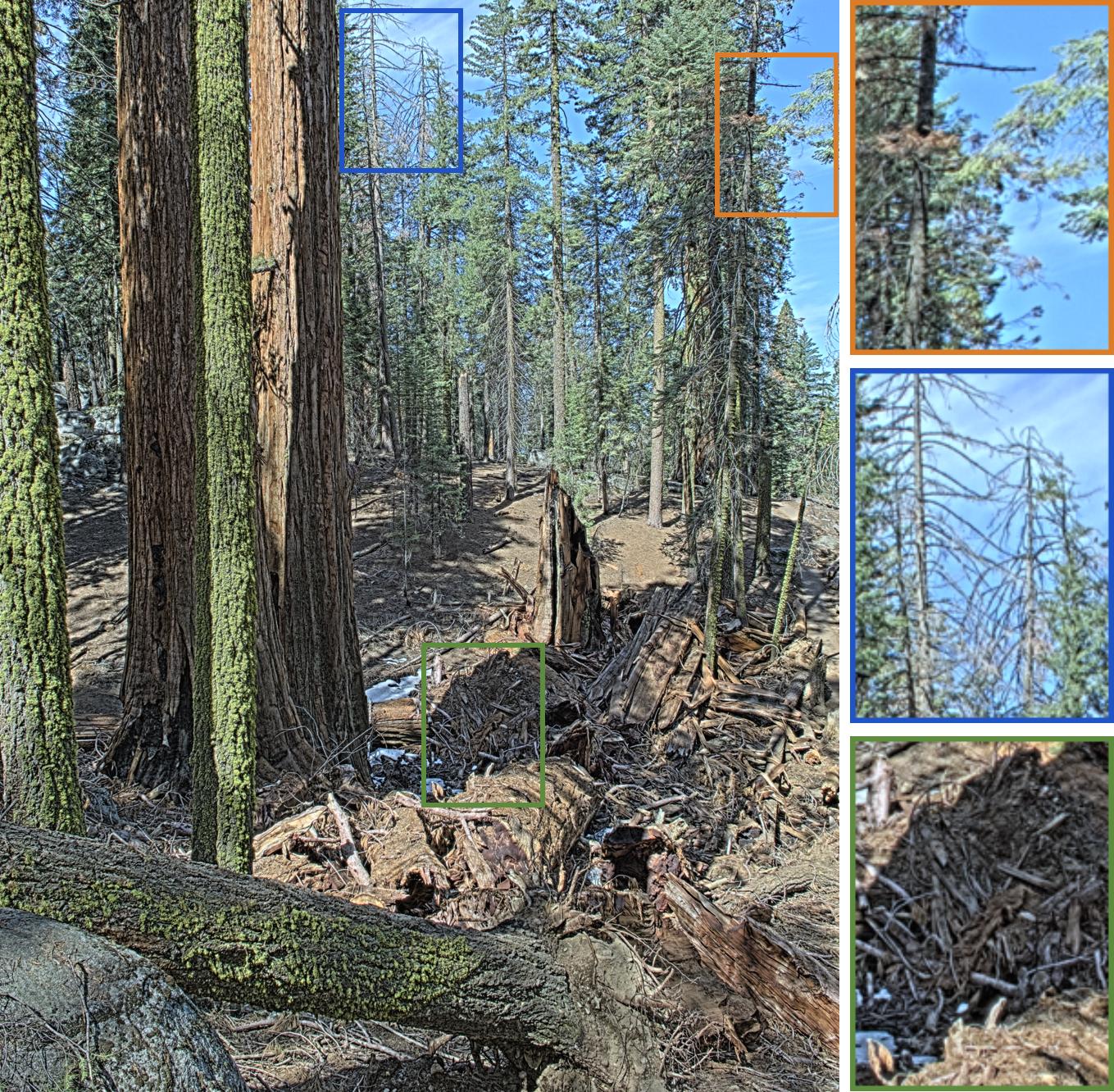}} \hskip0.3em
    \subfloat[Liang18]{\includegraphics[width=0.24\linewidth]{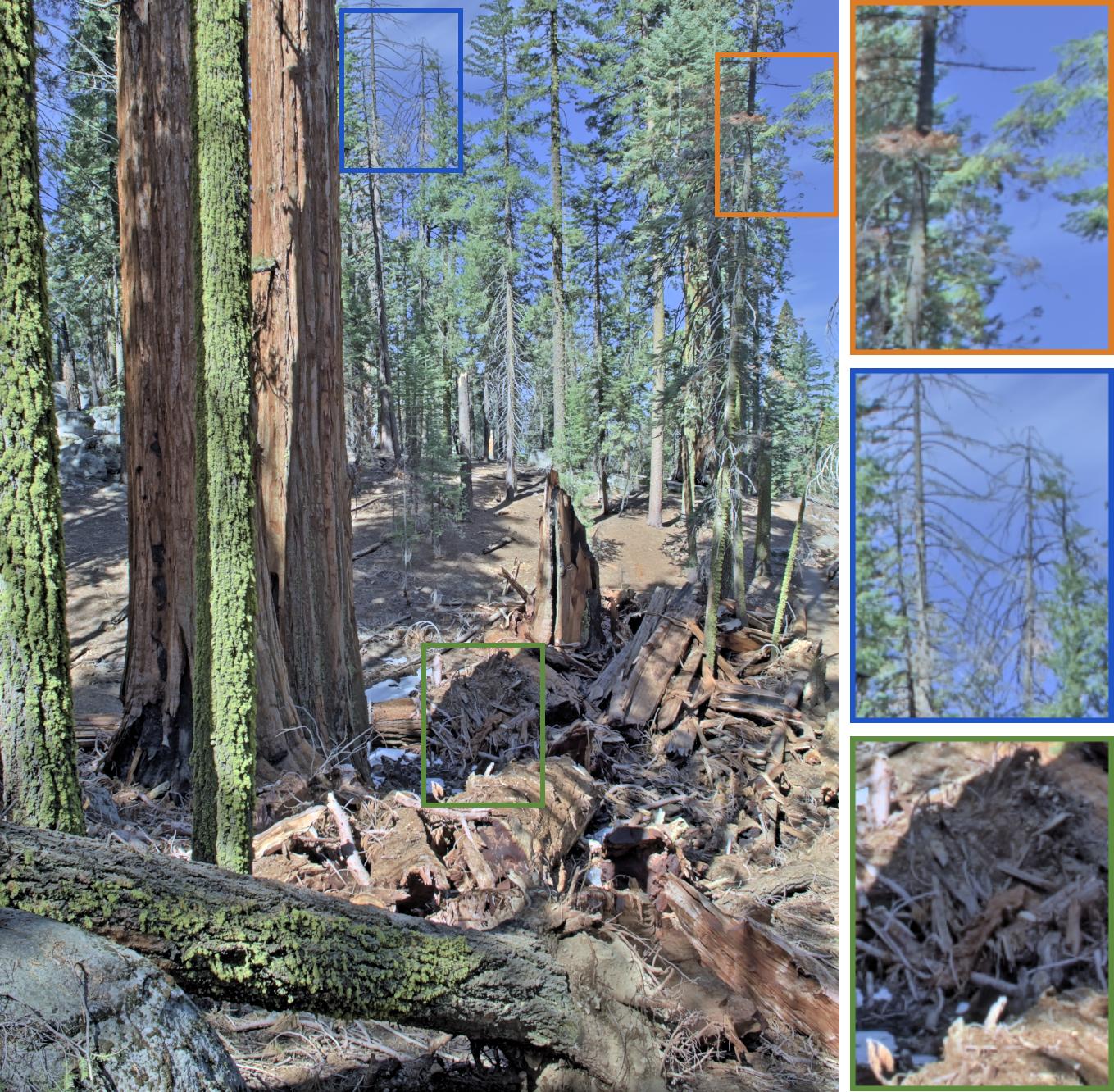}} \\\vspace{-0.5em}  
    \subfloat[Zhang20]{\includegraphics[width=0.24\linewidth]{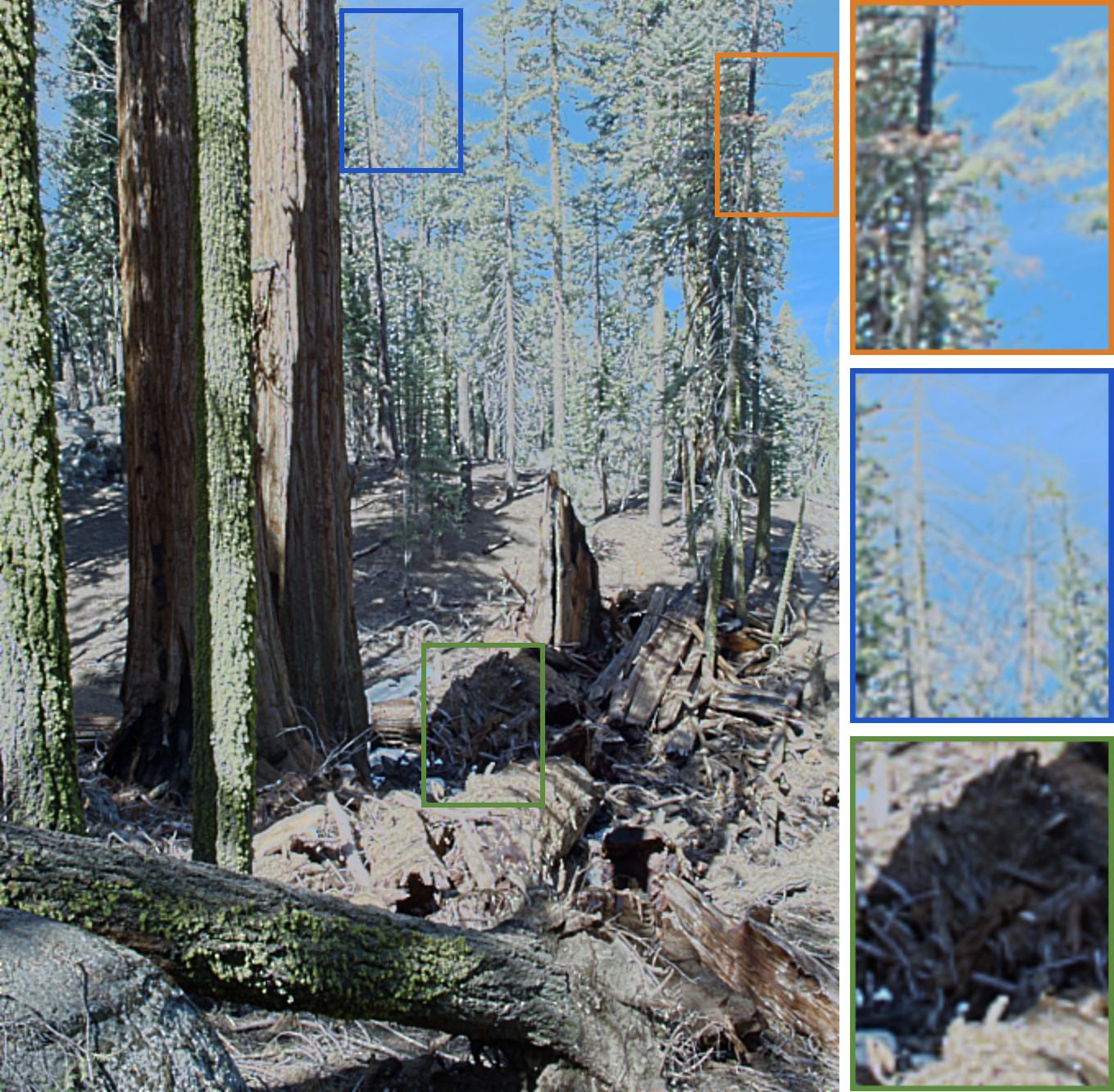}} \hskip0.3em
    \subfloat[Zhang21]{\includegraphics[width=0.24\linewidth]{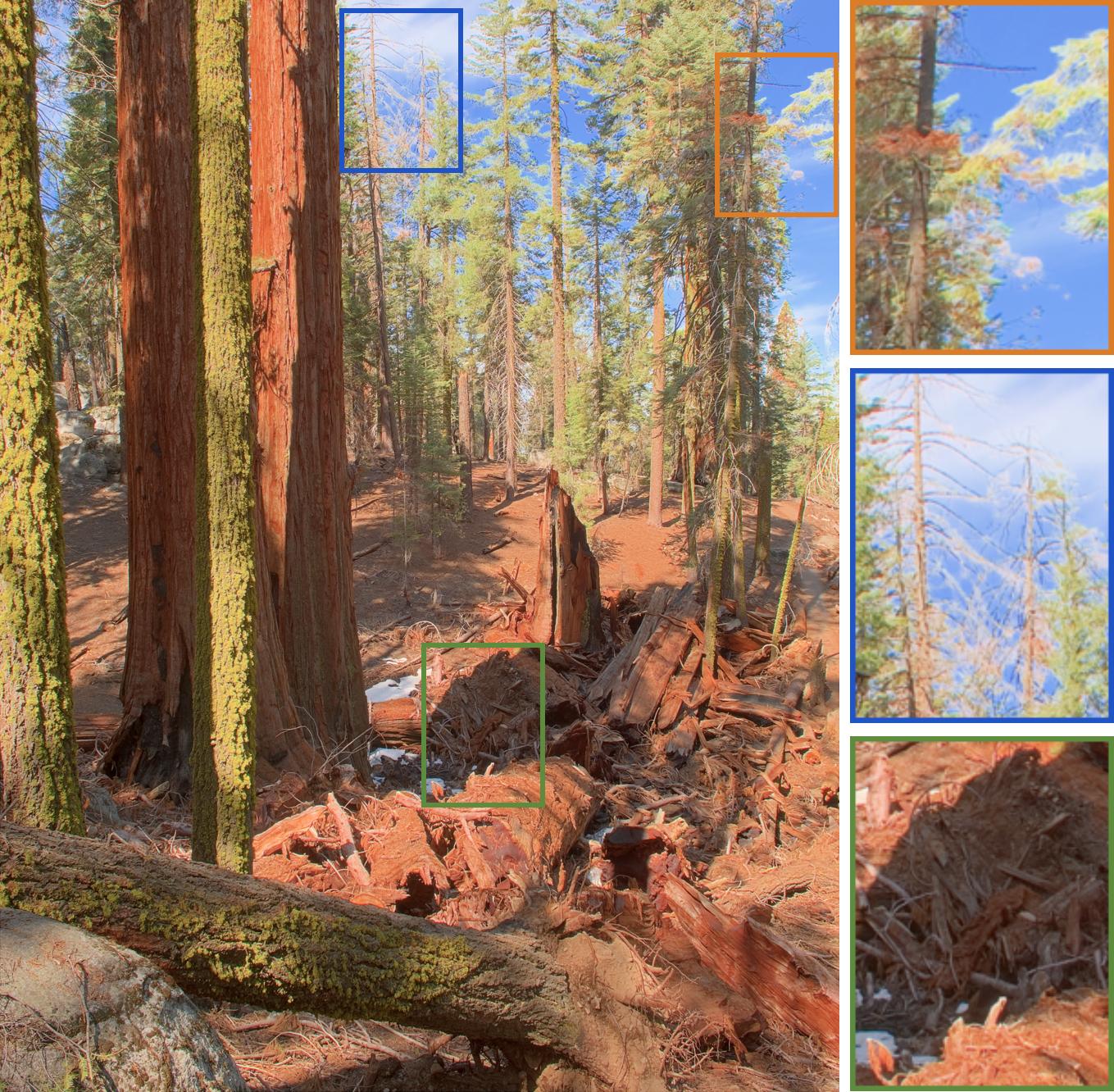}} \hskip0.3em
    \subfloat[Vinker21]{\includegraphics[width=0.24\linewidth]{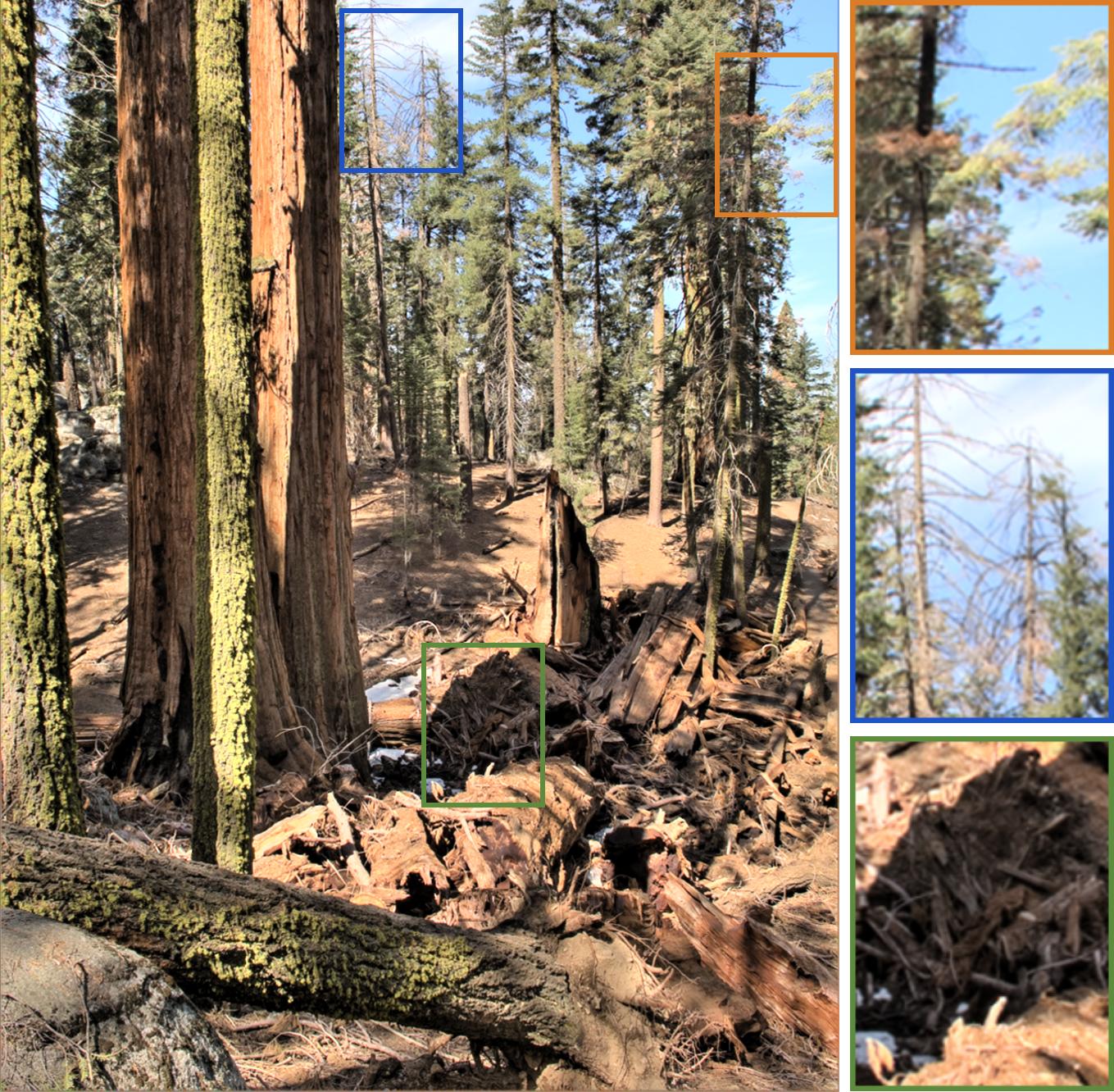}} \hskip0.3em
    \subfloat[PS-TMO]{\includegraphics[width=0.24\linewidth]{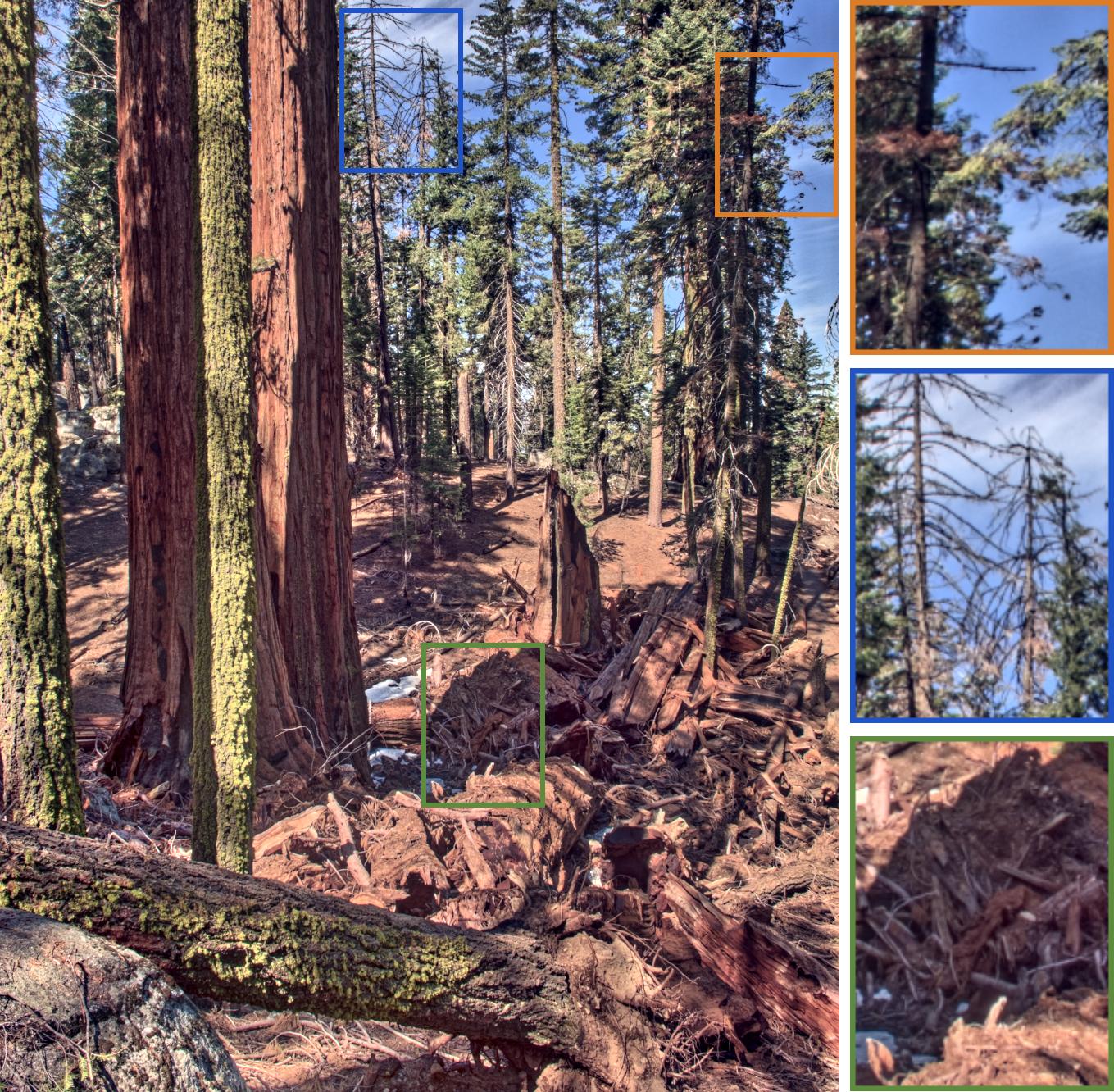}} \\ 
  \caption{Comparison of PS-TMO with Kim08~\cite{kim2008consistent}, WLS~\cite{farbman2008edge}, GR~\cite{shibata2016gradient}, Liang18~\cite{liang2018hybrid}, Zhang20~\cite{zhang2020retina}, Zhang21~\cite{Zhang2021}, and Vinker21~\cite{vinker2021unpaired} on a ``Forest" HDR scene.} 
  \label{fig:structure}
  \vspace{-1em} 
\end{figure*}

\begin{figure*}[t]
  \centering
    \subfloat[LLF]{\includegraphics[width=0.24\linewidth]{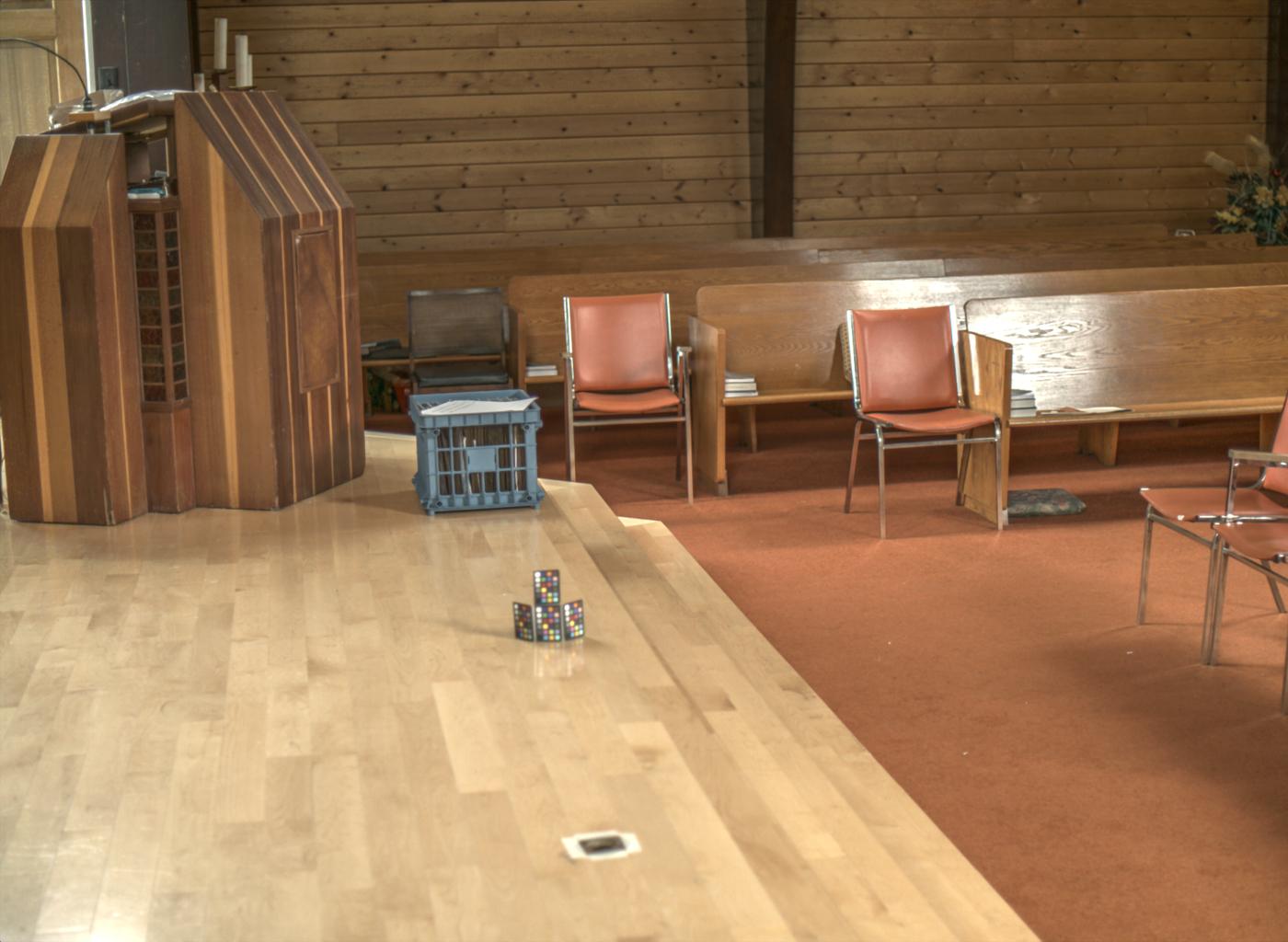}} \hskip0.3em
    \subfloat[NLPD-Opt]{\includegraphics[width=0.24\linewidth]{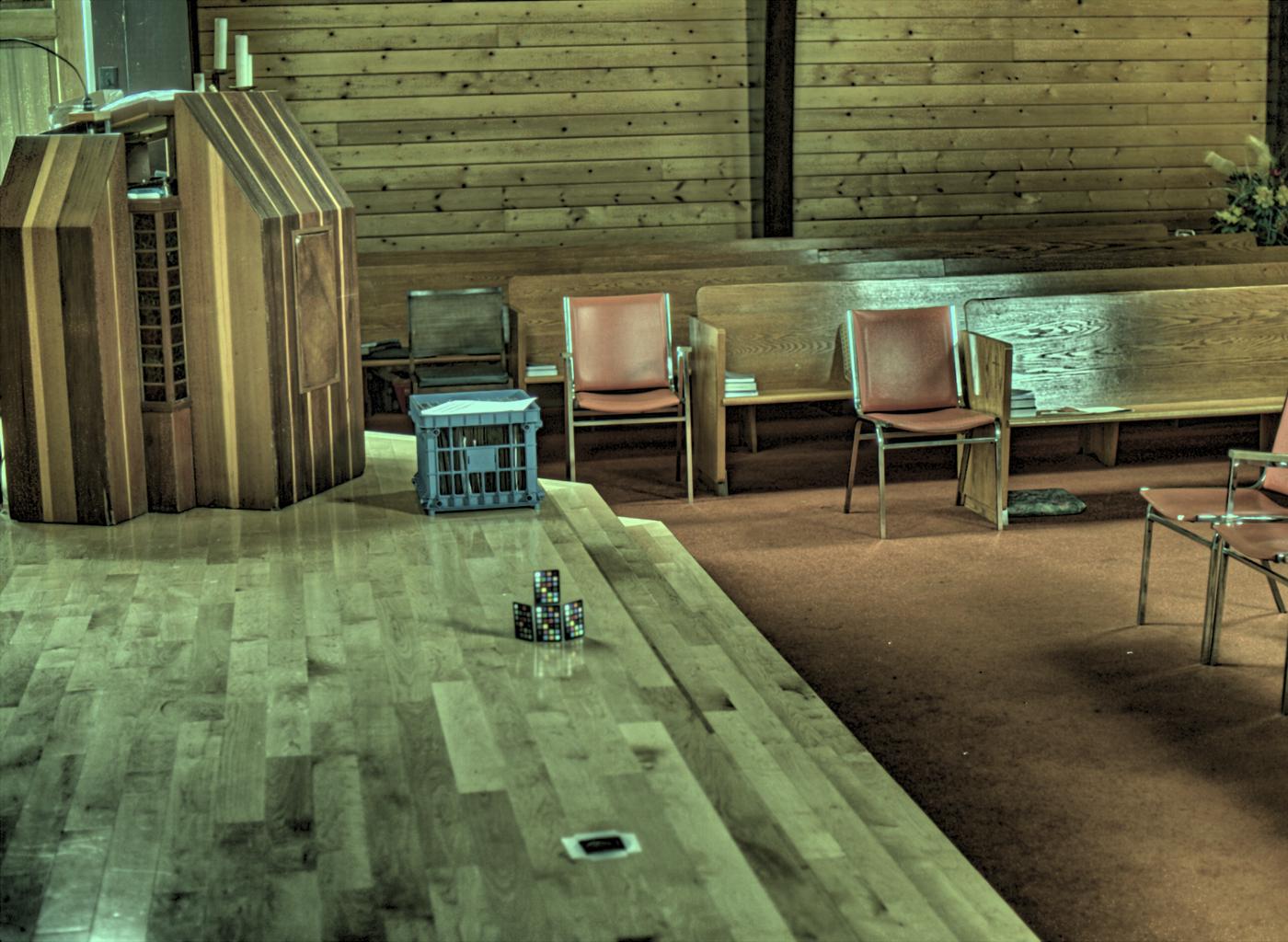}} \hskip0.3em
    \subfloat[Khan18]{\includegraphics[width=0.24\linewidth]{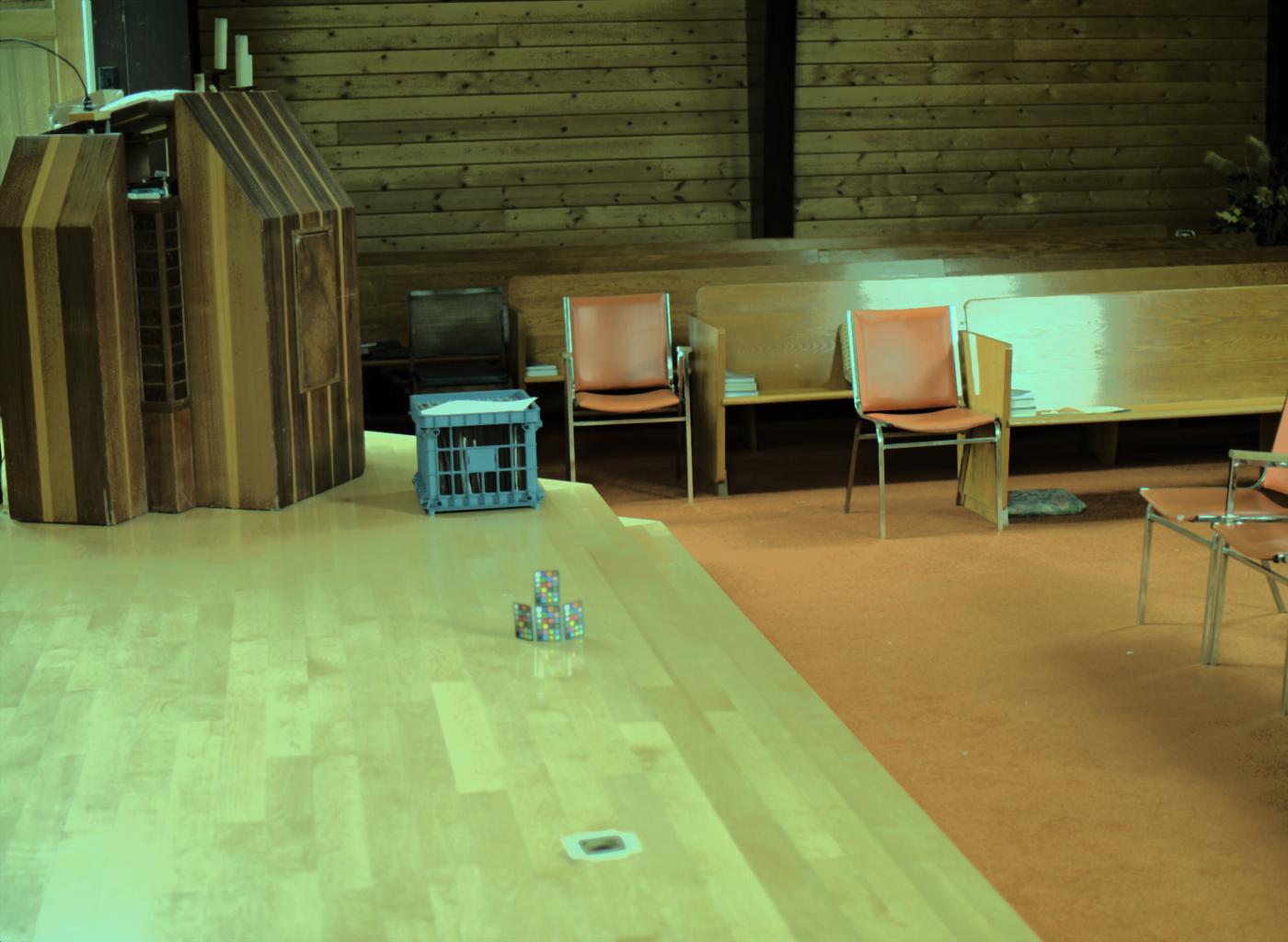}} \hskip0.3em
    \subfloat[Zhang20]{\includegraphics[width=0.24\linewidth]{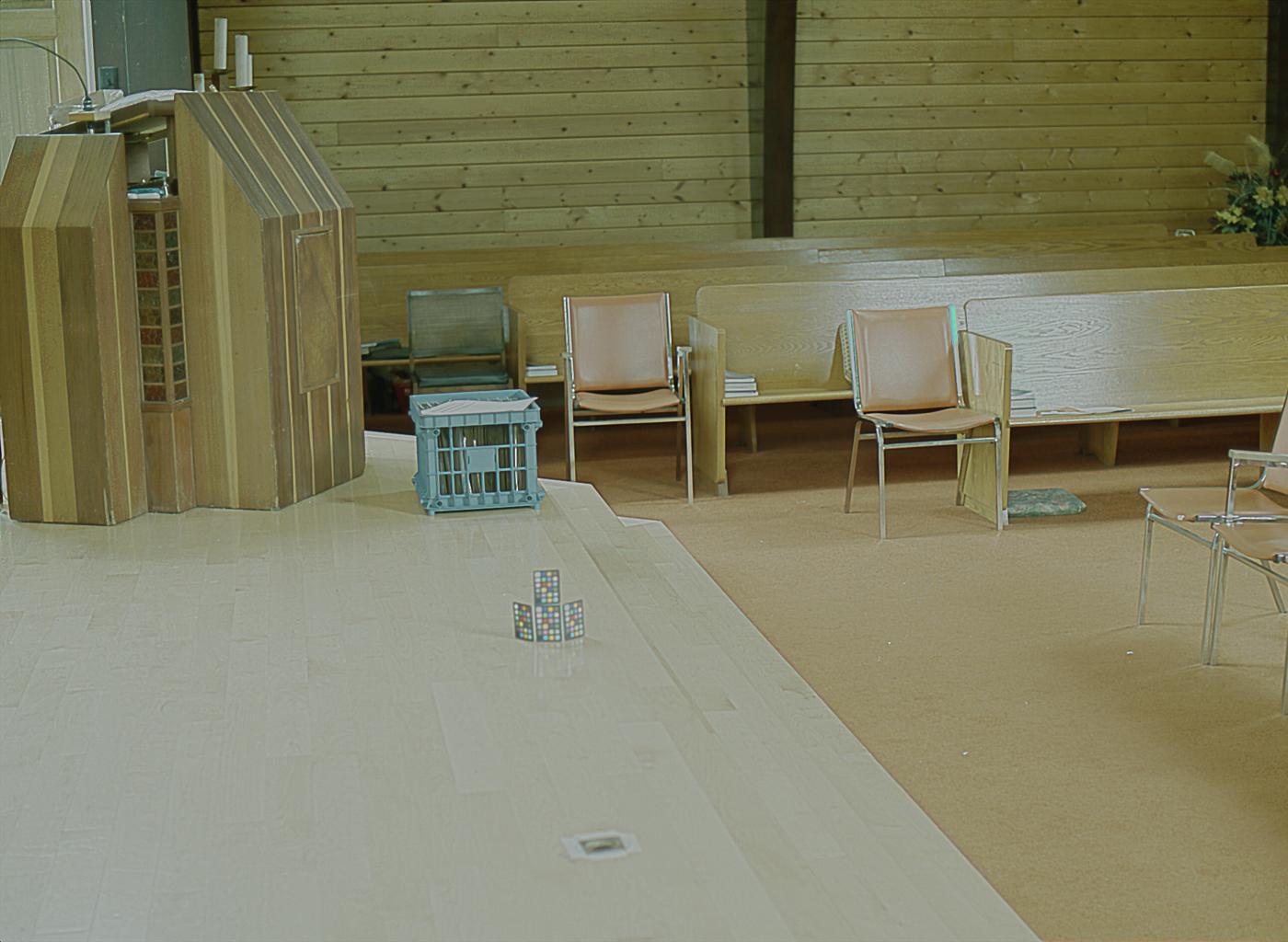}} \\\vspace{-0.5em}  
    \subfloat[Zhang21]{\includegraphics[width=0.24\linewidth]{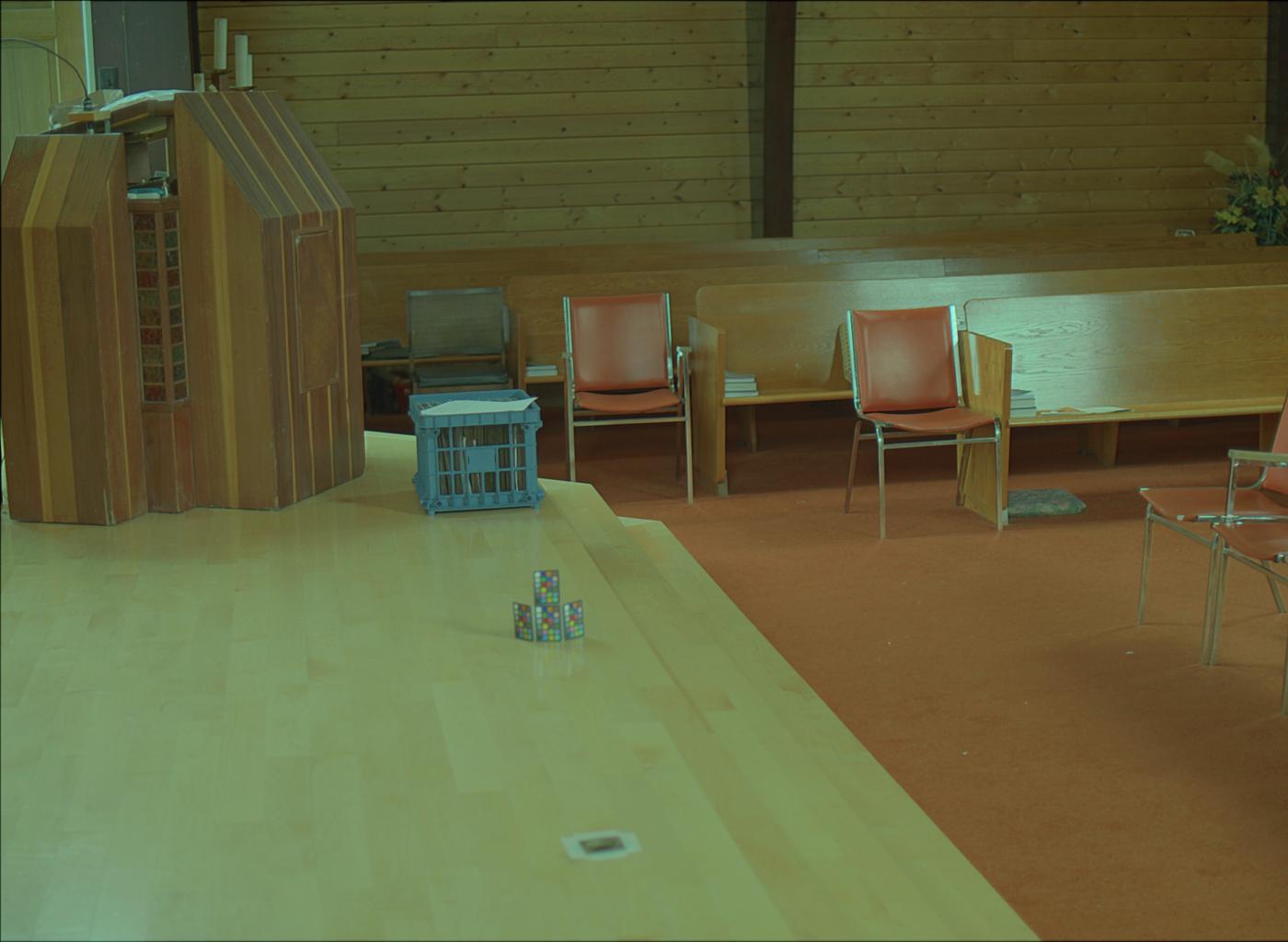}} \hskip0.3em
    \subfloat[Vinker21]{\includegraphics[width=0.24\linewidth]{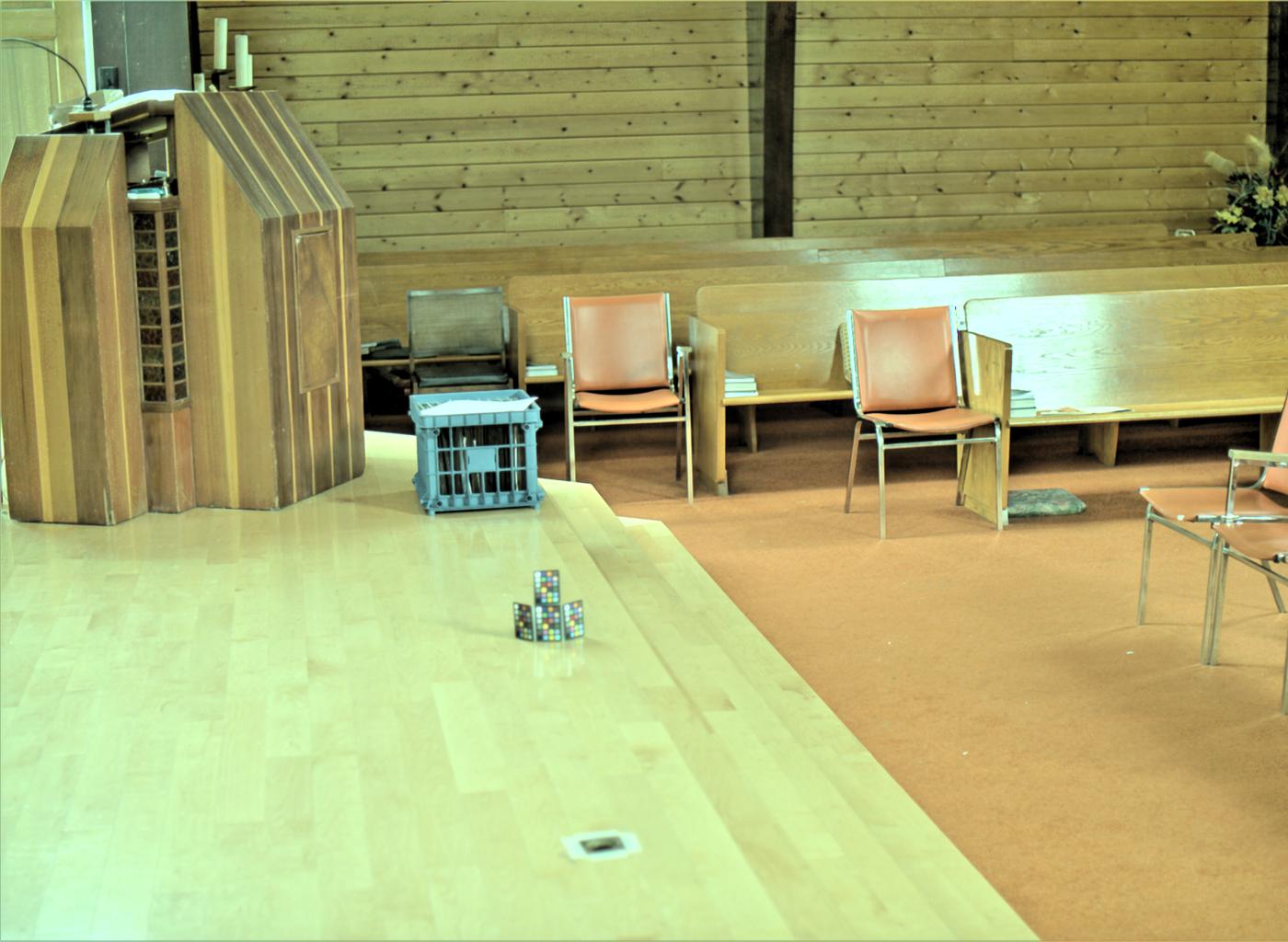}} \hskip0.3em
    \subfloat[Yang21]{\includegraphics[width=0.24\linewidth]{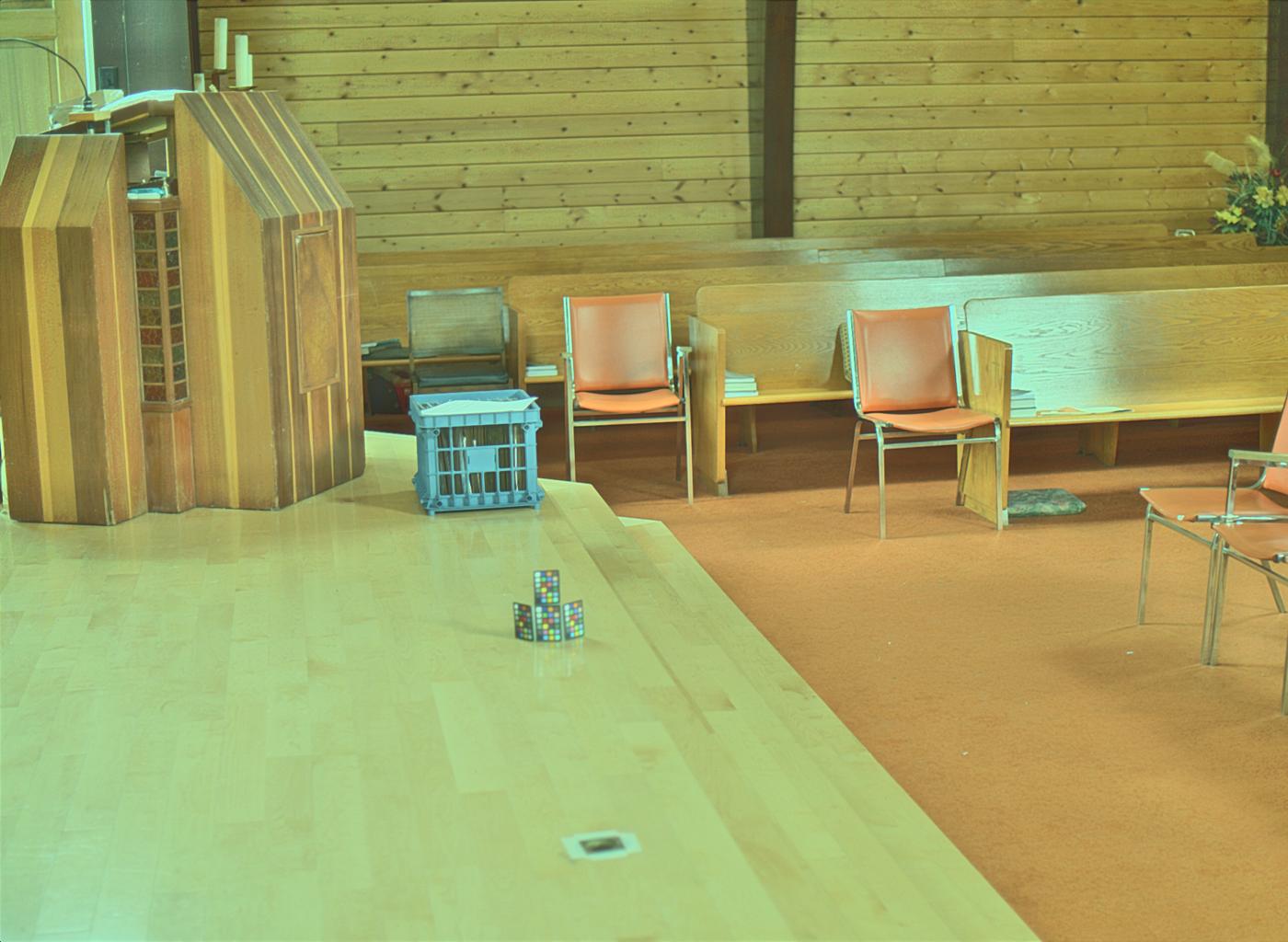}} \hskip0.3em
    \subfloat[PS-TMO]{\includegraphics[width=0.24\linewidth]{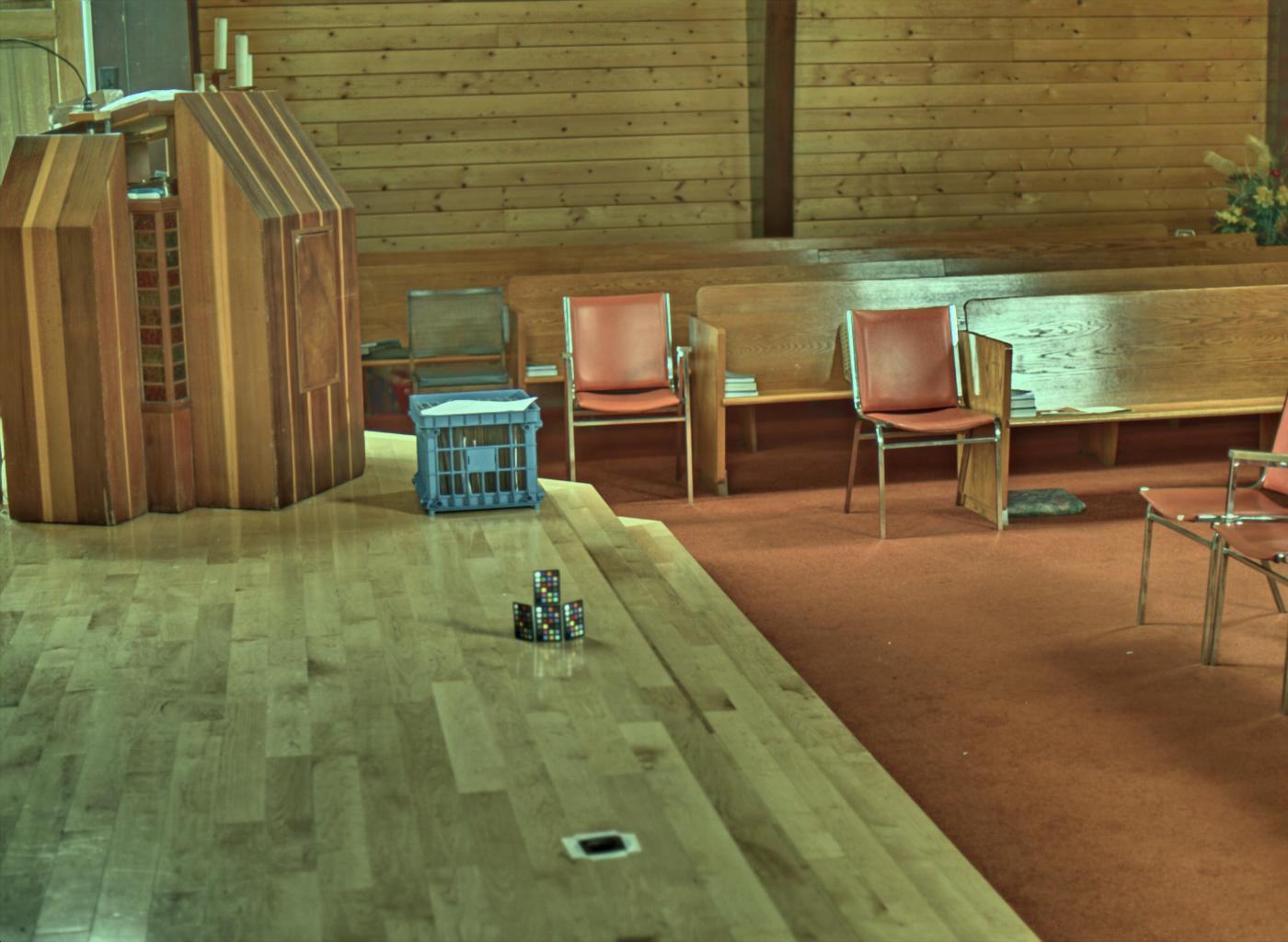}} \\ 
  \caption{Comparison of PS-TMO with LLF~\cite{paris2011local}, NLPD-Opt~\cite{laparra2017perceptually}, Khan18~\cite{Khan2018}, Zhang20~\cite{zhang2020retina}, Zhang21~\cite{Zhang2021}, Vinker21~\cite{vinker2021unpaired}, and Yang21~\cite{yang2021deep} on a ``Classroom" HDR scene.} 
  \label{fig:color}
  \vspace{-1em} 
\end{figure*}

\begin{figure*}[t]
  \centering
    \subfloat[Drago03]{\includegraphics[width=0.24\linewidth]{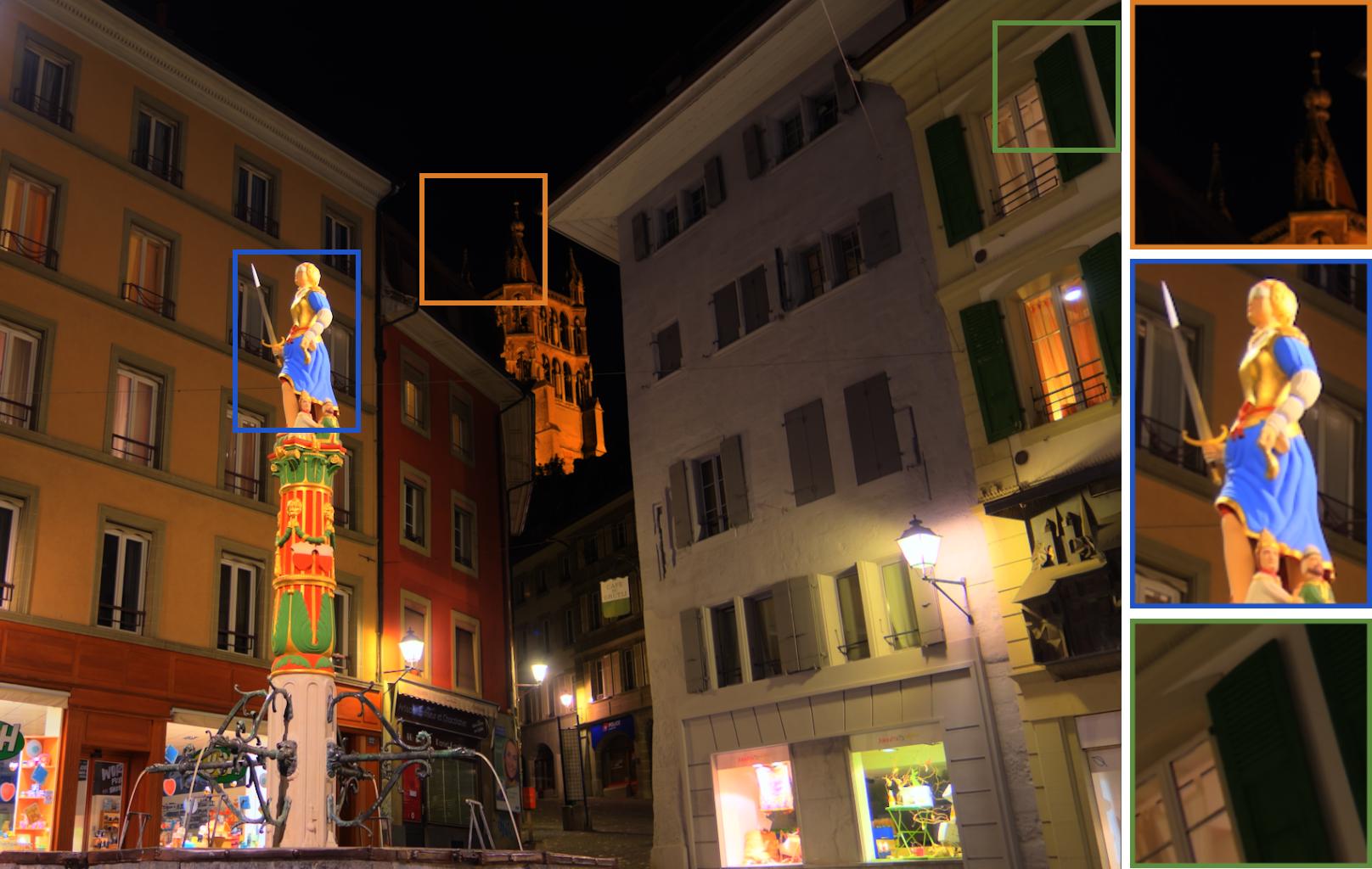}} \hskip0.3em
    \subfloat[Reinhard05]{\includegraphics[width=0.24\linewidth]{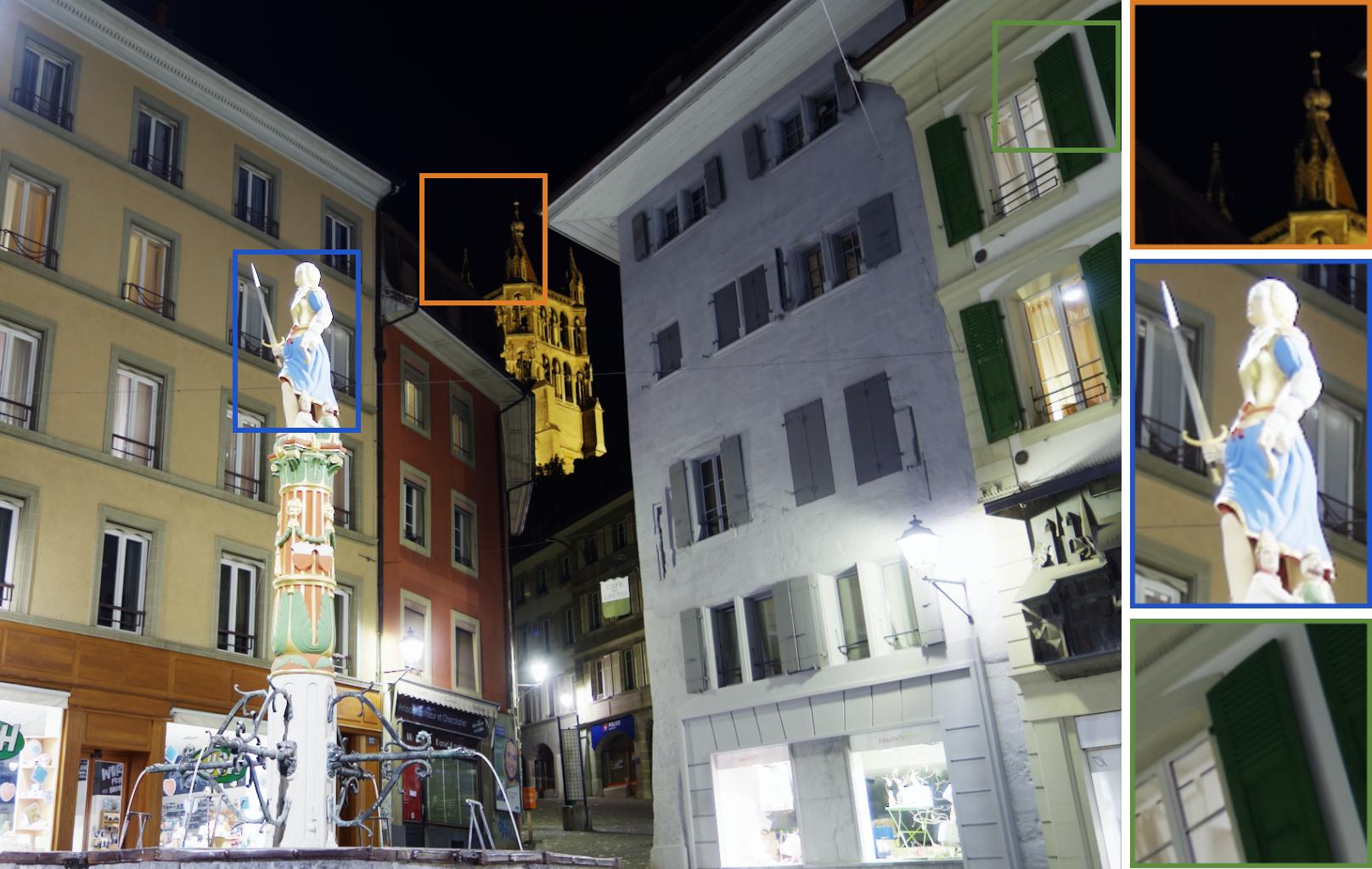}} \hskip0.3em
    \subfloat[Kim08]{\includegraphics[width=0.24\linewidth]{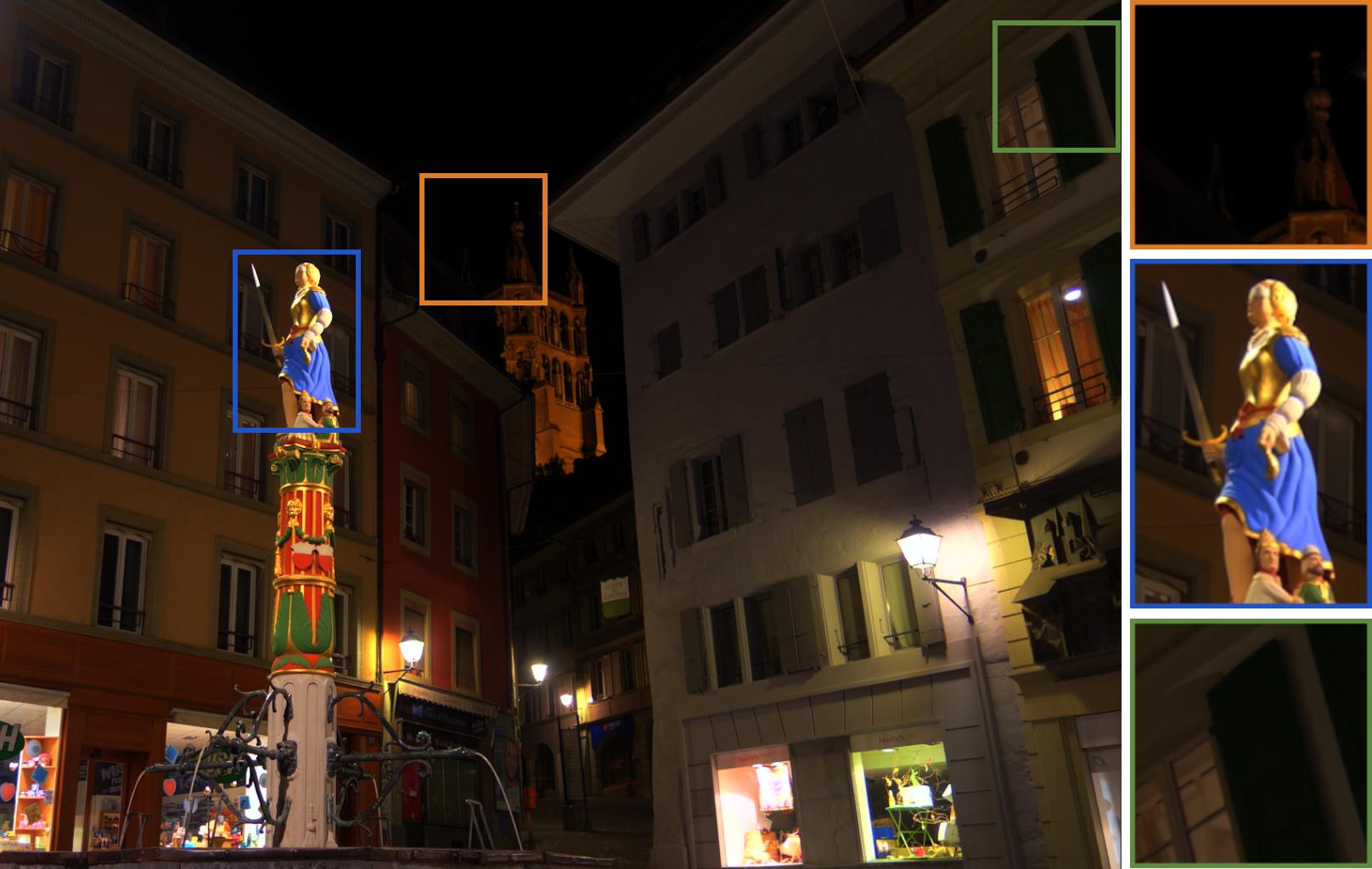}} \hskip0.3em
    \subfloat[Khan18]{\includegraphics[width=0.24\linewidth]{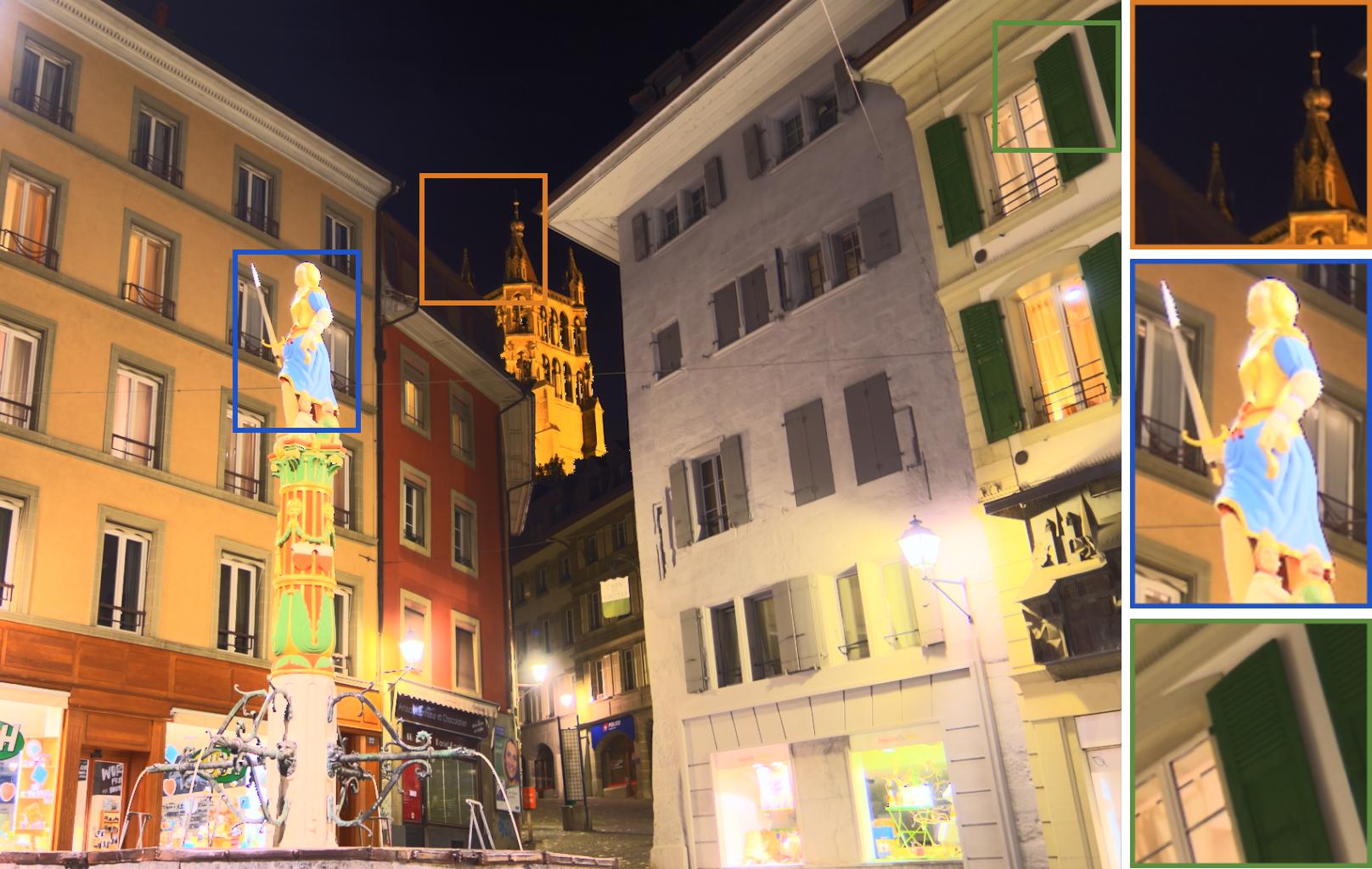}} \\\vspace{-0.5em}  
    \subfloat[NLPD-Opt]{\includegraphics[width=0.24\linewidth]{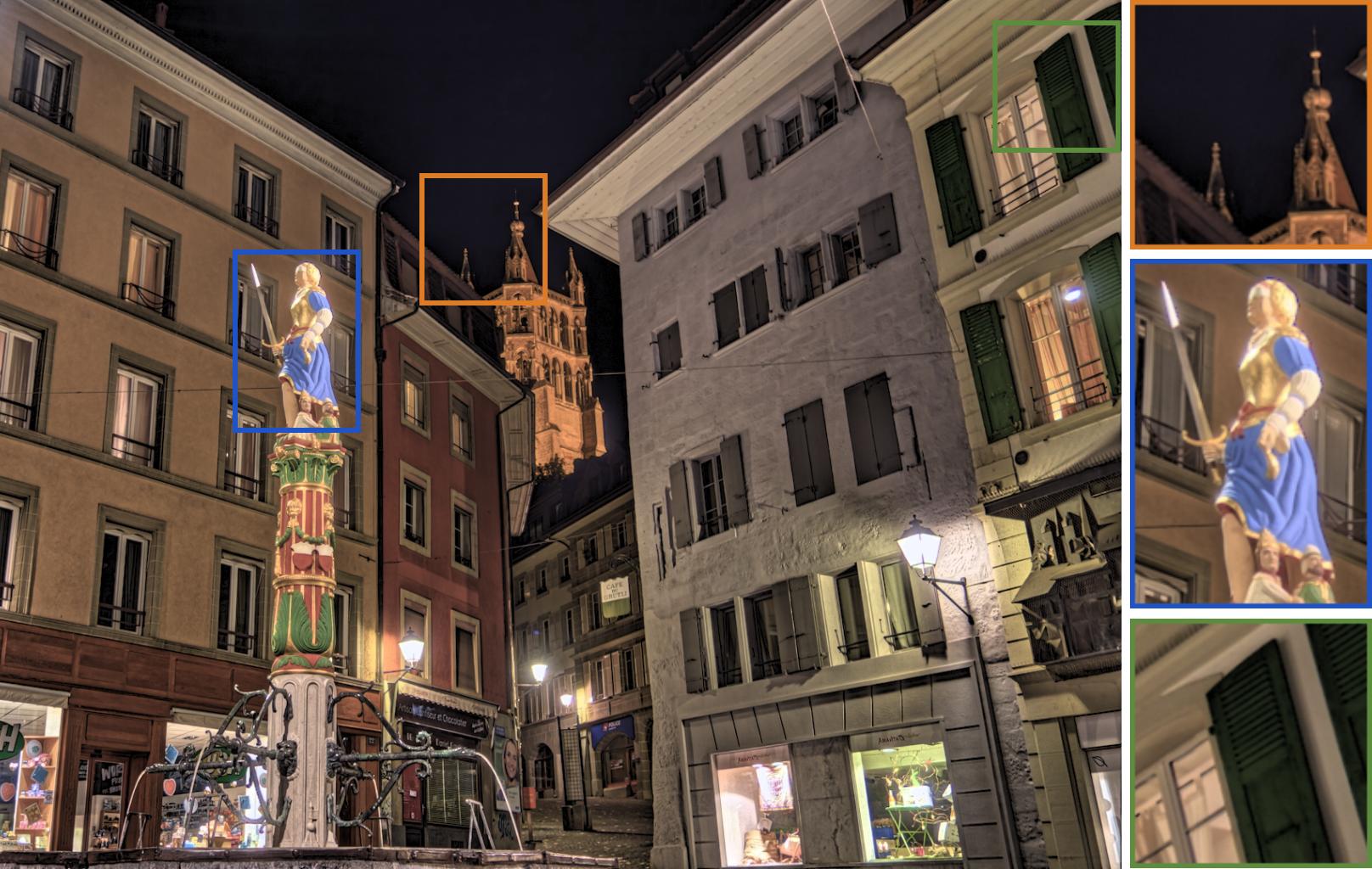}} \hskip0.3em
    \subfloat[Zhang20]{\includegraphics[width=0.24\linewidth]{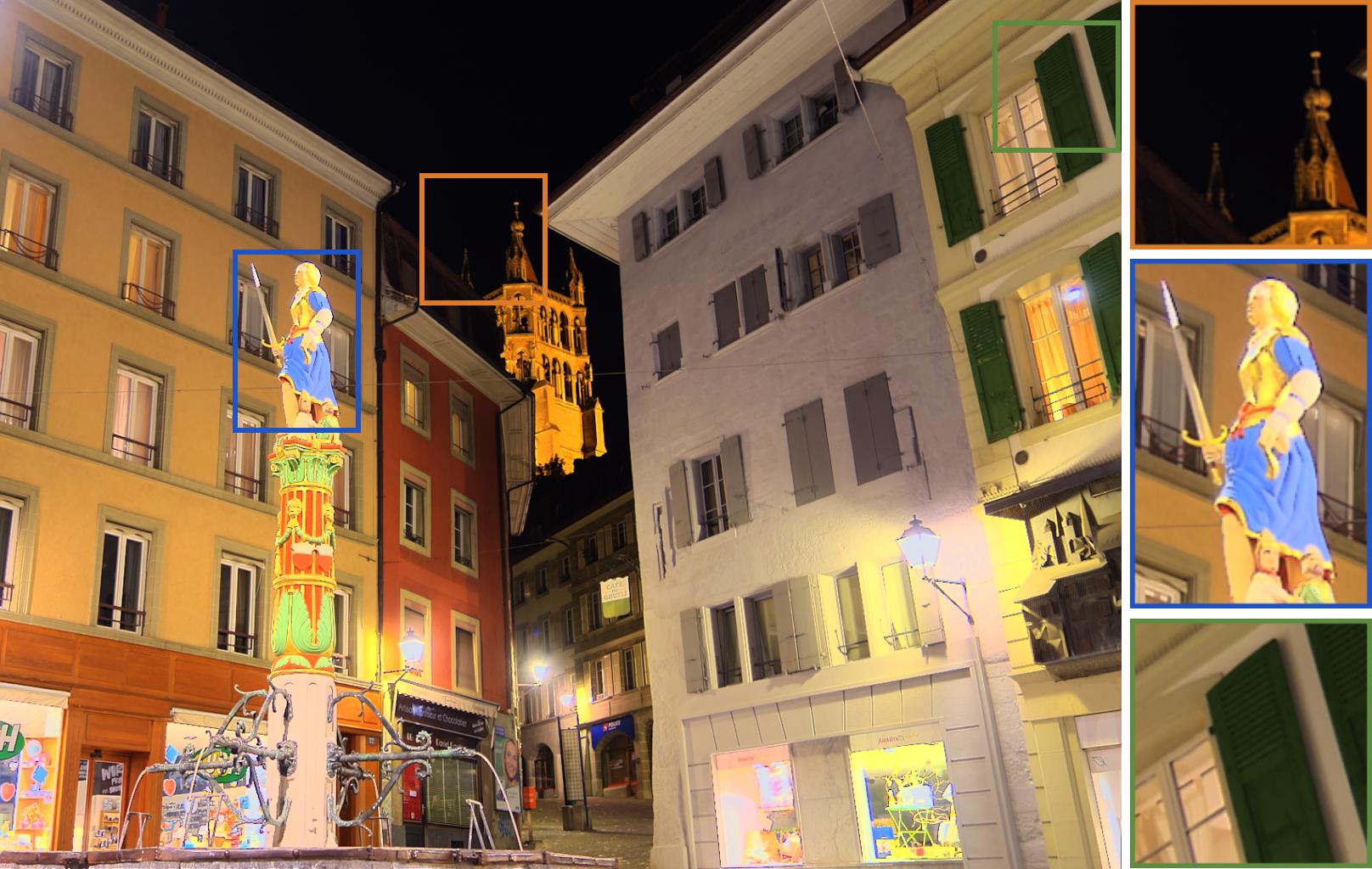}} \hskip0.3em
    \subfloat[Yang21]{\includegraphics[width=0.24\linewidth]{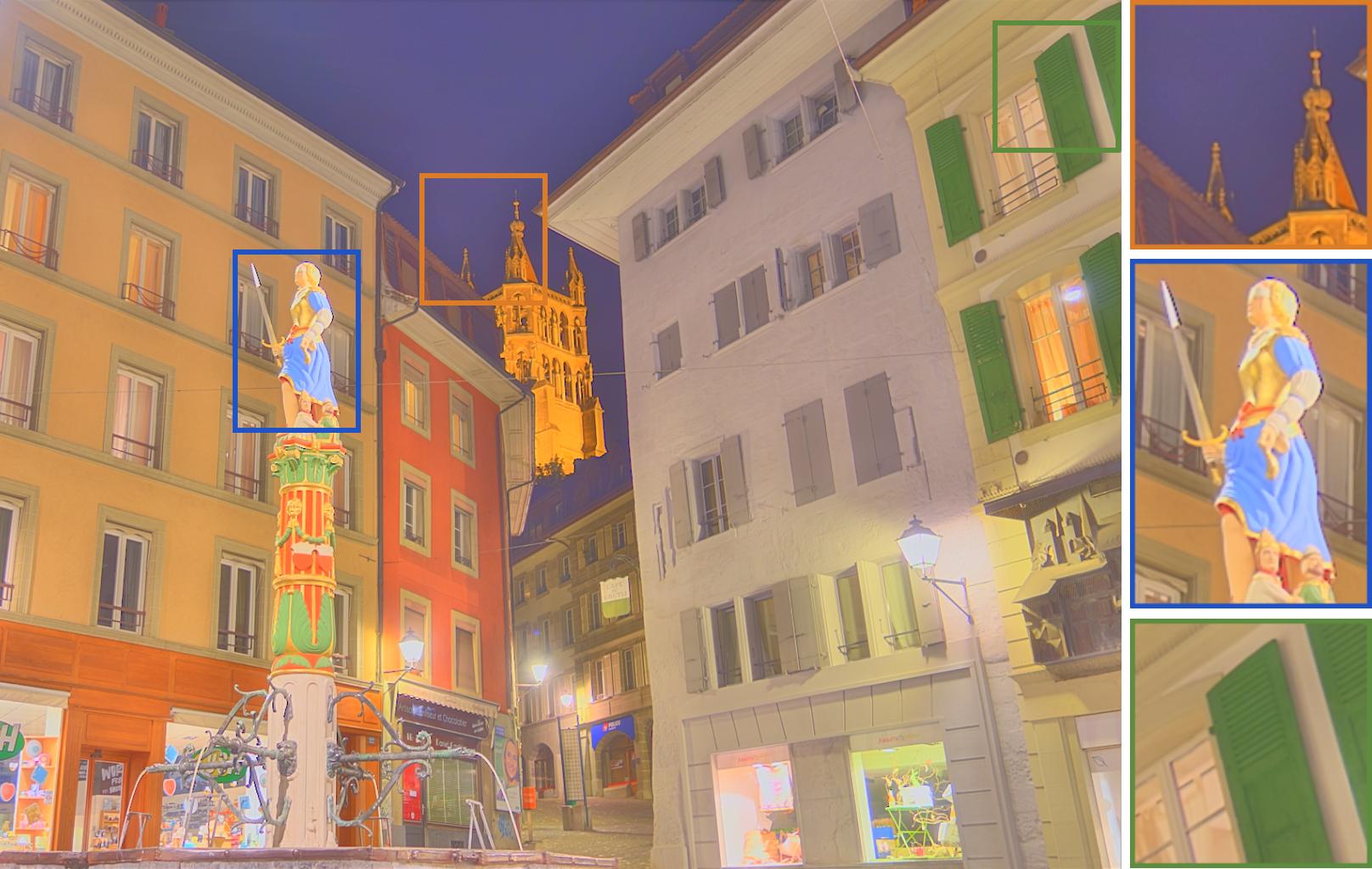}} \hskip0.3em
    \subfloat[PS-TMO]{\includegraphics[width=0.24\linewidth]{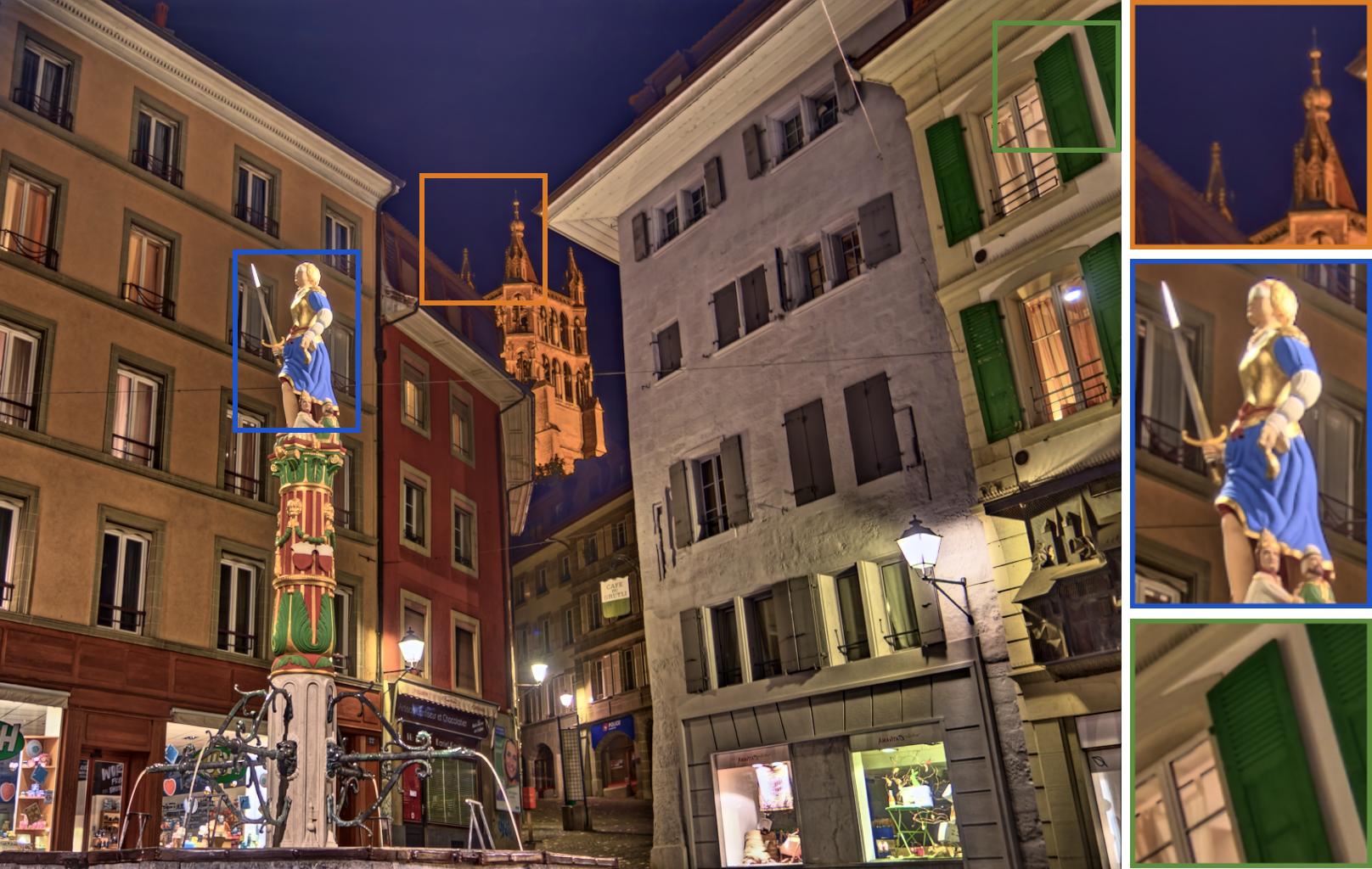}} \\ 
  \caption{Comparison of PS-TMO with Drago03~\cite{drago2003adaptive}, Reinhard05~\cite{reinhard2005dynamic}, Kim08~\cite{kim2008consistent}, Khan18~\cite{Khan2018}, NLPD-Opt~\cite{laparra2017perceptually}, Zhang20~\cite{zhang2020retina}, and Yang21~\cite{yang2021deep} on a ``Night Street" HDR scene.} 
  \label{fig:night}
  \vspace{-1em} 
\end{figure*}

\begin{figure*}[t]
  \centering
    \subfloat[Reinhard05]{\includegraphics[width=0.24\linewidth]{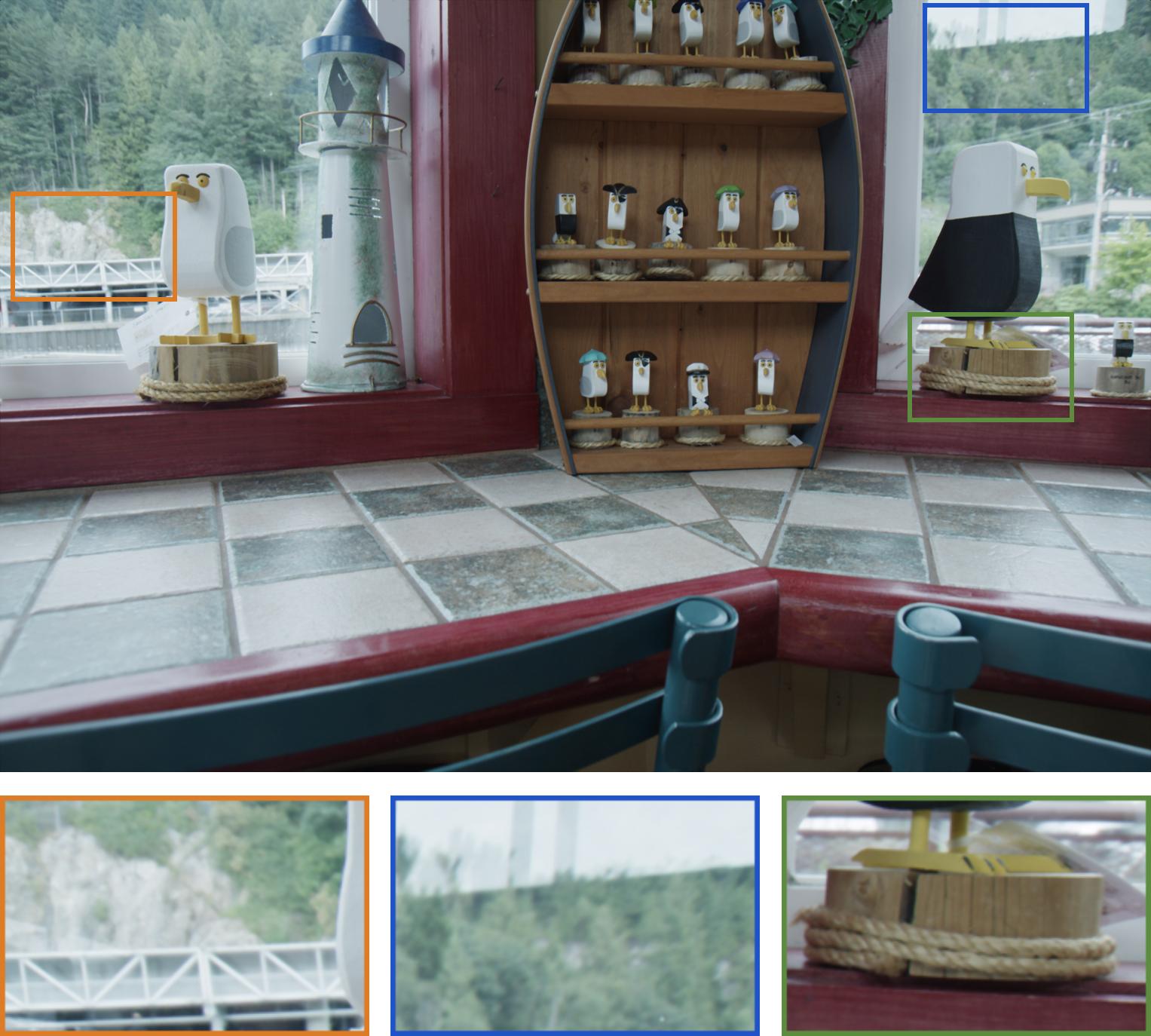}} \hskip0.3em
    \subfloat[WLS]{\includegraphics[width=0.24\linewidth]{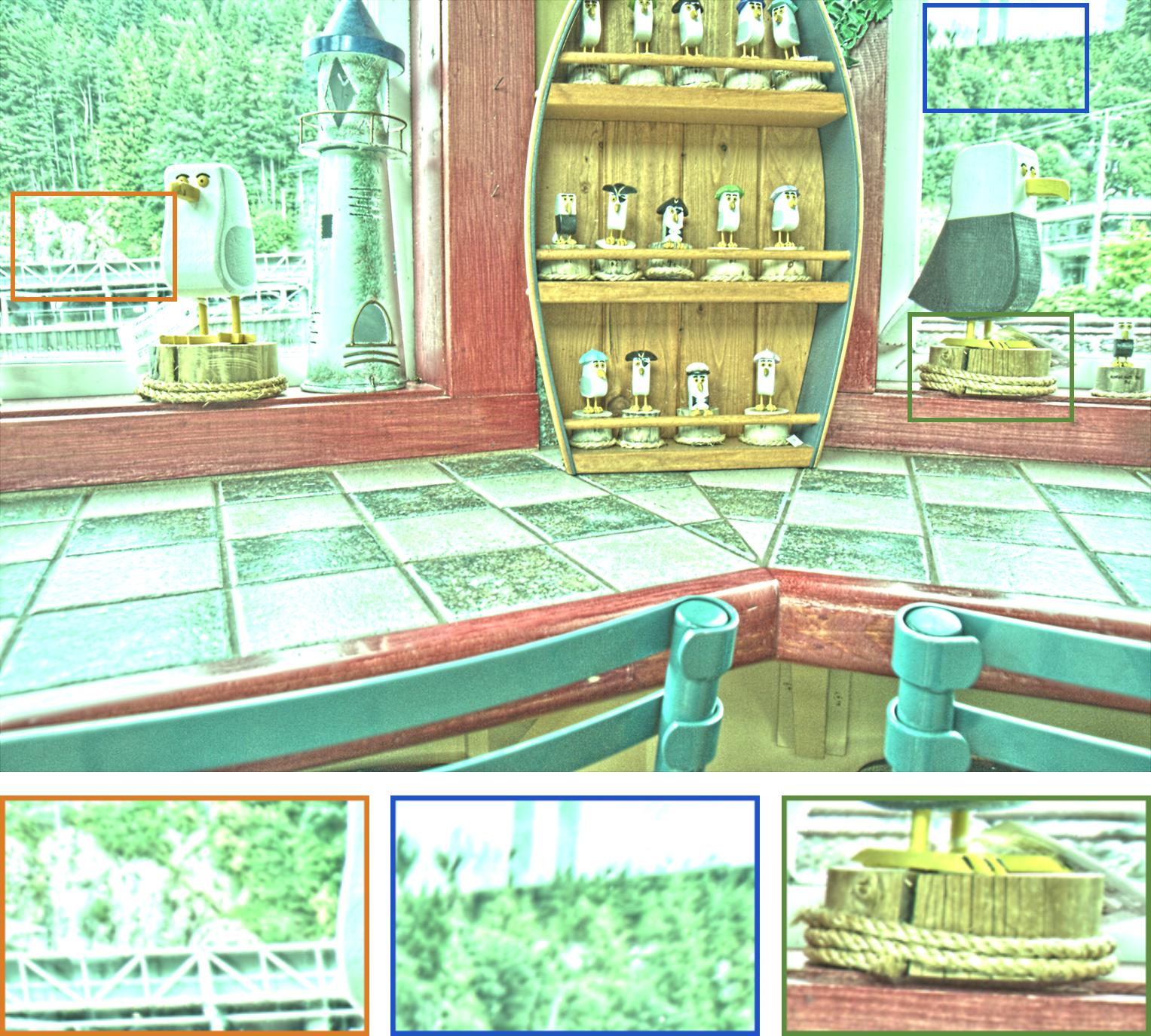}} \hskip0.3em
    \subfloat[LLF]{\includegraphics[width=0.24\linewidth]{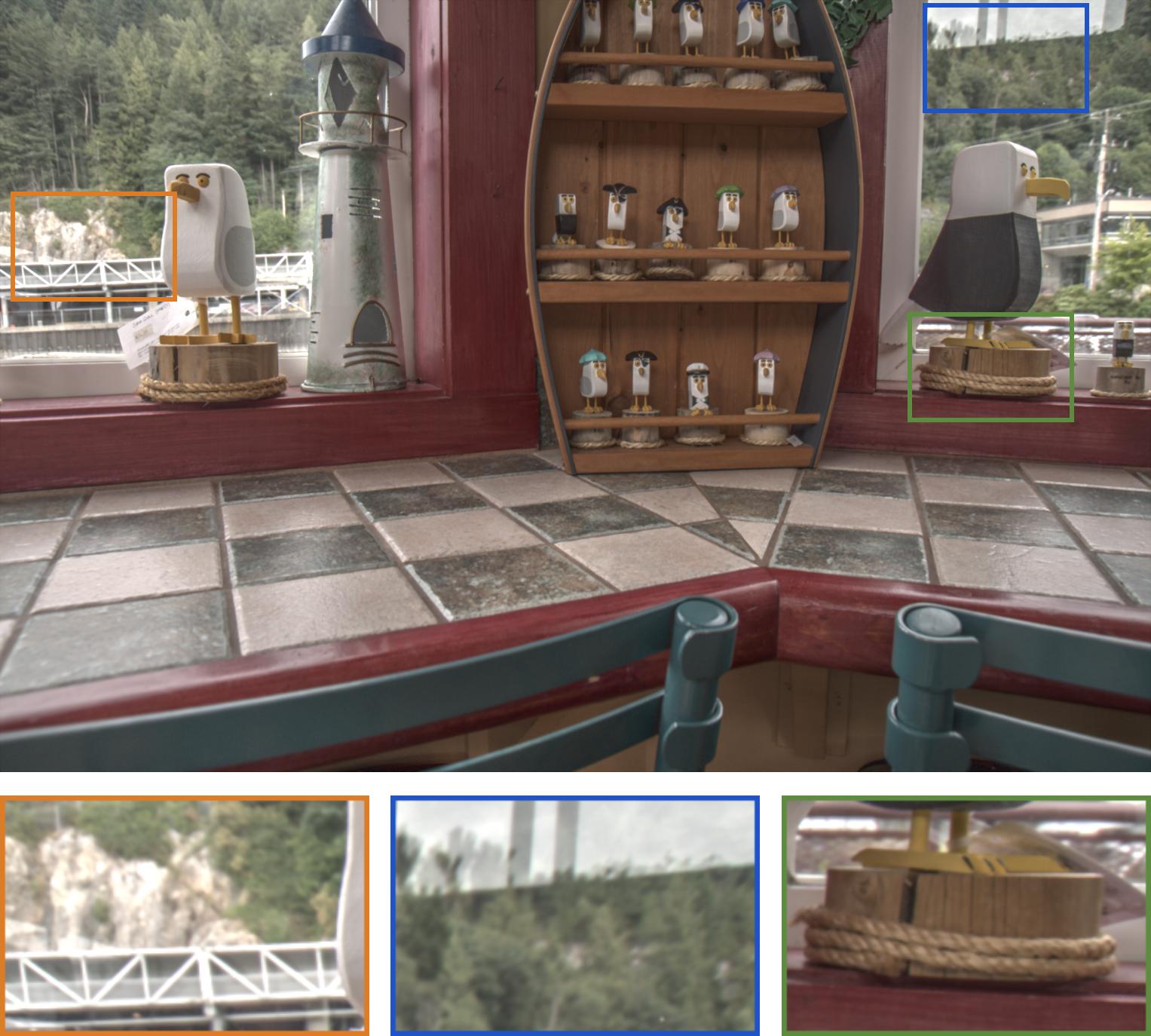}} \hskip0.3em
    \subfloat[Bruce14]{\includegraphics[width=0.24\linewidth]{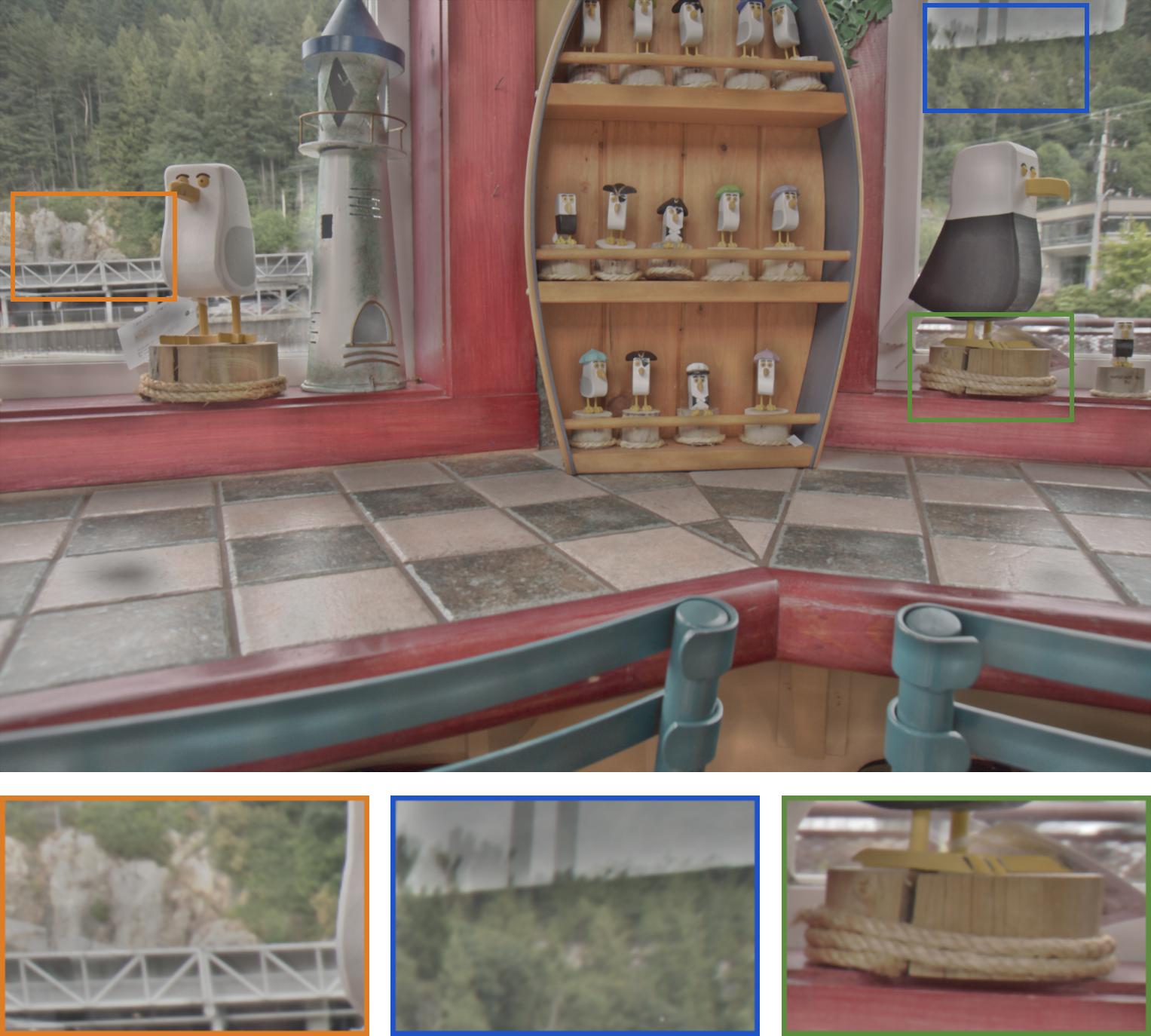}} \\\vspace{-0.5em}  
    \subfloat[Liang18]{\includegraphics[width=0.24\linewidth]{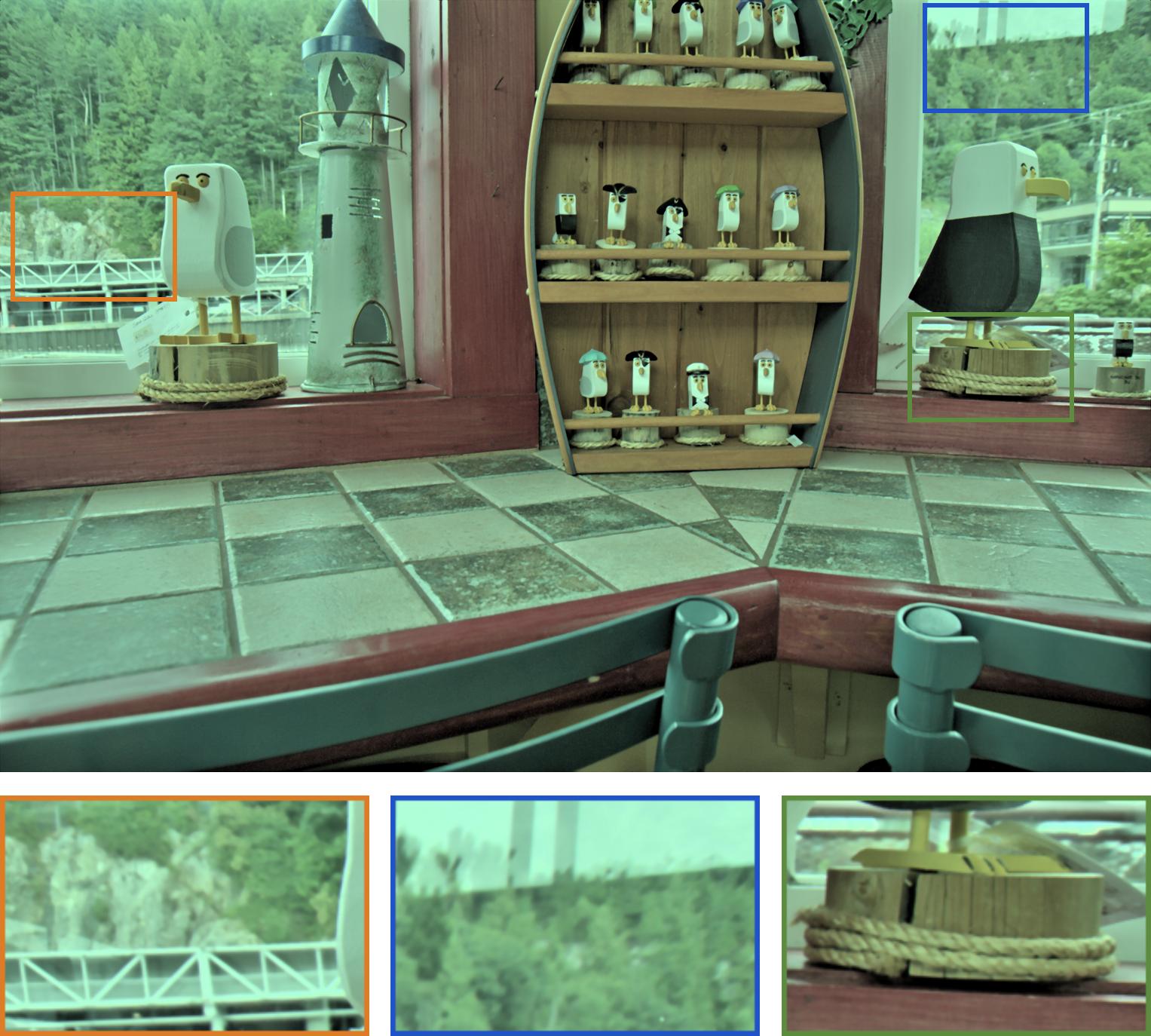}} \hskip0.3em
    \subfloat[Vinker21]{\includegraphics[width=0.24\linewidth]{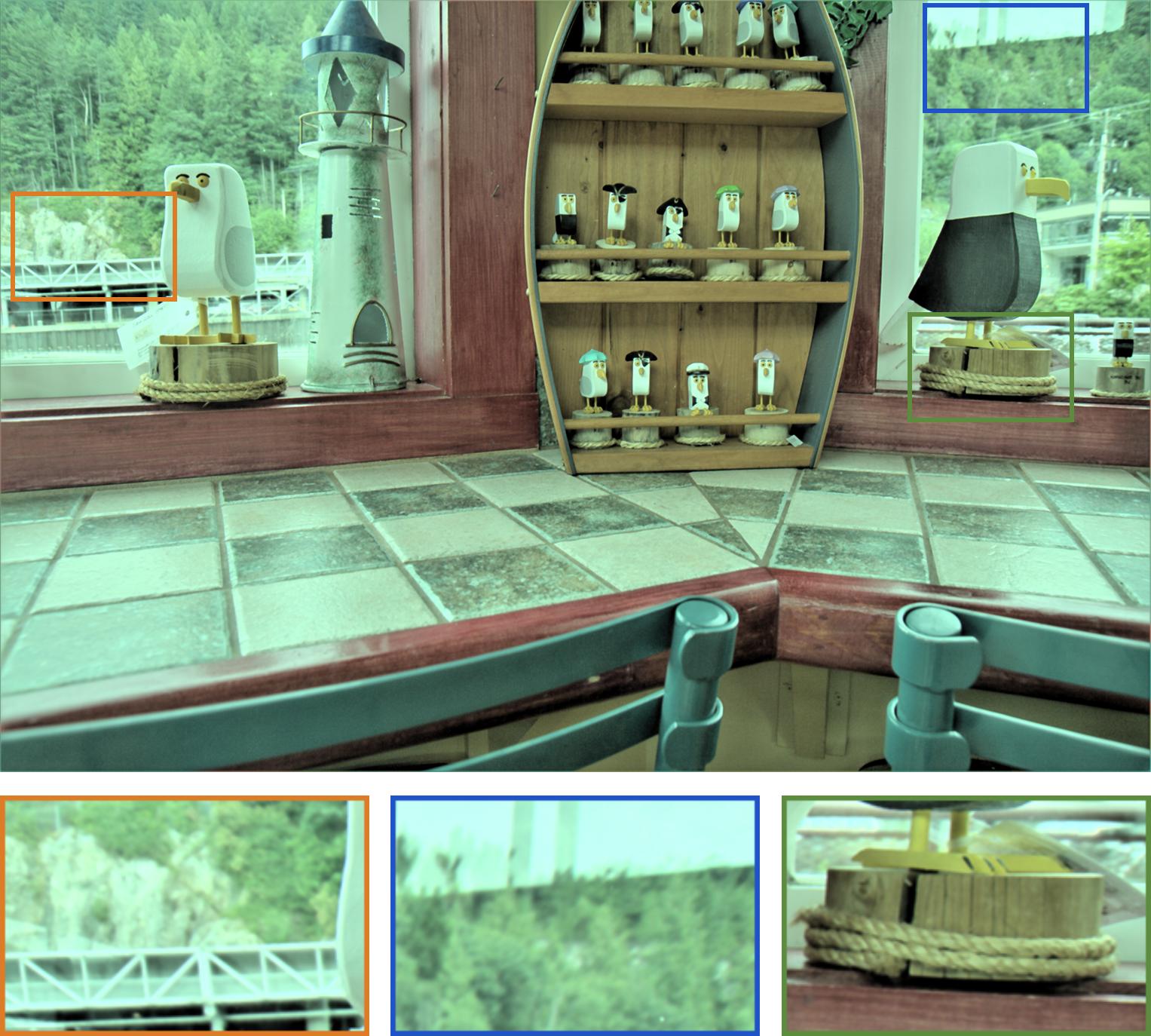}} \hskip0.3em
    \subfloat[Yang21]{\includegraphics[width=0.24\linewidth]{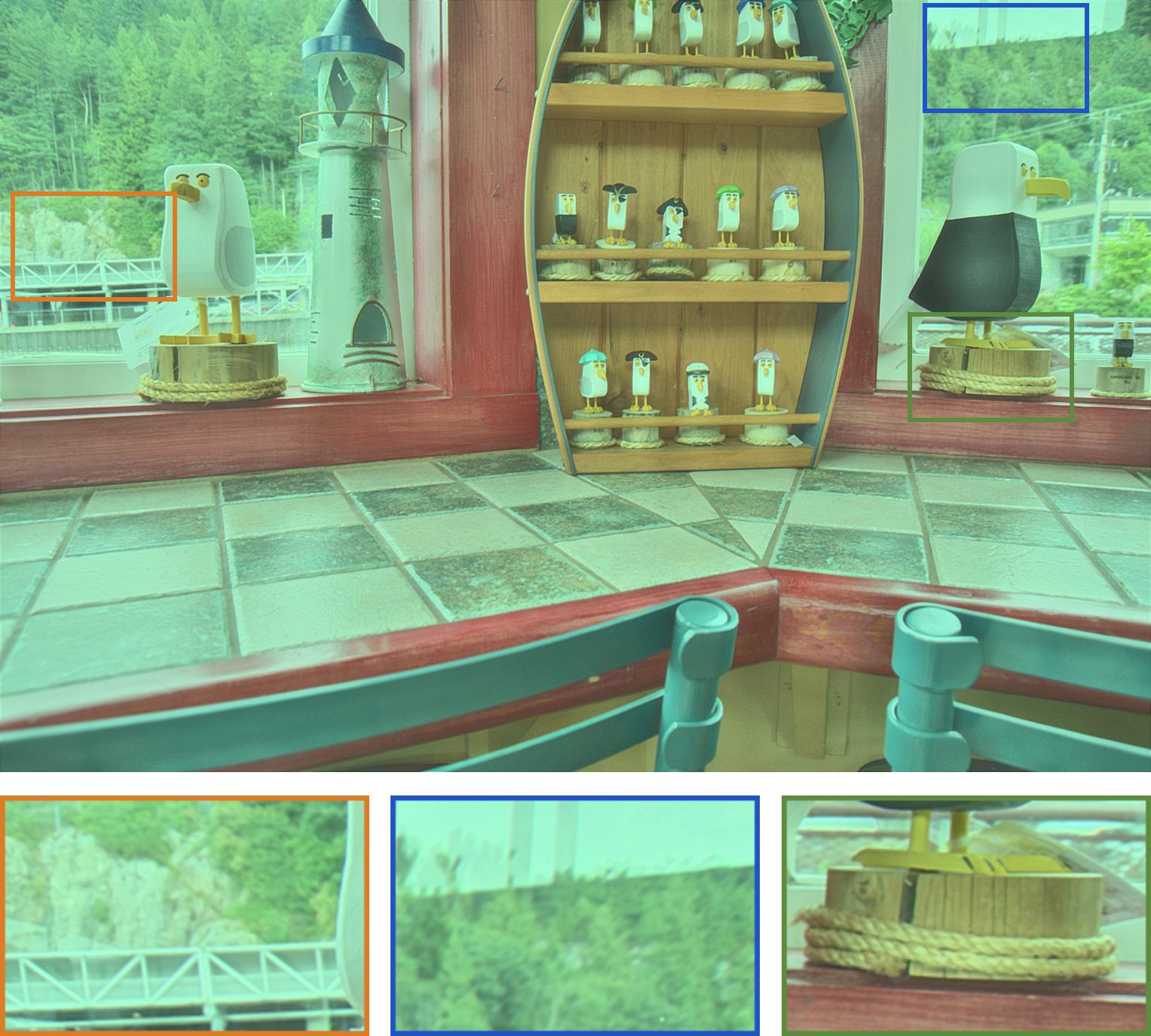}} \hskip0.3em
    \subfloat[PS-TMO]{\includegraphics[width=0.24\linewidth]{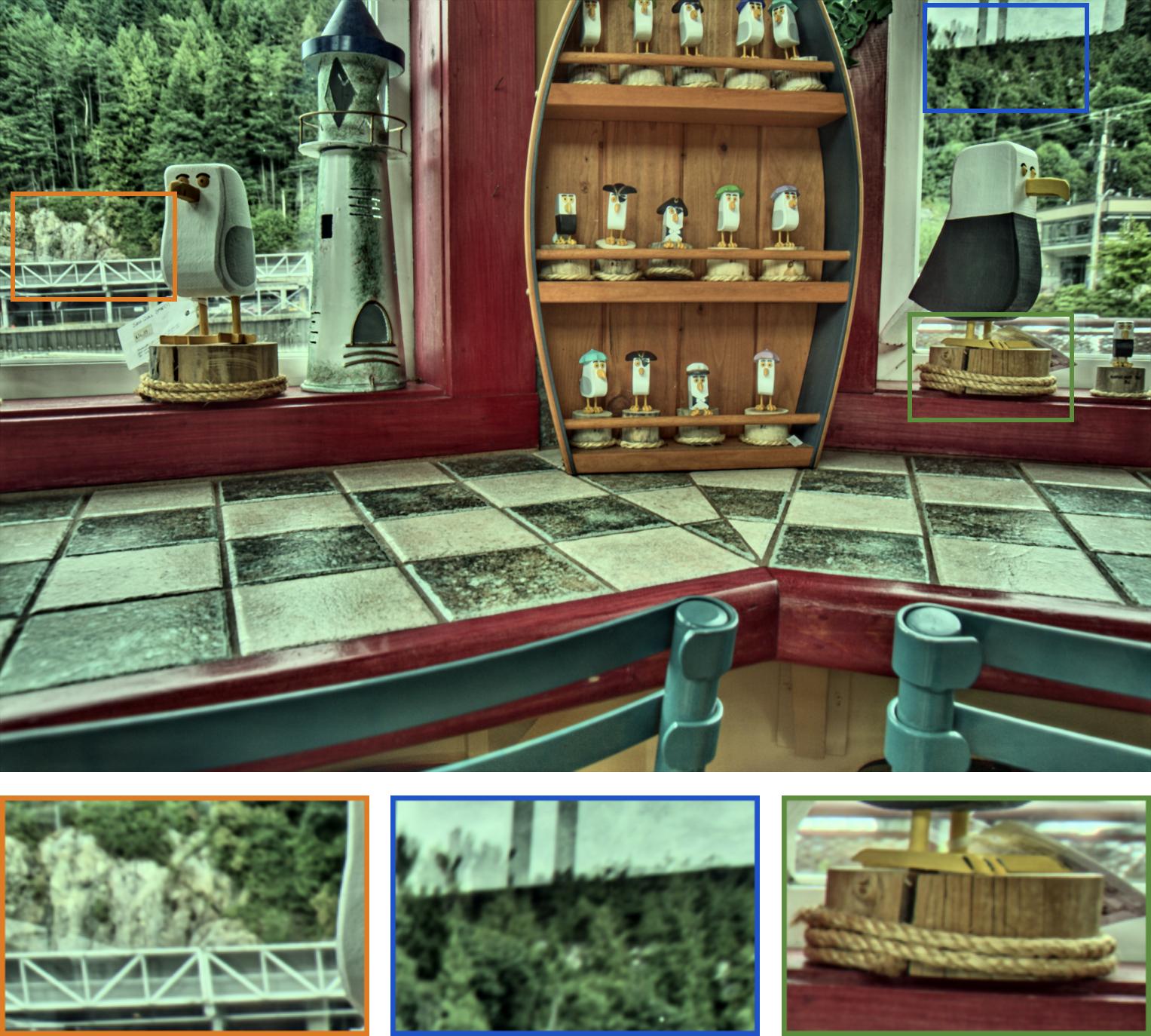}} \\ 
  \caption{Comparison of PS-TMO with Reinhard05~\cite{reinhard2005dynamic}, WLS~\cite{farbman2008edge}, LLF~\cite{paris2011local}, Bruce14~\cite{BRUCE201412}, Liang18~\cite{liang2018hybrid}, Vinker21~\cite{vinker2021unpaired}, and Yang21~\cite{yang2021deep} on a ``Windowsill" HDR scene.} 
  \label{fig:indoor}
  \vspace{-1em} 
\end{figure*}

\section{Experiments}\label{sec:experiment}
In this section, we compare PS-TMO with traditional and recent DNN-based TMOs in terms of subjective quality, objective quality (by TMQI~\cite{yeganeh2012objective} and NLPD~\cite{laparra2017perceptually}), and computational time. Moreover, we carry out a debiased subjective experiment~\cite{cao2021debiased} to verify the perceptual gains obtained by the proposed PS-TMO. In addition, we conduct a series of ablation experiments to justify each design choice of PS-TMO. 

We compare PS-TMO with fifteen existing TMOs, including Drago03~\cite{drago2003adaptive}, Reinhard05~\cite{reinhard2005dynamic}, Kim08~\cite{kim2008consistent}, WLS~\cite{farbman2008edge}, LLF~\cite{paris2011local}, Bruce14~\cite{BRUCE201412}, GR~\cite{shibata2016gradient}, NLPD-Opt~\cite{laparra2017perceptually}, Khan18~\cite{Khan2018}, Liang18~\cite{liang2018hybrid}, Zhang20~\cite{zhang2020retina}, Zhang21~\cite{Zhang2021}, Vinker21~\cite{vinker2021unpaired}, Yang21~\cite{yang2021deep}, and Le21~\cite{chenyang2021}. Zhang21, Vinker21, and Yang21 are DNN-based TMOs, while the others are conventional operators, among which Drago03 Reinhard05 and Kim08 are global operators, and the rest are local operators. 

Drago03 relies on an adaptive logarithmic mapping, while Reinhard05 uses a practical S-shaped curve. Kim08 improves upon Drago03 by refining the logarithmic curve from the perspective of  photosensitive material characteristics. WLS casts HDR tone mapping into a weighted least squares problem, and LLF involves ingenious manipulation of Laplacian pyramid coefficients. Bruce14 achieves tone mapping via MEF. GR and Liang18 are based on the two-layer decomposition in the gradient domain. Zhang20 is a retina-inspired TMO (by modeling retina horizontal and bipolar cells). NLPD-Opt compresses the HDR image by directly minimizing the NLPD metric in the image space. Thus, given sufficient iterations, NLPD-Opt is regarded as the lower bound for all TMOs in terms of NLPD. All DNN-based methods except Vinker21 require paired images for supervision. Zhang21 acquires the 
ground-truth LDR images from expert manipulation, while Yang21 selects the best LDR image produced by a list of existing TMOs subjectively (via human inspection). Zhang21 introduces a semi-supervised strategy to further leverage  the real-world LDR image distribution as a form of regularization. Like PS-TMO, Vinker21 does not need the ground-truth LDR images for training, which is, however, achieved by a rather empirical combination of several loss functions. Le21 was published in our conference version~\cite{chenyang2021}, which only includes the tone mapping network in Stage one of PS-TMO, and needs manual specification of a working maximum luminance for each test HDR scene. All algorithms are implemented either by Banterle~\etal~\cite{Banterle2017} in their great MATLAB Toolbox\footnote{\url{https://github.com/banterle/HDR Toolbox}} or by the respective author. We test them with the default settings.

\subsection{Main Results}
\subsubsection{Qualitative Comparison}

Fig.~\ref{fig:table} compares the tone mapping results of linear rescaling, Drago03~\cite{drago2003adaptive}, Liang18~\cite{liang2018hybrid}, Vinker21~\cite{vinker2021unpaired}, Zhang21~\cite{Zhang2021}, and PS-TMO on an ``Outdoor Table" HDR scene. The linear rescaling creates an over-exposed image in most local regions.  Drago03 generates a relatively dark appearance with reduced global contrast. The results produced by Zhang21 and Vinker21
look pale with over-saturation appearances; Liang18 performs better than the two methods with enhanced local details. In contrast, PS-TMO significantly outperforms the competing methods in terms of color and detail reproduction, giving rise to a more perceptually appealing overall appearance. 

Fig.~\ref{fig:structure} compares the tone mapping results of Kim08~\cite{kim2008consistent}, WLS~\cite{farbman2008edge}, GR~\cite{shibata2016gradient}, Liang18~\cite{liang2018hybrid}, Zhang20~\cite{zhang2020retina}, Zhang21~\cite{Zhang2021}, Vinker21~\cite{vinker2021unpaired}, and PS-TMO on a ``Forest" HDR scene.
The global TMO Kim08~\cite{kim2008consistent} contains noticeable under-exposed areas. Local TMOs like WLS~\cite{farbman2008edge} and GR~\cite{shibata2016gradient} focus on local detail enhancement, and ultimately lead to edge-related artifacts. Such over-enhancement is less pronounced for Liang18~\cite{liang2018hybrid} due to $\ell_0$ flattening and Zhang20~\cite{zhang2020retina}. The result by DNN-based  Zhang21~\cite{Zhang2021} is slightly over-saturated with fewer details. Vinker21~\cite{vinker2021unpaired} generates a natural-looking LDR image similar to that of PS-TMO, despite that the former is weaker at reproducing warm colors. 

Fig.~\ref{fig:color} compares the tone mapping results of LLF~\cite{paris2011local}, NLPD-Opt~\cite{laparra2017perceptually}, Khan18~\cite{Khan2018}, Zhang20~\cite{zhang2020retina}, Zhang21~\cite{Zhang2021}, Vinker21~\cite{vinker2021unpaired}, Yang21~\cite{yang2021deep}, and PS-TMO on a ``Classroom'' HDR scene. PS-TMO gives a more faithful color reproduction for the chairs and the carpet. Moreover, it does an excellent job in recovering the fine structures of the wood floor and the soft shadow. On the contrary, most competing methods suffer from the problems of reduced global contrast, color cast, and detail loss. Of particular interest, NLPD-Opt~\cite{laparra2017perceptually} tends to overshoot the details, and is slightly noisy with a manually optimized $S_\mathrm{max}= 10^6$ $\rm cd/m^2$ for perceptual quality (not physical plausibility\footnote{In terms of physical plausibility, a maximum luminance of $S_\mathrm{max} = 10^6$ $\rm cd/m^2$ would be too high for the ``Classroom'' HDR scene with no direct light sources.}). Comparison with NLPD-Opt provides strong evidence that PS-TMO automatically combines the best perceptual aspects of the LDR stack for improved tone mapping.

Fig.~\ref{fig:night} compares the tone mapping results of  Drago03~\cite{drago2003adaptive}, Reinhard05~\cite{reinhard2005dynamic}, Kim08~\cite{kim2008consistent}, Khan18~\cite{Khan2018}, NLPD-Opt~\cite{laparra2017perceptually} ($S_\mathrm{max} = 10^5$ $\rm cd/m^2$), Zhang20~\cite{zhang2020retina}, Yang21~\cite{yang2021deep}, and PS-TMO on a ``Night Street" HDR scene. Global TMOs - Drago03~\cite{drago2003adaptive}, Reinhard05~\cite{reinhard2005dynamic}, and Kim08~\cite{kim2008consistent} lose many  details in the regions of the dark sky and around the bright light sources. Local TMOs - Khan18~\cite{Khan2018}, NLPD-Opt~\cite{laparra2017perceptually}, and Zhang20~\cite{zhang2020retina} perform better in detail preservation. Nevertheless, the colors reproduced by Khan18~\cite{Khan2018} and Zhang20~\cite{zhang2020retina} are unnatural. The result by Yang21~\cite{yang2021deep} looks over-exposed and over-saturated. In contrast, PS-TMO gives a more vivid color appearance with balanced local contrast and details.

 \begin{table}[t]
 \renewcommand{\arraystretch}{1.25}
	\begin{center}
	\caption{Quantitative results of TMOs in terms of TMQI~\cite{yeganeh2012objective} (and its two components structural fidelity $\mathrm{SF}$ and statistical naturalness $\mathrm{SN}$), and NLPD~\cite{laparra2017perceptually}. Le21 is the preliminary version of PS-TMO,  which only includes the tone mapping network in PS-TMO, and needs manual specification of a working maximum luminance for each test HDR scene. The top two results are highlighted in bold} \label{tab:tmqi_nlpd_time_score}
		\begin{tabular}{l|c c c c }
			\toprule
			TMO & TMQI$\uparrow$ & SF$\uparrow$ & SN$\uparrow$ & NLPD$\downarrow$ \\
			\hline
			Drago03 & 0.8362 & 0.9142 &  0.2037 & 0.2163 \\
			Reinhard05& 0.8162  &  0.8729 & 0.1590 & 0.2139 \\
			Kim08 & 0.8474 & 0.9177 & 0.2648 & 0.2151 \\
			\hline
			WLS &  0.8295 & 0.8754 & 0.2454 & 0.2360  \\
			LLF &  0.9300 & {\bfseries 0.9436 } & 0.6439  &  0.2224 \\
			Bruce14 & 0.8648 & 0.8654 & 0.4104 &  0.2352  \\
			GR & 0.8808 & 0.8794 & 0.4619 & 0.2334 \\
			NLPD-Opt & 0.8870 & 0.9133 & 0.4529 & {\bfseries 0.2003}   \\
			Khan18 & 0.9233 & 0.9132 &  0.6562 &  0.2295 \\
			Liang18 & 0.9128 & 0.8992 & 0.6135 & 0.2301 \\
			Zhang20 & 0.9275 & 0.8774 & 0.7327 & 0.2386 \\
			Zhang21 & 0.8579 & 0.8747 & 0.3538 & 0.2405 \\
			Vinker21 &  0.9121 & 0.9083 & 0.5948 & 0.2179 \\
			Yang21 & 0.9006 & 0.8915 & 0.5486 & 0.2350 \\
			\hline
                Le21 & {\bfseries 0.9432 } & {\bfseries 0.9279 } & {\bfseries 0.7480 } & 0.2101 \\
			PS-TMO& {\bfseries 0.9509 } & 0.9145  & {\bfseries 0.8157 } & {\bfseries 0.2059} \\
		\bottomrule
		\end{tabular}
	\end{center}
 \vspace{-1em} 
\end{table}

Fig.~\ref{fig:indoor} compares the tone mapping results of  Reinhard05~\cite{reinhard2005dynamic}, WLS~\cite{farbman2008edge}, LLF~\cite{paris2011local}, Bruce14~\cite{BRUCE201412}, Liang18~\cite{liang2018hybrid}, Vinker21~\cite{vinker2021unpaired}, Yang21~\cite{yang2021deep}, and PS-TMO on a ``Windowsill'' HDR scene. Most TMOs are incapable of fully reproducing the details outside the windows. Among them, LLF does a better job in this at the cost of a relatively dark appearance. The result by Liang18~\cite{liang2018hybrid} 
is a little blurry due to the $\ell_{0}$-flattening. DNN-based methods Vinker21~\cite{vinker2021unpaired} and Yang21~\cite{yang2021deep} suffer from both color and contrast problems. In contrast, the result of PS-TMO looks more natural and engaging with rich details.

\subsubsection{Quantitative Comparison}
To evaluate the performance of the competing TMOs quantitatively, we adopt two objective metrics: TMQI~\cite{yeganeh2012objective} and NLPD~\cite{laparra2017perceptually}. TMQI is specifically designed for cross-dynamic-range image quality evaluation. It combines structural fidelity (denoted by $\mathrm{SF}$) and statistical naturalness (denoted by $\mathrm{SN}$) measurements to assess a tone-mapped image with reference to the corresponding HDR image. NLPD can also be applied to the cross-dynamic-range scenario because of the divisive normalization step, which serves as a form of local gain control. A larger TMQI or a smaller NLPD value indicates  better predicted quality. Table~\ref{tab:tmqi_nlpd_time_score} shows the results, from which we have several interesting observations. First, local operators generally outperform global operators in terms of TMQI. This is not surprising because TMQI is biased towards comparing local structure similarity, which is the design focus of local TMOs. Such result discrepancy is less pronounced in terms of NLPD. Second, DNN-based methods are not necessarily better than conventional methods. This is also reasonable as it is generally difficult (and conceptually impossible) to specify ground-truth LDR images, otherwise, the problem of HDR image tone mapping is readily solvable. As a consequence, with a limited number of paired data (and a set of unpaired data)
 for supervised (and semi-supervised) training, the learned DNNs may be weak at generalizing to unseen challenging HDR scenes, producing unexpected visually annoying appearances. Third, as expected, NLPD-Opt achieves the best performance in terms of NLPD, followed by PS-TMO and Le21 (the preliminary version of PS-TMO with human specification of the working maximum luminance). Fourth, it is interesting to see that PS-TMO achieves the best performance measured by TMQI, which provides strong justifications for the perceptual (sub-)optimality of  PS-TMO.

\begin{figure}[t]
\centering
\subfloat{\includegraphics[width=0.98\linewidth]{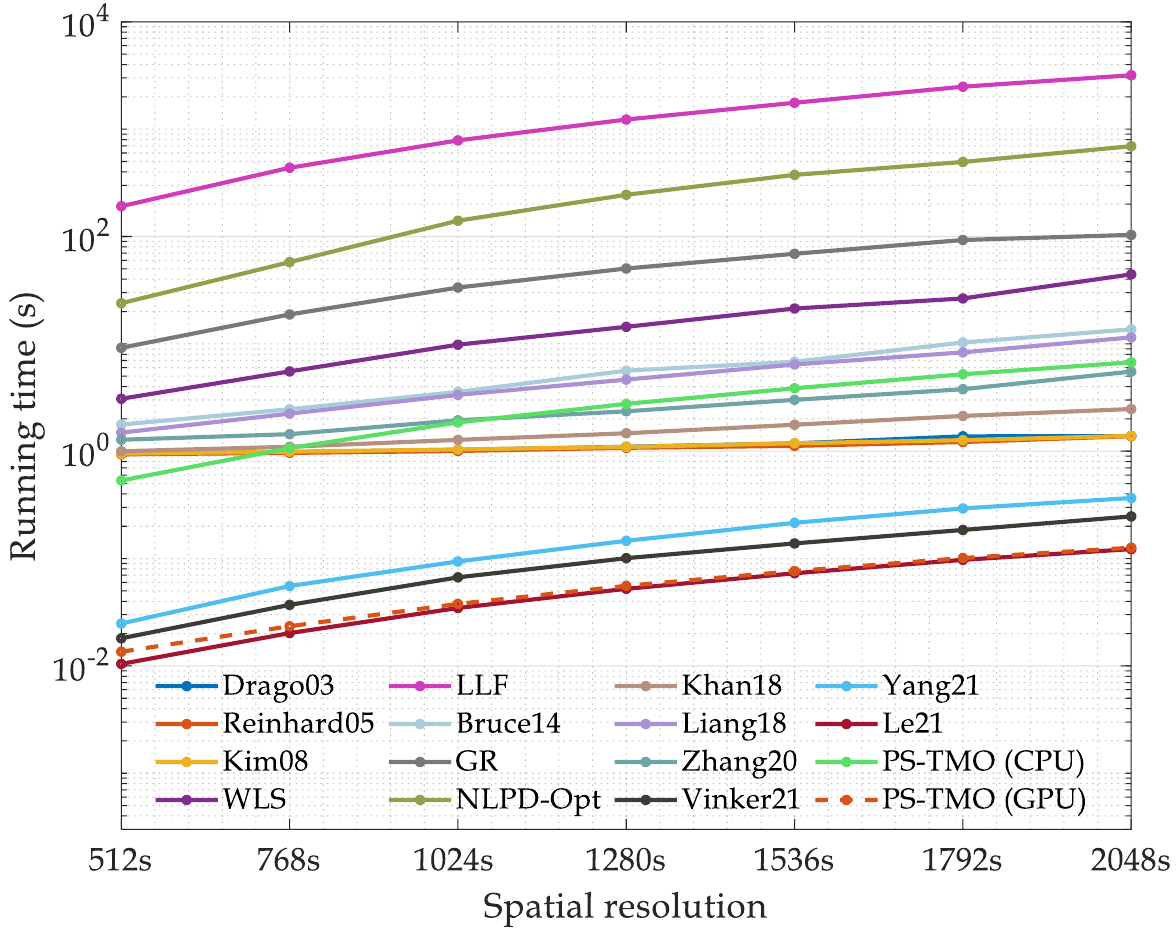}}\\
\vspace{-0.5em} 
   \caption{ Running time comparison across a wide range of resolutions. The ``s'' appending to each number in the horizontal axis means that the short side of the test HDR image is equal to (or resized to) the target resolution.}
\label{fig:time}
\vspace{-0.6em} 
\end{figure}

We test the computation time of PS-TMO with the fourteen TMOs on a computer with a 4.4GHz CPU, a 64G RAM, and an NVIDIA GTX 3080Ti GPU. All conventional methods are based on the MATLAB implementations~\cite{Banterle2017}, while the DNN-based methods are implemented using PyTorch (or Tensorflow for Yang21). It can be observed from Fig.~\ref{fig:time} that PS-TMO based on the CPU only runs the fastest among all local TMOs for resolutions ranging from $512$s to $1,024$s thanks to the manually optimized lightweight network architectures. Here, the ``s'' appending to the resolution number indicates that the short side of the test HDR image is equal to (or resized to) that target resolution.
Moreover, when equipped with the GPU, PS-TMO takes less than $0.13$ second to process images with resolutions ranging from $512$s to $2,048$s, which is faster than DNN-based Vinker21 and Yang21.

\subsubsection{Debiased Subjective Experiment}\label{sec:subjective}
In order to further validate that NLPD and MEF-SSIM optimization indeed result in perceptual gains of PS-TMO, we carry out a debiased subjective experiment~\cite{cao2021debiased} in a normal indoor office. To ensure a fair comparison (\ie, to avoid potential cherry-picking test results), we adopt the debiased subjective assessment method to select $15$ HDR images of diverse content variations and luminance ranges. 
After that, we invite $15$ subjects, including $8$ males and $7$ females with ages between $20$ and $30$, to participate in the subjective experiment. All subjects have general knowledge of image processing but are blind to the detailed purpose of this study. We adopt the two-alternative forced choice (2AFC) approach to gather human preferences  for several reasons. First, 2AFC involves a relatively simple experimental task, and is therefore well suited for non-expert participants. Second, it alleviates calibration issues, which are frequently encountered in cardinal measurements~\cite{tsukida2011analyze}. Third, it generally provides higher sensitivity and a lower measurement error when compared to cardinal rating~\cite{Shah2016}.
The image pairs for subjective testing are thus $\binom{16}{2}\times 15 = 1,800$, where $16$ is the number of the competing TMOs, including the proposed PS-TMO.
The subjects are free to zoom in to any portion of the images for more careful comparison, and are given unlimited time to look at the images and to make their decisions. 
Finally, we adopt the maximum likelihood for multiple options \cite{tsukida2011analyze} under the Thurstone's model \cite{thurstone1927law} to infer the global quality score.

\begin{figure}[!t]
\centering
\vspace{0.6em} 
\includegraphics[width=0.9\linewidth]{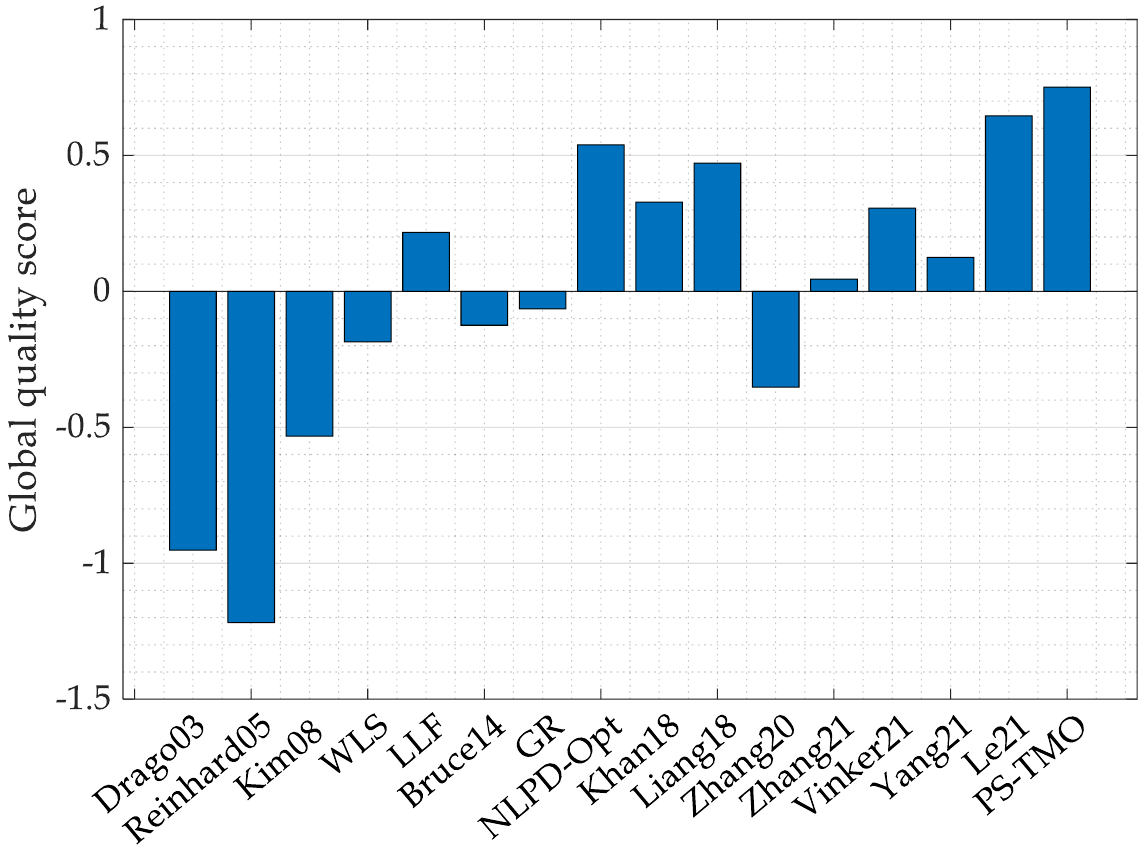}\\
   \caption{Comparison of the competing TMOs in our debiased subjective quality experiment.}
\label{fig:subjective}
\vspace{-0.9em} 
\end{figure}

\begin{figure}[!t]
  \centering
    \subfloat[One-level]{\includegraphics[width=0.48\linewidth]{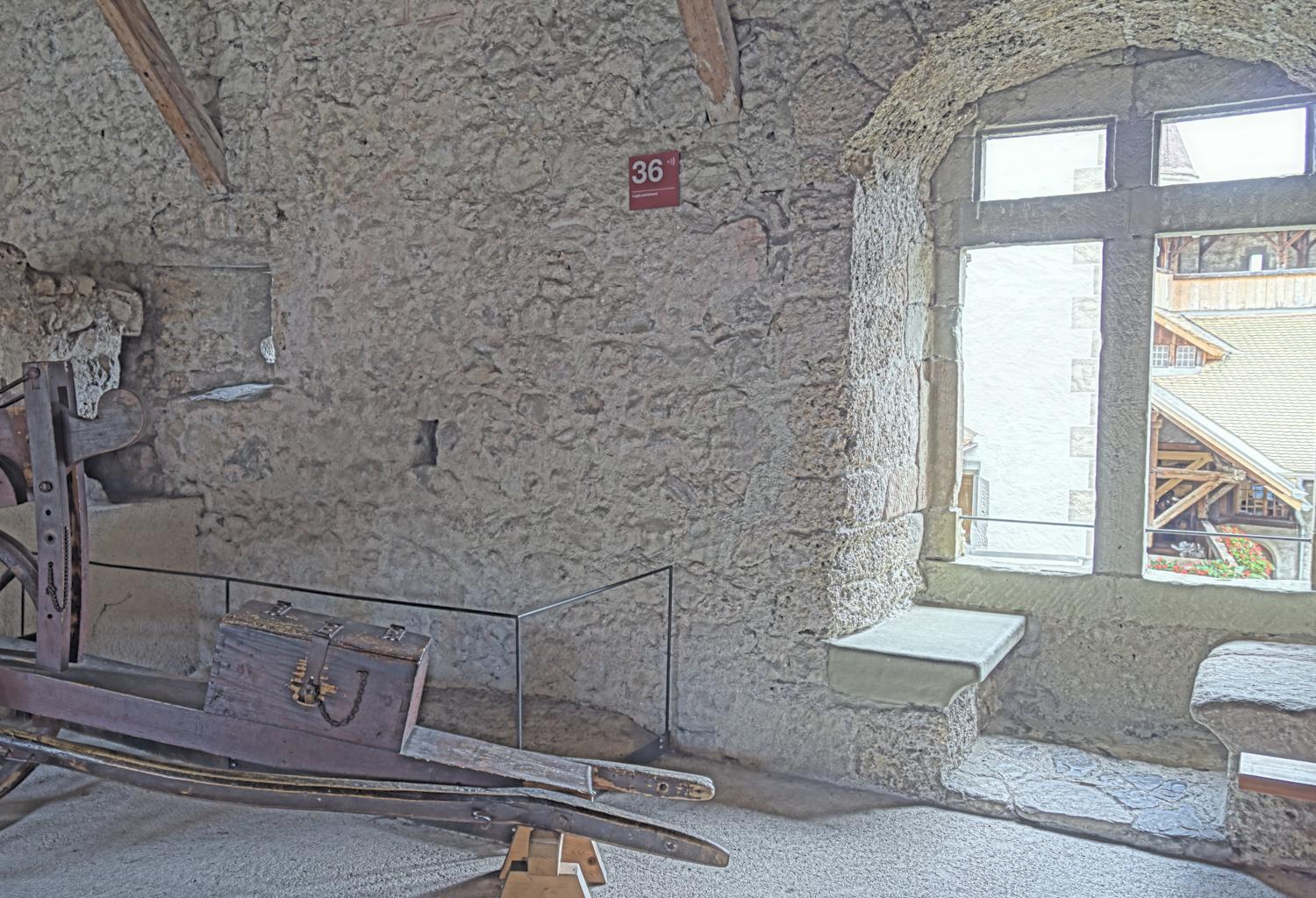}} \hskip0.3em
    \subfloat[Two-level]{\includegraphics[width=0.48\linewidth]{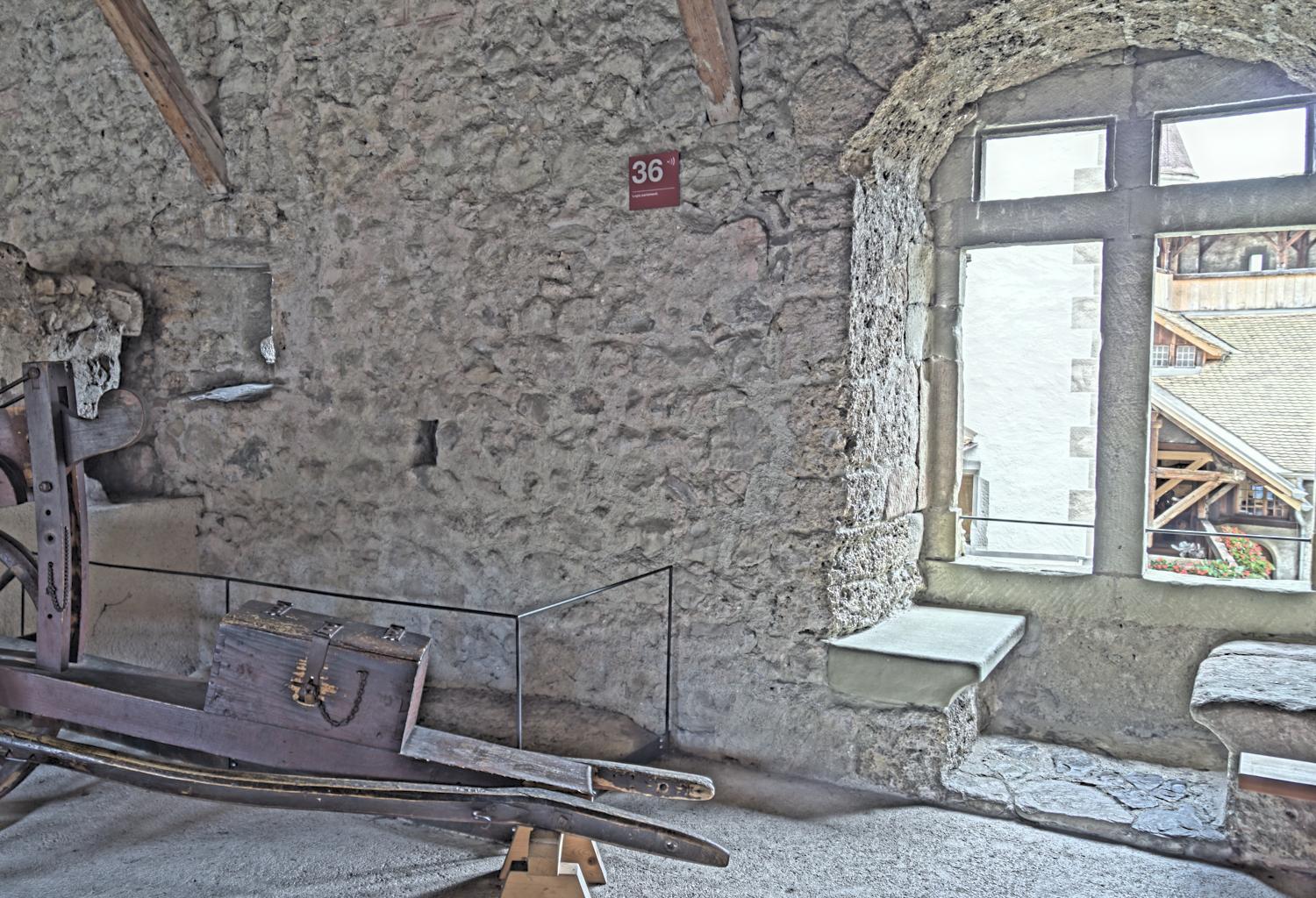}} \\ \vspace{-0.5em}
    \subfloat[Three-level]{\includegraphics[width=0.48\linewidth]{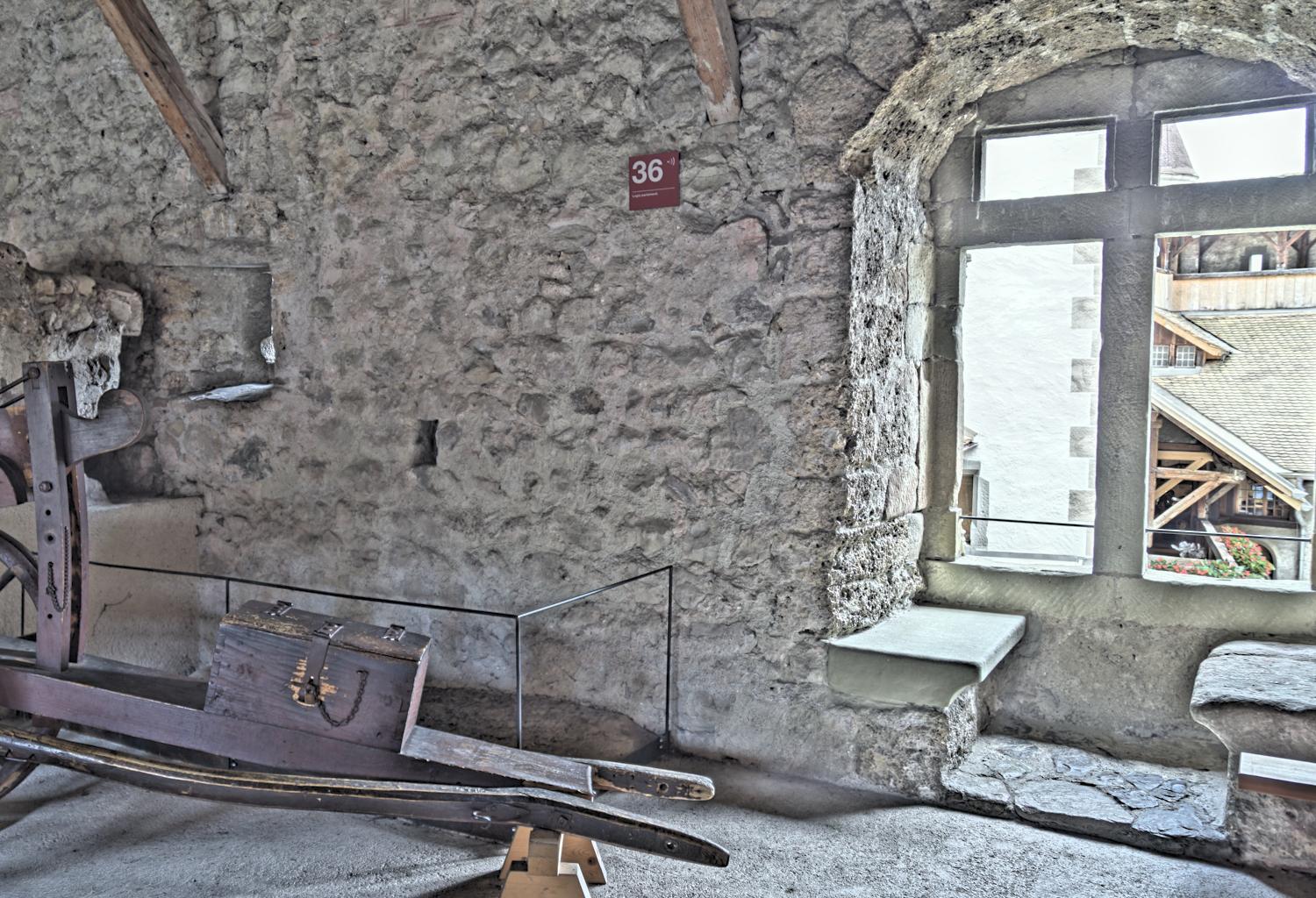}} \hskip0.3em
    \subfloat[Four-level]{\includegraphics[width=0.48\linewidth]{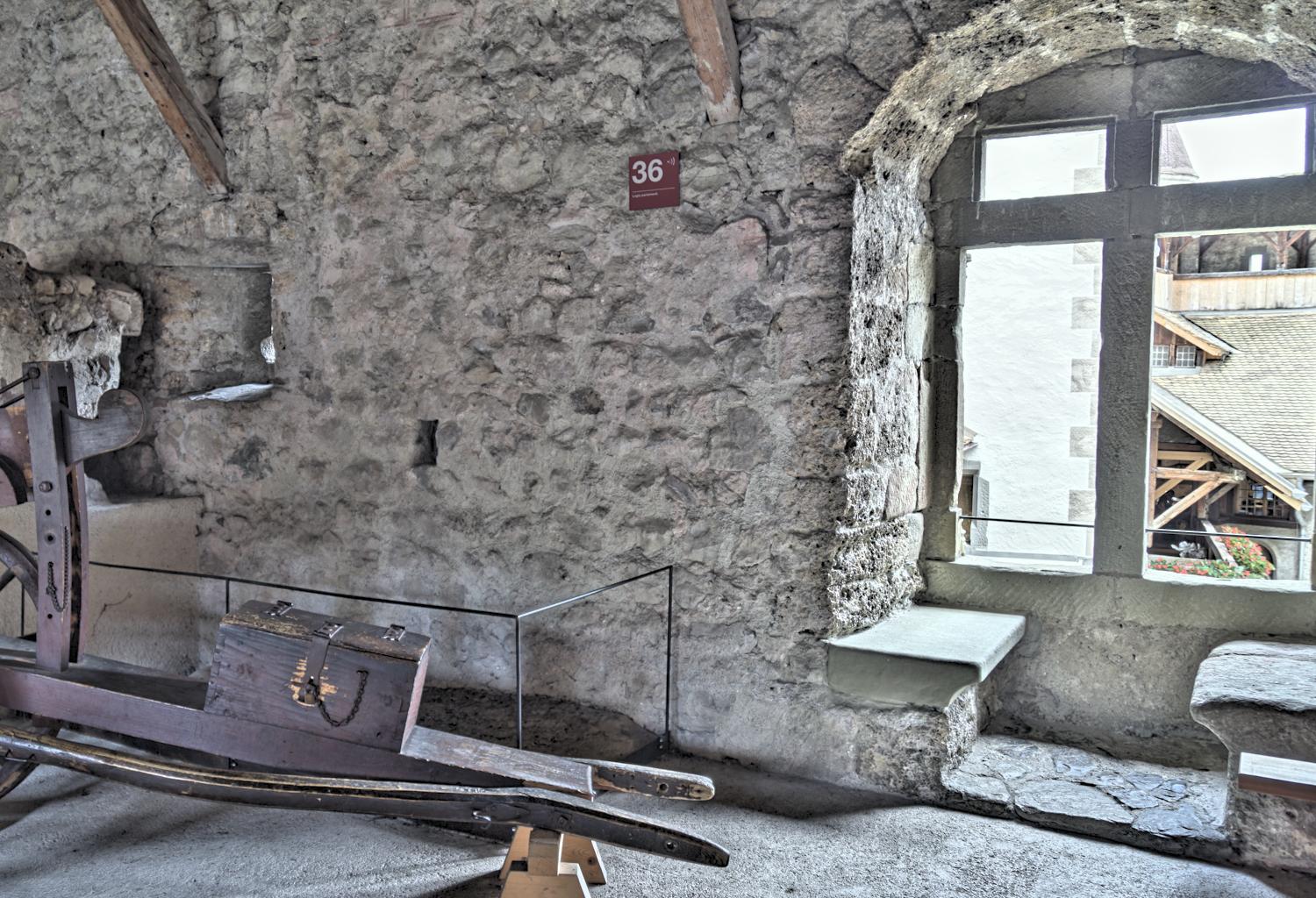}} \\  \vspace{-0.5em}
    \subfloat[\textbf{Five-level}]{\includegraphics[width=0.48\linewidth]{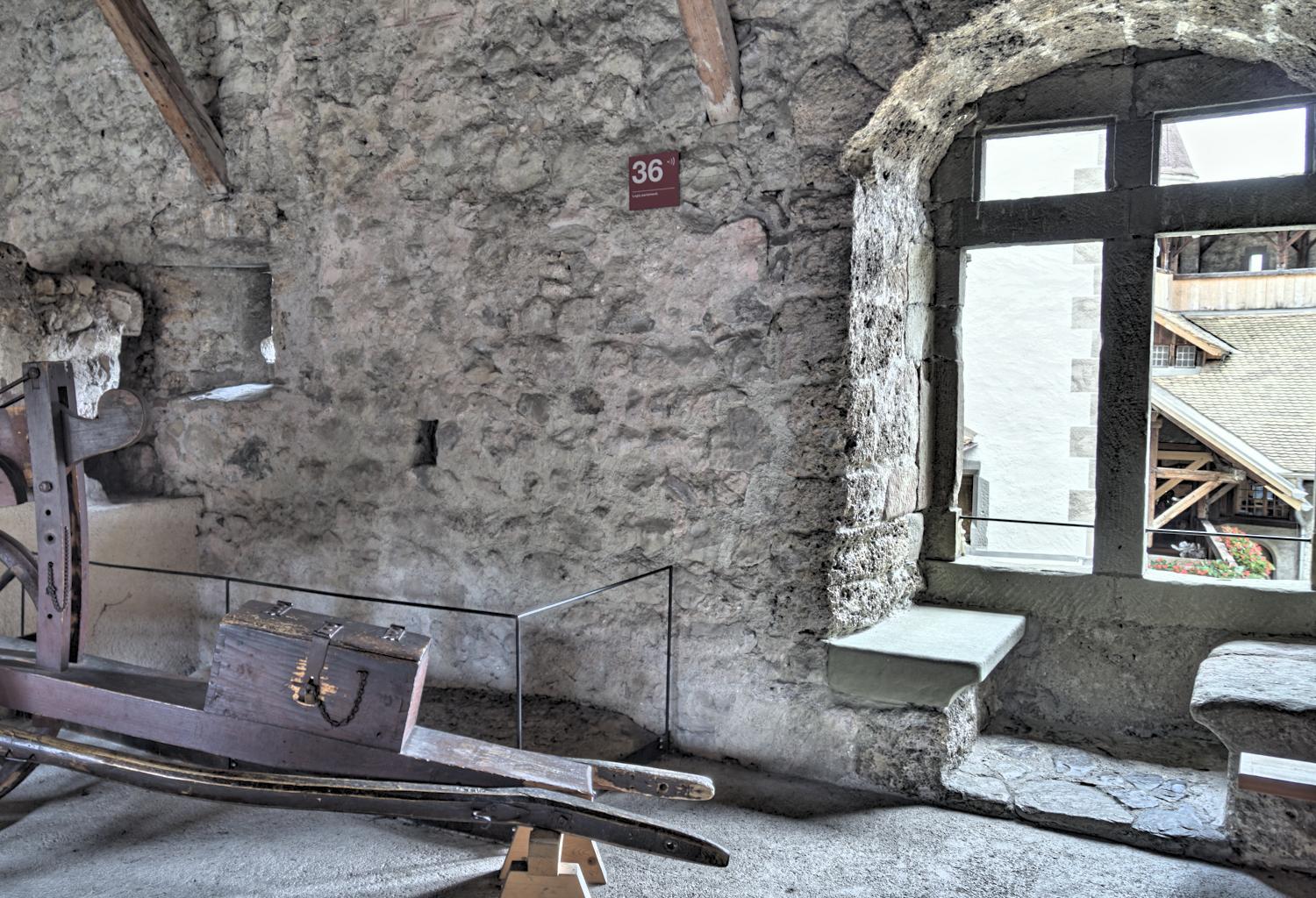}} \hskip0.3em
    \subfloat[Six-level]{\includegraphics[width=0.48\linewidth]{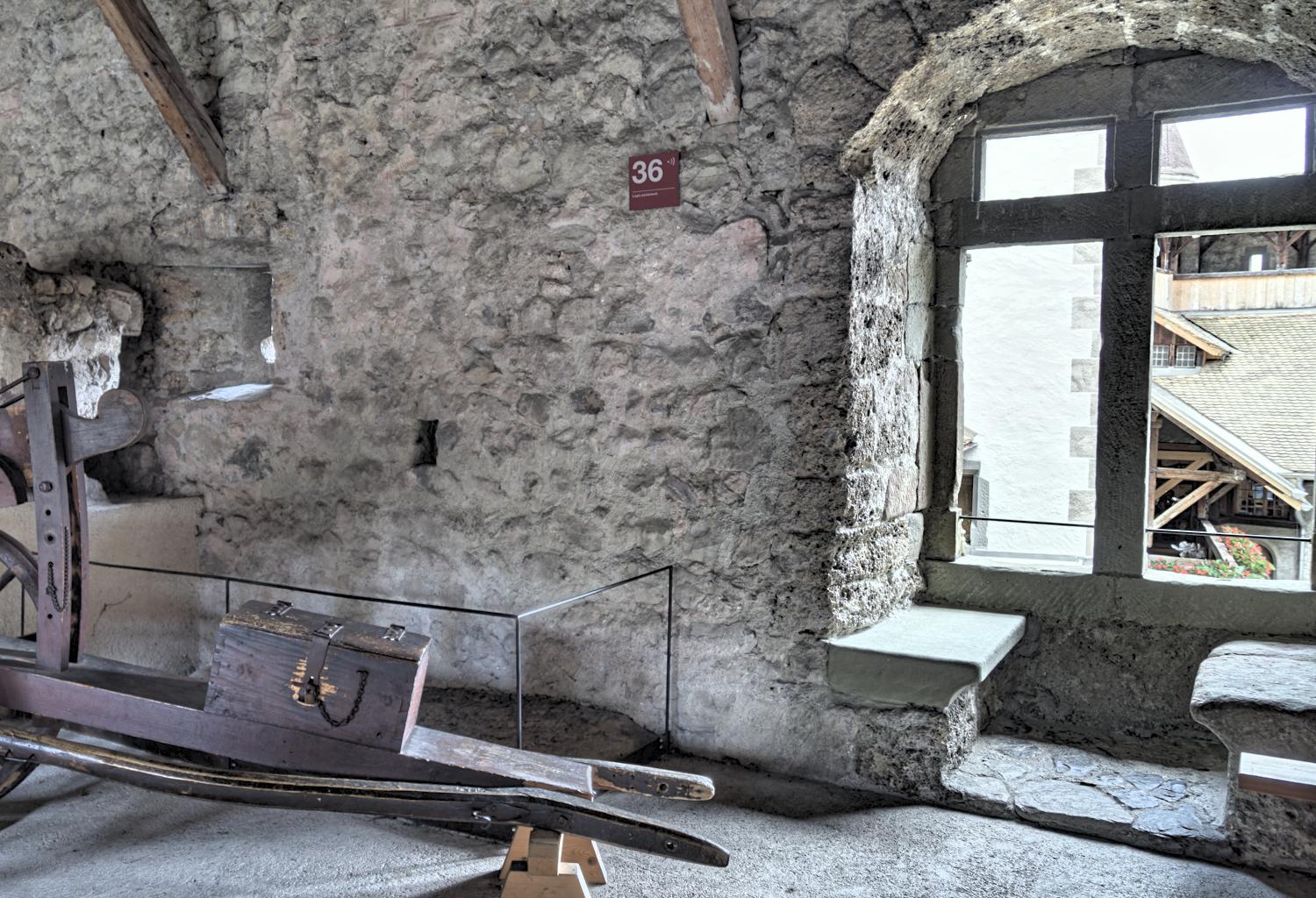}} \\ 
  \caption{Tone mapping results of the ``Workshop'' HDR scene with different input pyramid levels.} 
  \label{fig:level}
  \vspace{-0.5em} 
\end{figure}

We show the global quality aggregation results in Fig.~\ref{fig:subjective}, where we find that  
PS-TMO performs the best, followed by Le21, which even outperforms NLPD-Opt, optimizing for the same objective in the image space. We believe this arises because NLPD-Opt sometimes overfits NLPD during the single-example optimization, and creates an over-enhanced (and even noisy) appearance similar to GR~\cite{shibata2016gradient}. In Stage two, our MEF-SSIM-optimized fusion network is able to reduce the over-enhancement problem. Nevertheless, NLPD-Opt ranks third in our subjective experiment, verifying the suitability of NLPD as an objective quality measure for benchmarking existing TMOs and guiding the design of more perceptual TMOs. Moreover, some DNN-based methods achieve low rankings, and are far behind some conventional methods. We view this as caveats of the current ad-hoc combination of loss functions as the training objective without verifying the perceptual relevance. 

\begin{table}[t]
\renewcommand{\arraystretch}{1.25}
	\begin{center}
	\caption{Ablation results with different input pyramid levels. The default setting is highlighted in bold} \label{tab:scale}
    \resizebox{\linewidth}{!}{
		\begin{tabular}{l|c c c c c c}
			\toprule
			Pyramid Level & TMQI$\uparrow$ & SF$\uparrow$ & SN$\uparrow$ & NLPD$\downarrow$ & Time\\
			\hline
			One & 0.8955 & 0.7961 & 0.6292 & 0.2215 & 0.0634 \\
			Two & 0.9090 & 0.8354 & 0.6746 & 0.2178 & 0.0765 \\
			Three & 0.9294 & 0.8723 & 0.7294 & 0.2129 & 0.0807 \\
			Four & 0.9408 & 0.8897 & 0.7850 & 0.2093 & 0.0814 \\
			\textbf{Five} & 0.9509 & 0.9145 & 0.8157 & 0.2059 & 0.0820 \\
			Six & 0.9620 & 0.9408 & 0.8230 & 0.2046 & 0.0827 \\
		\bottomrule
		\end{tabular}}
	\end{center}
 \vspace{-1em} 
\end{table}

\begin{figure*}[t]
  \centering
    \subfloat[MAE-optimized]{\includegraphics[width=0.24\linewidth]{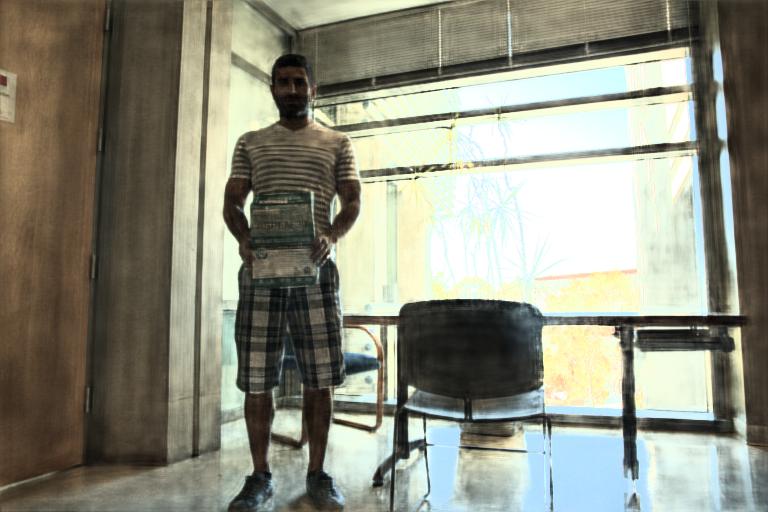}} \hskip0.3em
    \subfloat[SSIM-optimized]{\includegraphics[width=0.24\linewidth]{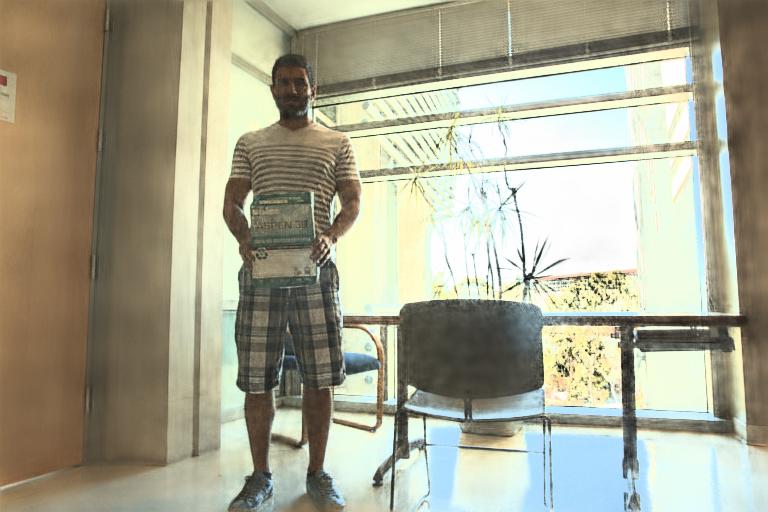}} \hskip0.3em
    \subfloat[TMQI-optimized]{\includegraphics[width=0.24\linewidth]{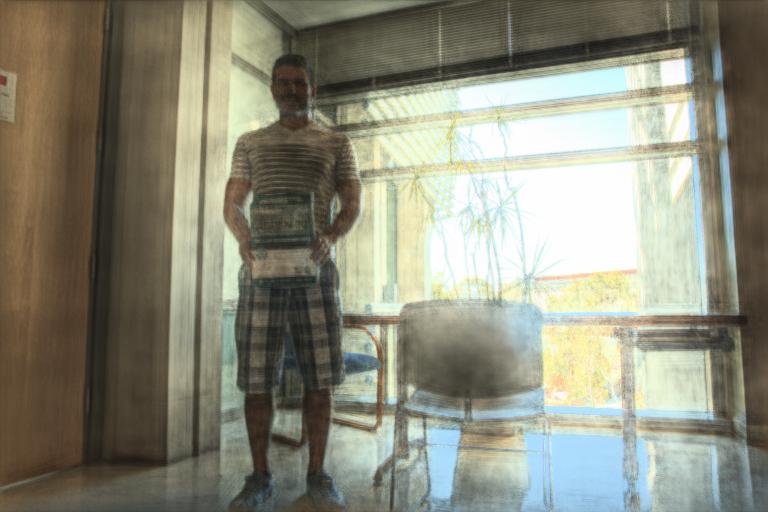}} \hskip0.3em
    \subfloat[\textbf{NLPD-optimized}]{\includegraphics[width=0.24\linewidth]{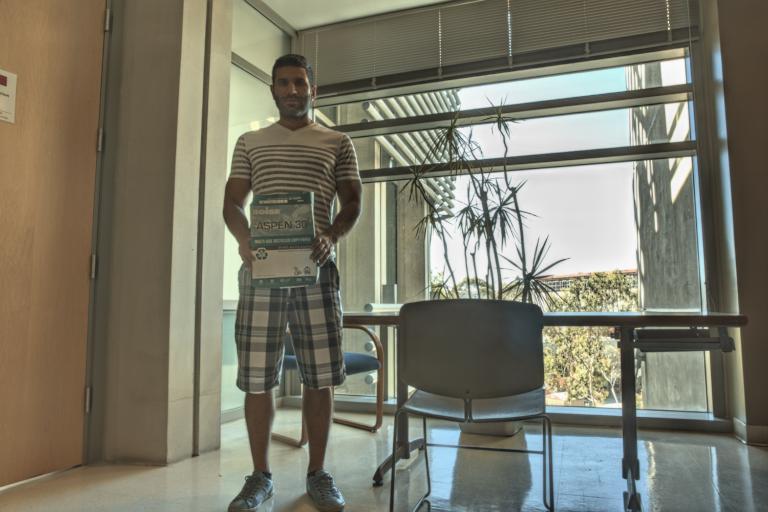}} \\ 
  \caption{Tone mapping results of the ``Man'' HDR scene with different objective functions.} 
  \label{fig:optimization}
  \vspace{-1em} 
\end{figure*}

\begin{figure}[t]
  \centering
  \addtocounter{subfigure}{0}
    \subfloat[]{\includegraphics[width=0.32\linewidth]{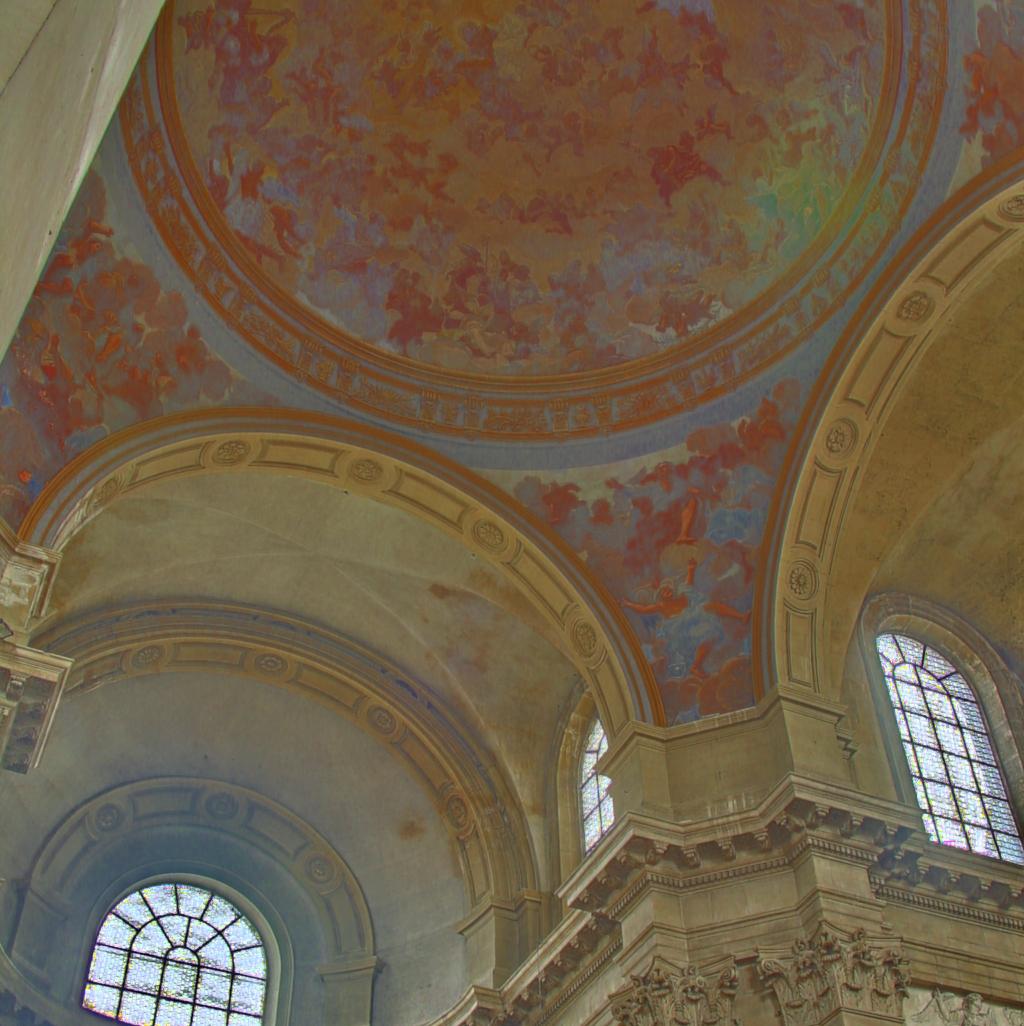}} \hskip0.3em
    \subfloat[]{\includegraphics[width=0.32\linewidth]{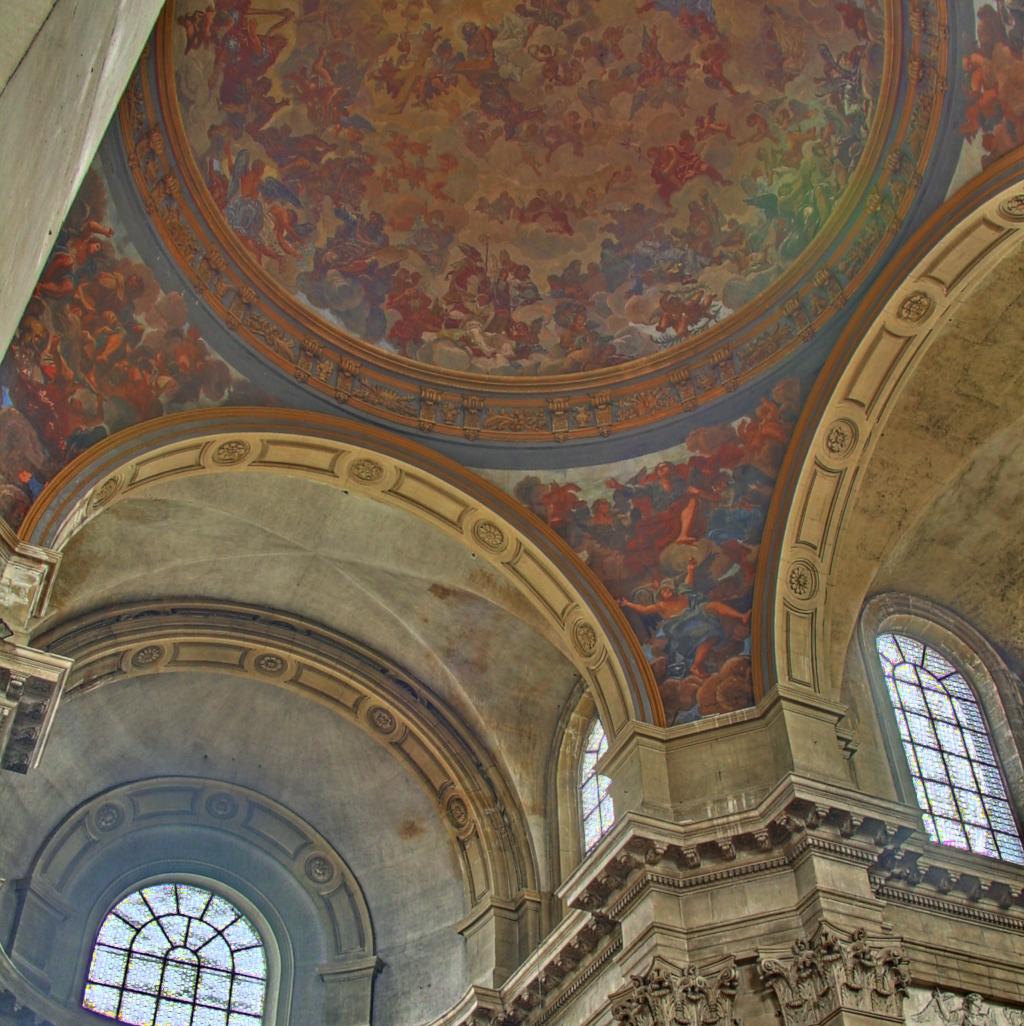}} \hskip0.3em
    \subfloat[]{\includegraphics[width=0.32\linewidth]{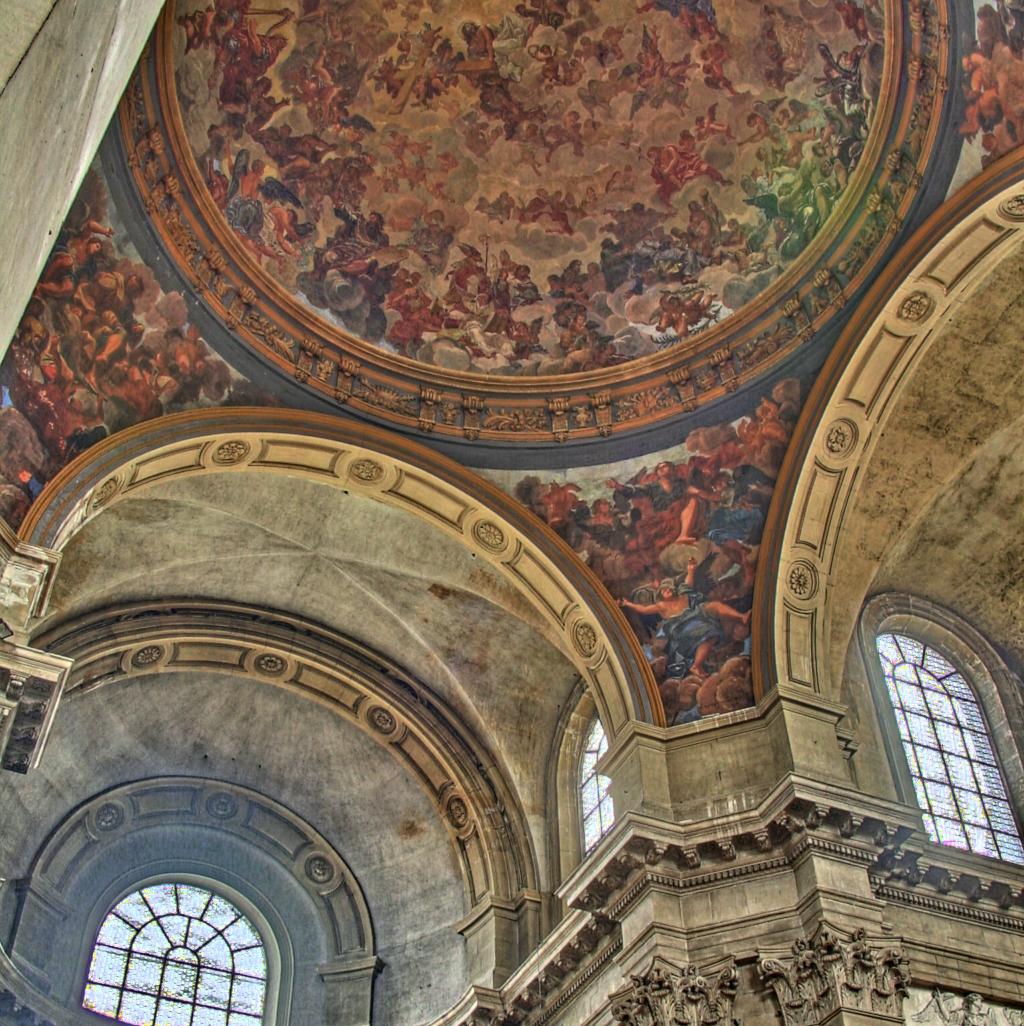}} \\ \vspace{-0.5em}
    \subfloat[]{\includegraphics[width=0.32\linewidth]{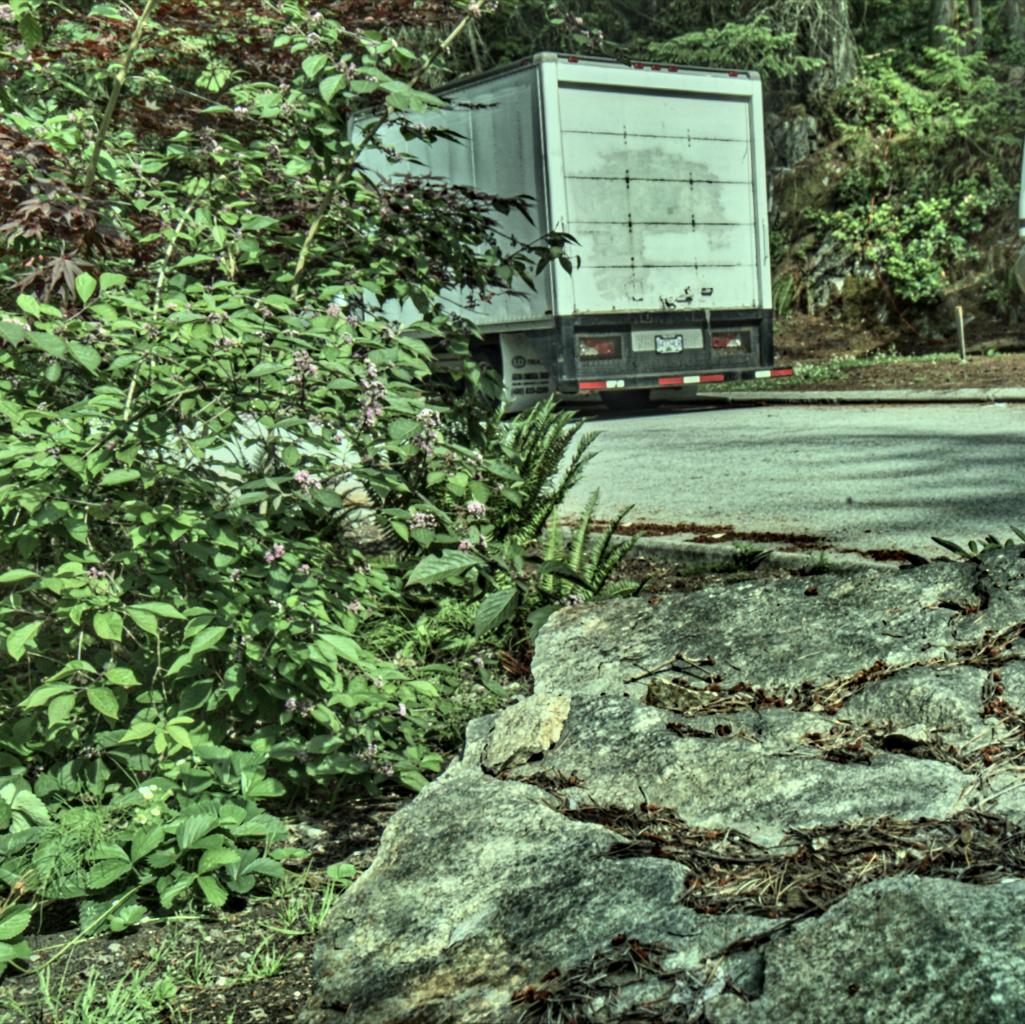}} \hskip0.3em
    \subfloat[]{\includegraphics[width=0.32\linewidth]{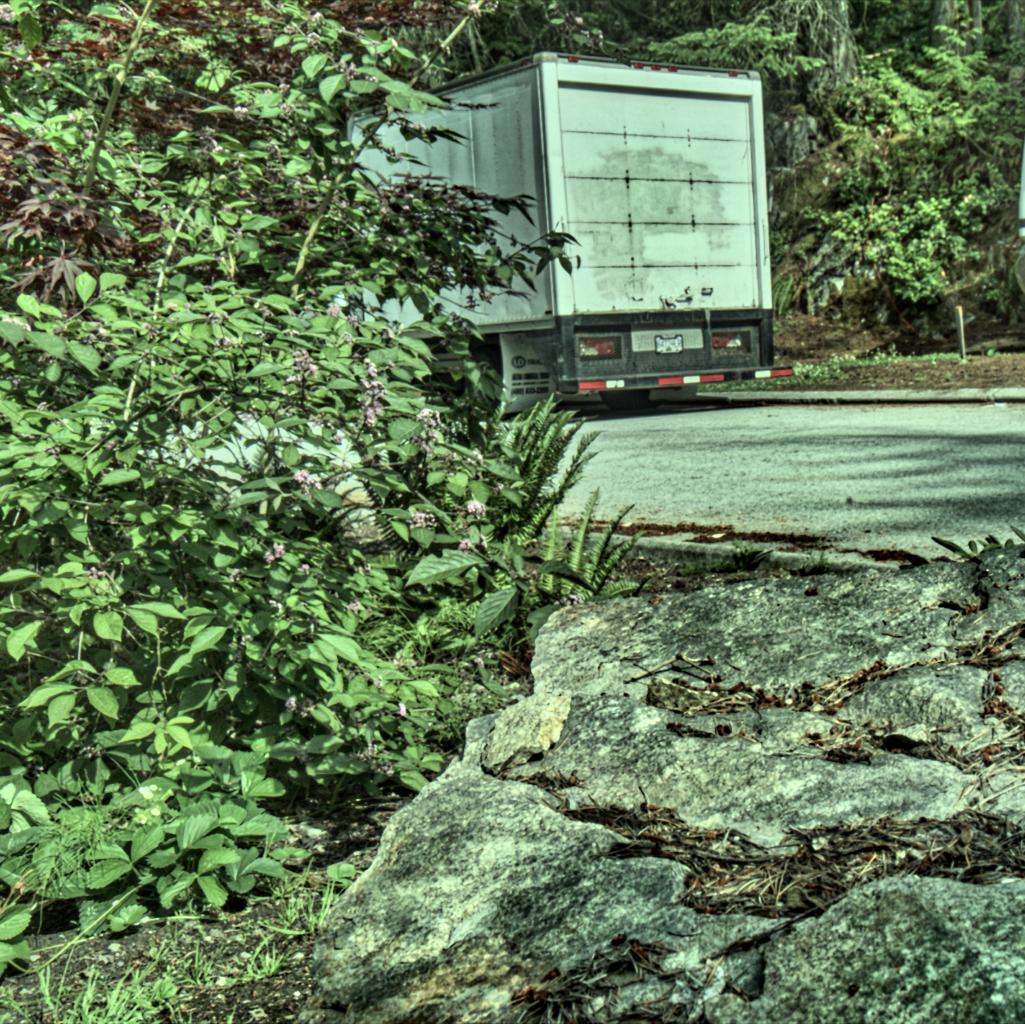}} \hskip0.3em
    \subfloat[]{\includegraphics[width=0.32\linewidth]{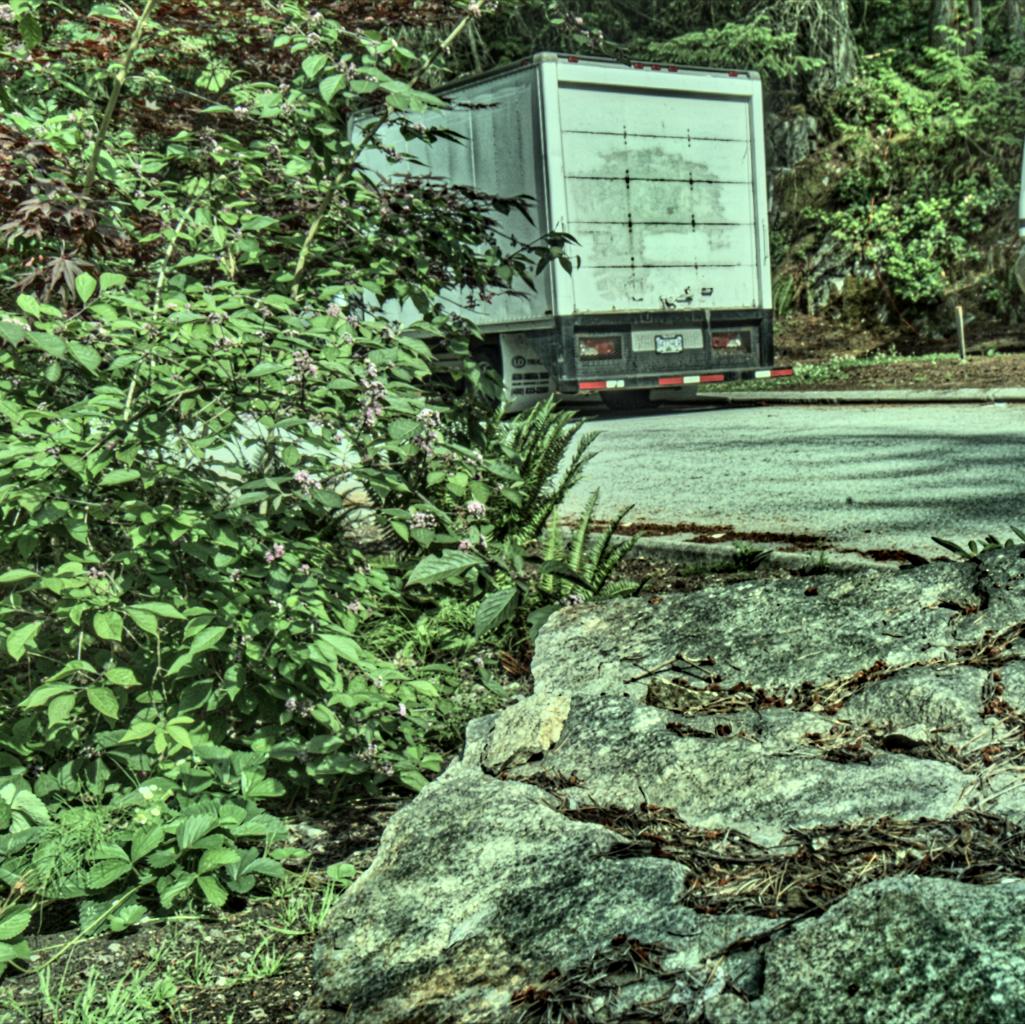}} \\ 
  \caption{Tone mapping results of the ``Arched Roof'' (top) and the ``Road'' (bottom) HDR scenes with different sets of maximum luminances for self-calibration. \textbf{(a)} and \textbf{(d)}: $[10^{3}, 10^{5}]$ $\rm cd/m^2$.  \textbf{(b)} and \textbf{(e)}: $[10^{3},10^{6}]$ $\rm cd/m^2$. \textbf{(c)} and \textbf{(f)}: $\mathbf{[10^{3},10^{7}]}$ $\rm cd/m^2$.} 
  \label{fig:range}
  \vspace{-1em} 
\end{figure}

\subsection{Ablation Experiments}\label{sec:ablation experiment}
We conduct a series of ablation experiments to single out the contributions of the algorithm design (\ie, normalized Laplace decomposition) and the perceptual optimization (\ie, NLPD for the tone mapping network and MEF-SSIM for the fusion network).

We first analyze the effect of the \textit{input pyramid level} on final visual quality. It is noteworthy that one level corresponds to directly feeding the raw HDR image into a single network for tone mapping.
As shown in Fig.~\ref{fig:level}, more levels lead to improved detail reproduction at the cost of increased computational complexity, which is also evidenced by the quantitative results in Table~\ref{tab:scale}. The default five-level pyramid keeps a good balance between visual quality and computational speed.

We then disable the fusion network, and \textit{switch NLPD to three other objective functions}: mean absolution error (MAE), SSIM~\cite{wang2004image}, and TMQI~\cite{yeganeh2012objective}, while fixing the tone mapping network architecture. Fig.~\ref{fig:optimization} shows
the optimization results, which are optimal under their respective objectives. As can be seen, the NLPD-optimized image better preserves structures outside the window with few artifacts. Qualitatively, we find these results consistent across a wide range of HDR scenes.

During the self-calibration of PS-TMO, we sample $K=5$ maximum luminances uniformly (in the logarithmic scale) from  the range of $[10^3, 10^7]$ $\rm cd/m^2$, which covers the maximum luminances of most challenging HDR scenes\footnote{For example, a luminance of $10^7$ $\rm cd/m^2$ corresponds to the filament of a clear incandescent lamp. See \url{https://en.wikipedia.org/wiki/Orders_of_magnitude_(luminance)} for more information.}. Here we fix $K$ and compare the tone mapping results  self-calibrated by three different maximum luminance ranges - $[10^{3}, 10^{5}]$, $[10^{3},10^{6}]$, and $[10^{3},10^{7}]$ $\rm cd/m^2$. The results are shown in Fig.~\ref{fig:range}, where we find that the ``Arched Roof'' scene with a higher dynamic range benefits from a higher $S_\mathrm{max}$, and the tone-mapped image is well-saturated and more detailed. For the ``Road'' with a lower dynamic range, the tone mapping result is relatively insensitive to the setting of $S_\mathrm{max}$. To make PS-TMO more widely applicable, it is preferred to work with a wider maximum luminance range, and let the fusion network decide which pseudo-exposures to rely on (see Fig.~\ref{fig:fusion}). 

\begin{figure}[t]
  \centering
    \subfloat[Three-exposure]{\includegraphics[width=0.32\linewidth]{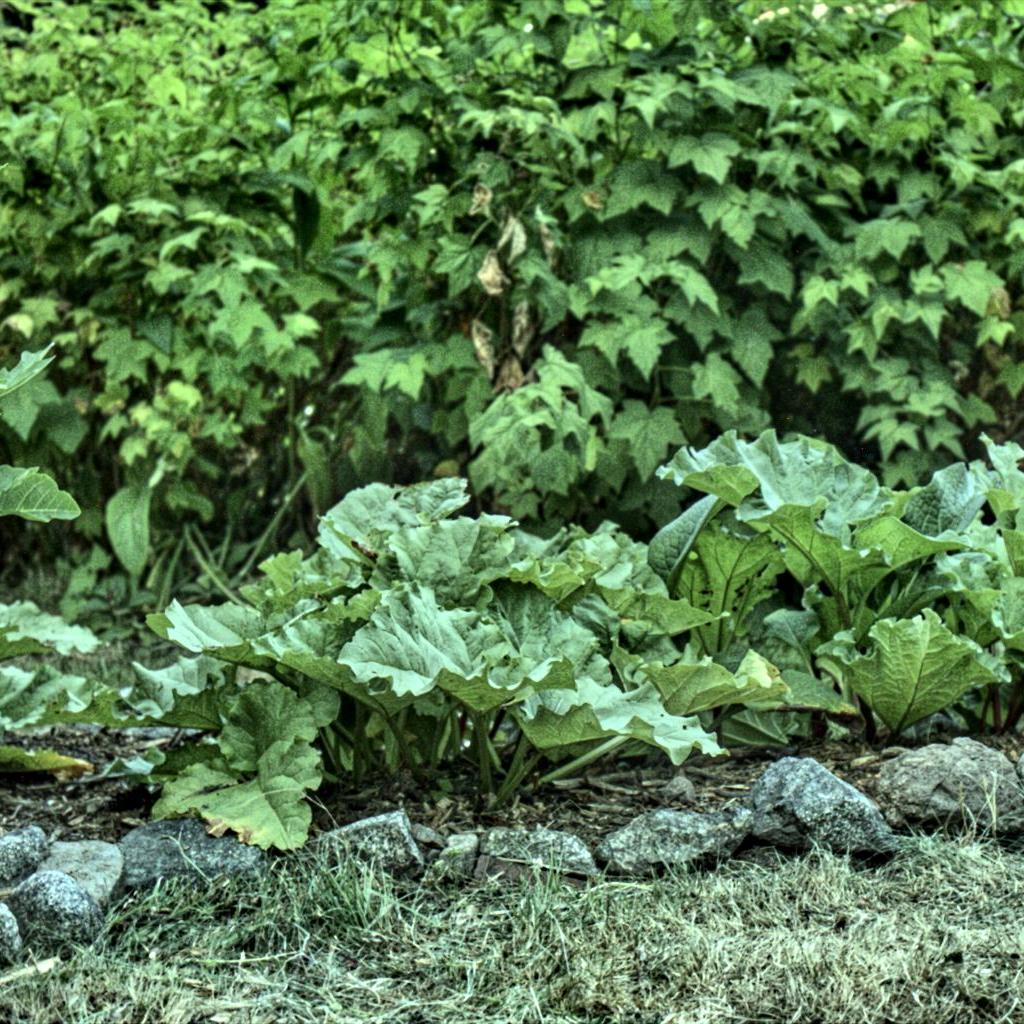}} \hskip0.3em
    \subfloat[\textbf{Five-exposure}]{\includegraphics[width=0.32\linewidth]{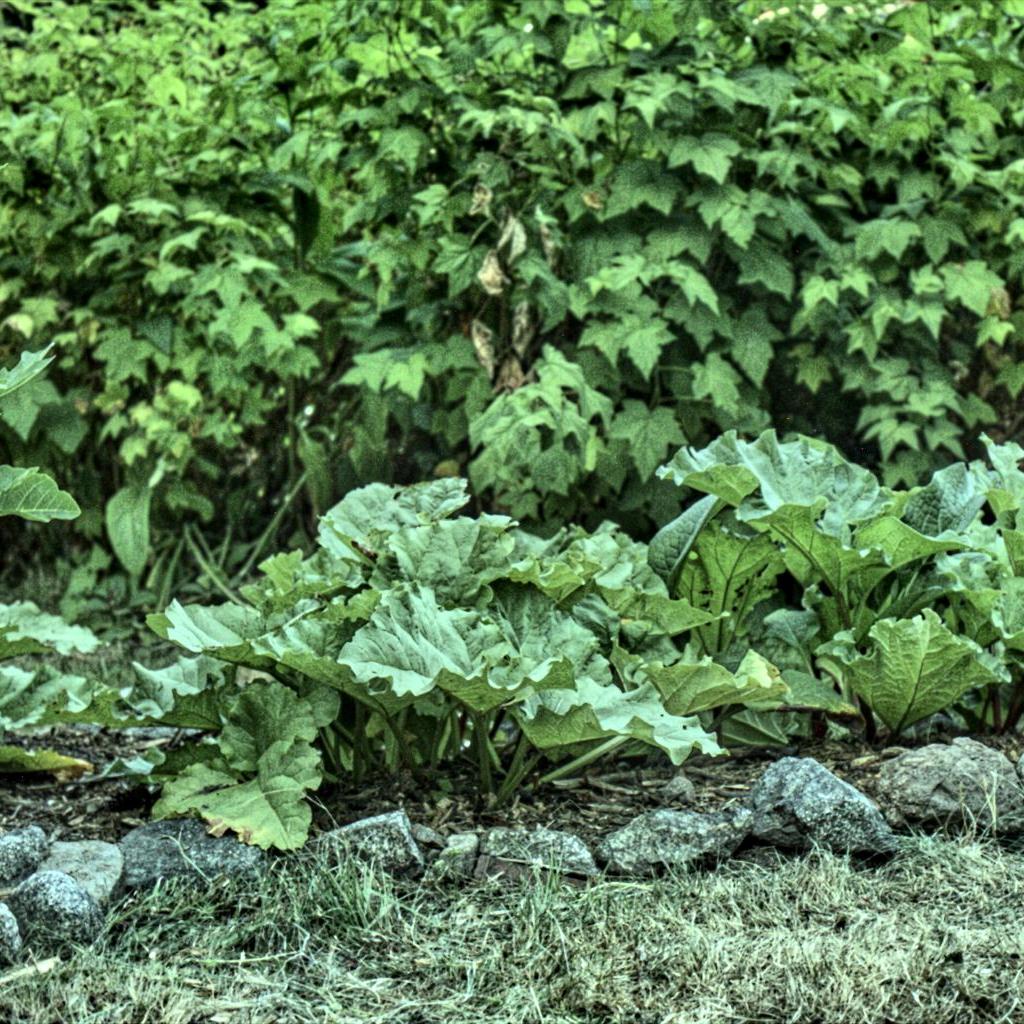}} \hskip0.3em
    \subfloat[Seven-exposure]{\includegraphics[width=0.32\linewidth]{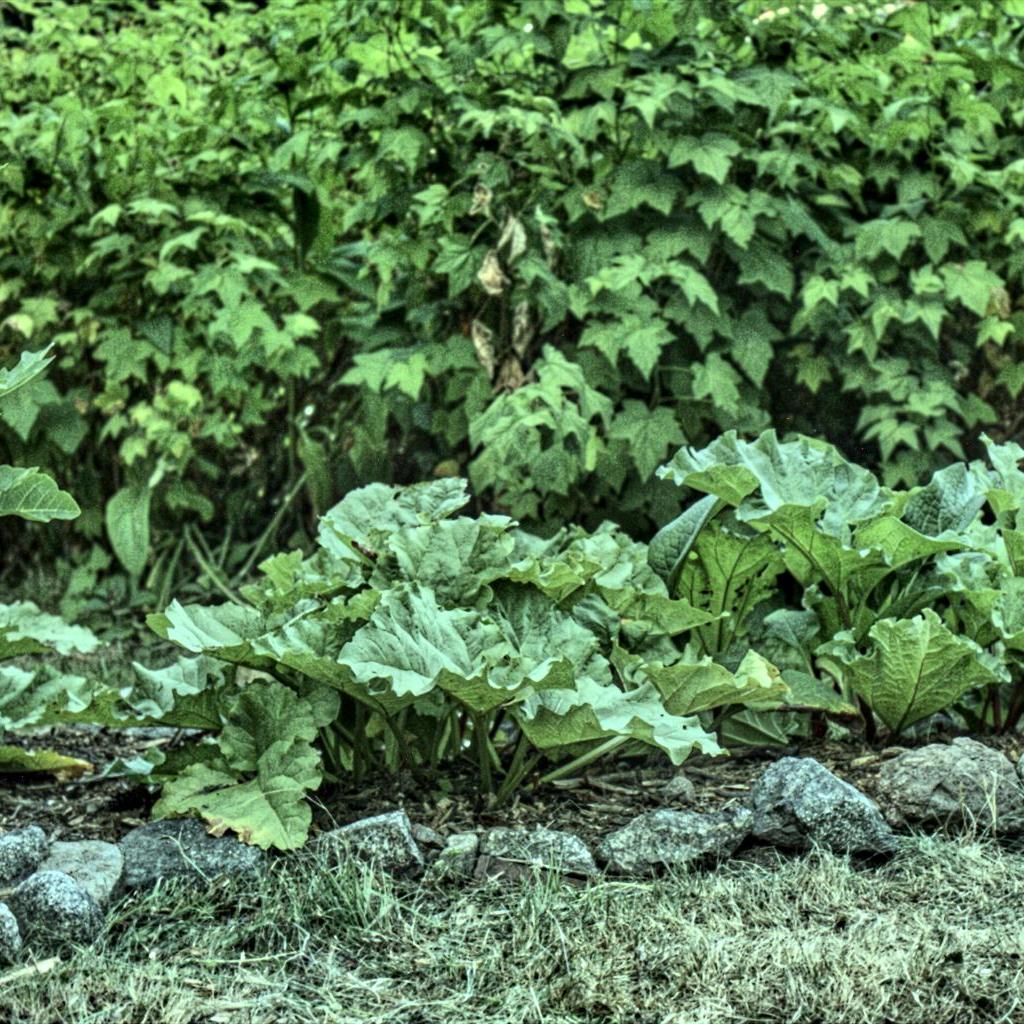}} \\\vspace{-0.5em}  
    \subfloat[Three-exposure]{\includegraphics[width=0.32\linewidth]{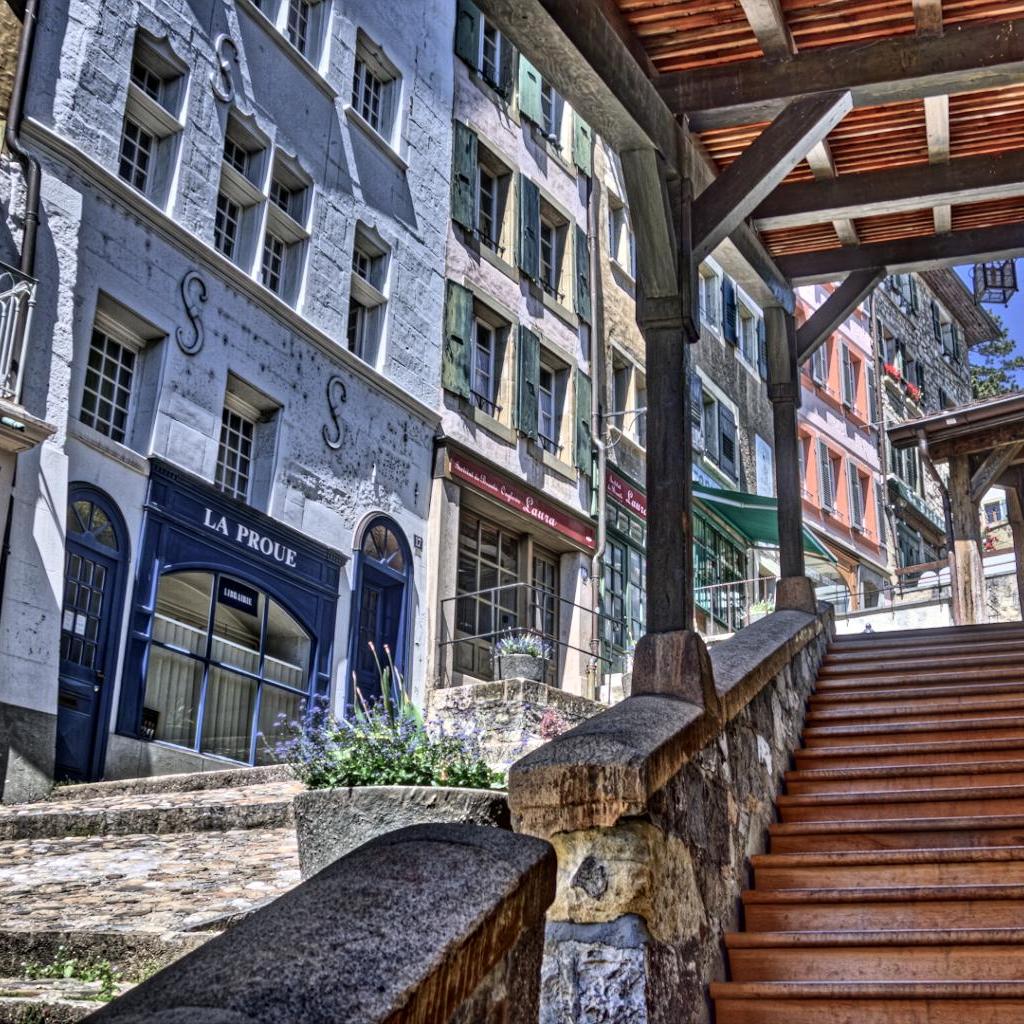}} \hskip0.3em
    \subfloat[\textbf{Five-exposure}]{\includegraphics[width=0.32\linewidth]{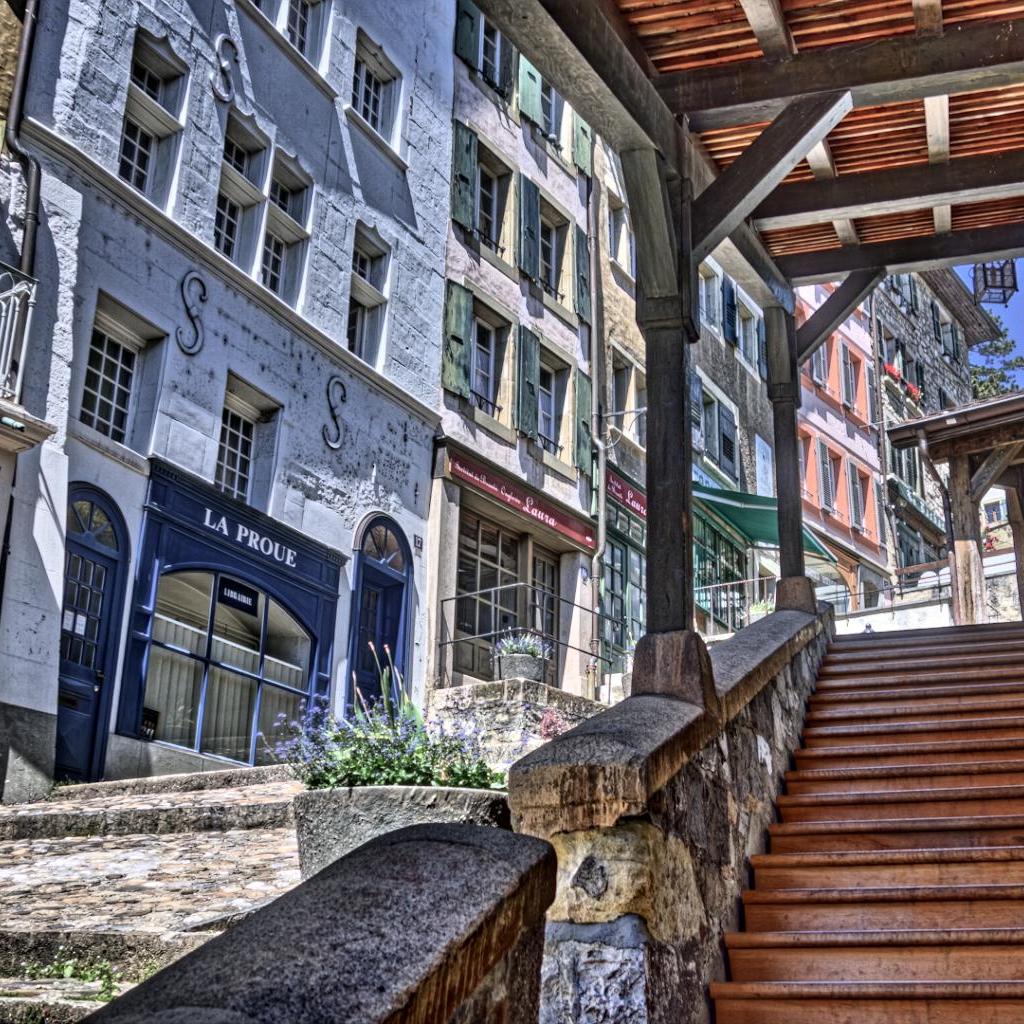}} \hskip0.3em
    \subfloat[Seven-exposure]{\includegraphics[width=0.32\linewidth]{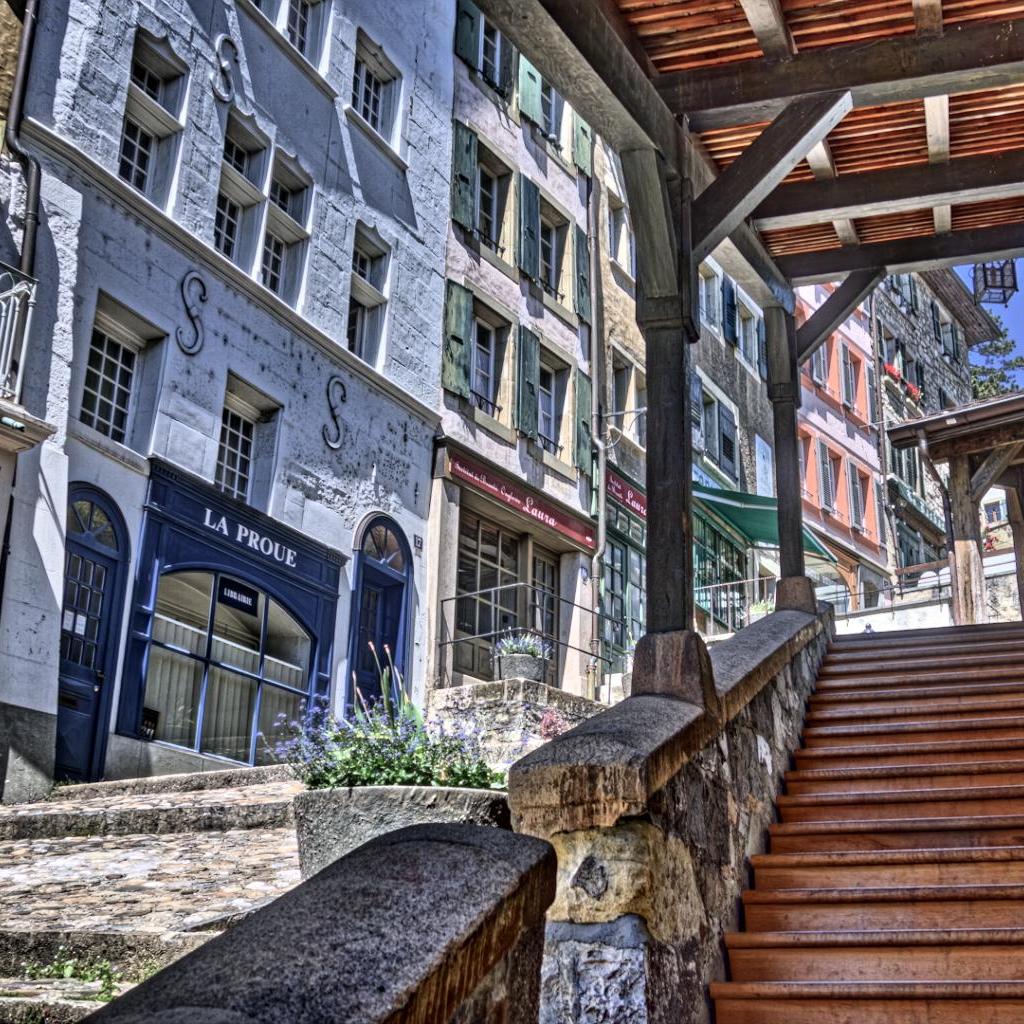}} \\ 
  \caption{Tone mapping results of (a)-(c) the ``Leafy Plant'' and the (d)-(f) the ``Outdoor Corridor'' HDR scenes with pseudo-multi-exposure image stacks of different lengths. \textbf{(a)}: TMQI = $0.9689$, NLPQ = $0.0737$. \textbf{(b)}: TMQI = $0.9696$, NLPQ = $0.0728$. \textbf{(c)}: TMQI = $0.9701$, NLPQ = $0.0721$.\textbf{(d)}: TMQI = $0.9005$, NLPQ = $0.2320$. \textbf{(e)}: TMQI = $0.9009$, NLPQ = $0.2311$. \textbf{(f)}: TMQI = $0.9013$, NLPQ = $0.2310$. } 
  \label{fig:exp_num}
  \vspace{-1em} 
\end{figure}

We next fix the maximum luminance range to $[10^3, 10^7]$ $\rm cd/m^2$ during self-calibration, and vary $K$, \textit{the length of the pseudo-multi-exposure image stack}, to probe the robustness of PS-TMO. We generate three image stacks consisting of three, five, and seven pseudo-multi-exposure images, respectively. The results are shown in Fig.~\ref{fig:exp_num}, where we observe that although TMQI and NLPD improve slightly with the length of the image stack, such improvements are barely noticeable by the human eye.

\begin{figure}[t]
  \centering
  \addtocounter{subfigure}{0}
    \subfloat[$\rho=0.4$ ]{\includegraphics[width=0.32\linewidth]{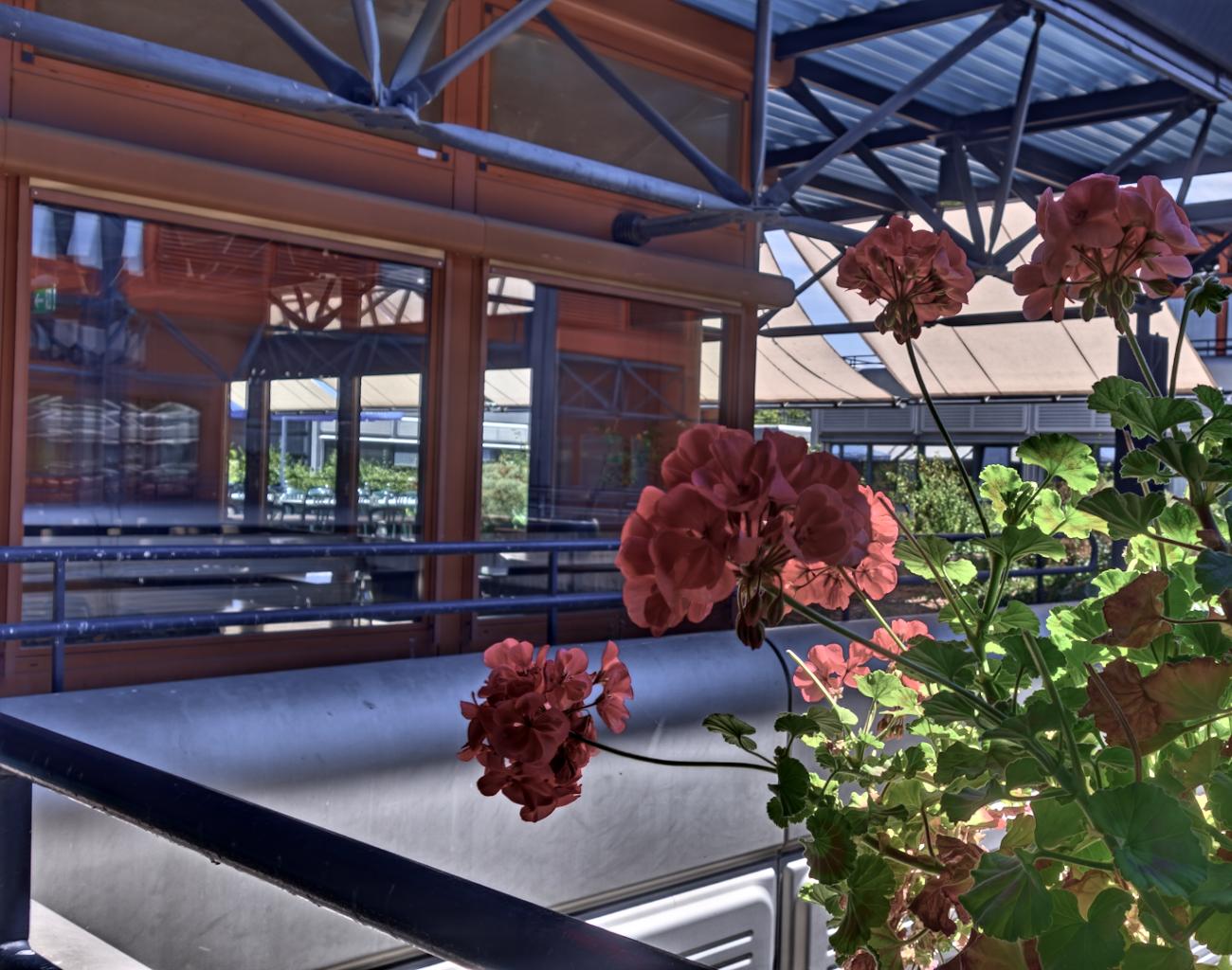}} \hskip0.3em
    \subfloat[$\mathbf{\rho=0.6}$ ]{\includegraphics[width=0.32\linewidth]{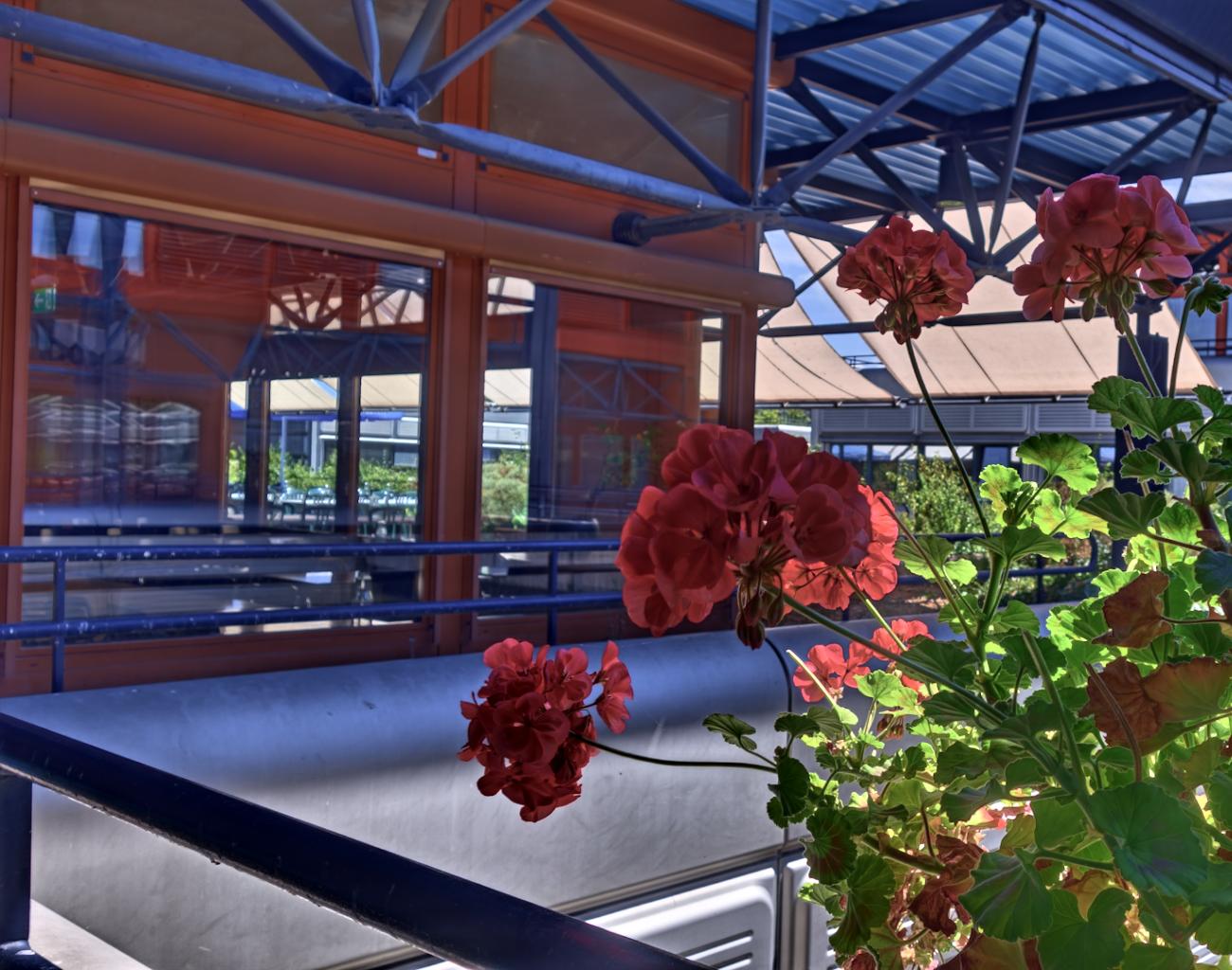}} \hskip0.3em
    \subfloat[$\rho=0.8$ ]{\includegraphics[width=0.32\linewidth]{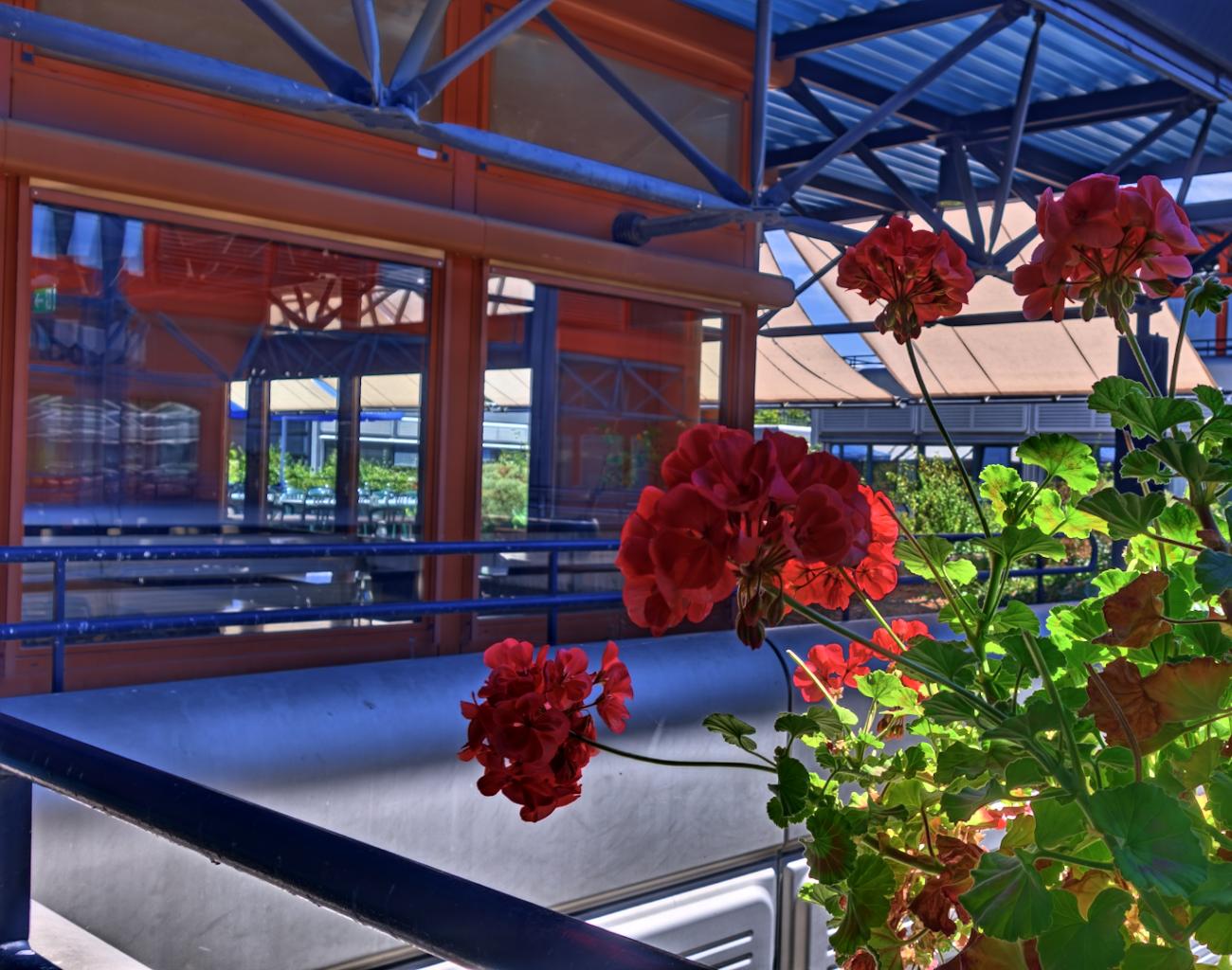}} \\ \vspace{-0.5em}
    \subfloat[$\rho=0.4$ ]{\includegraphics[width=0.32\linewidth]{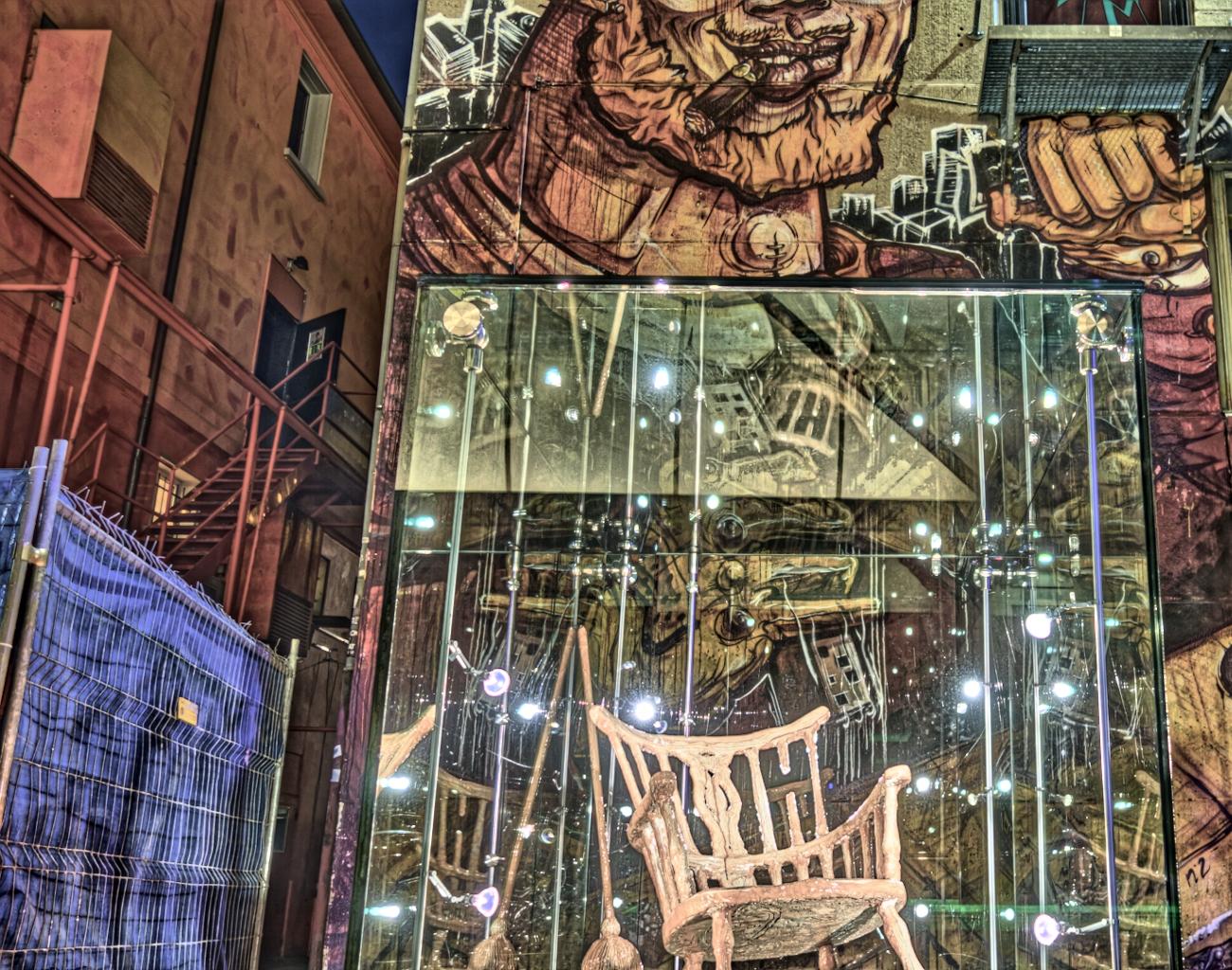}} \hskip0.3em
    \subfloat[$\mathbf{\rho=0.6}$ ]{\includegraphics[width=0.32\linewidth]{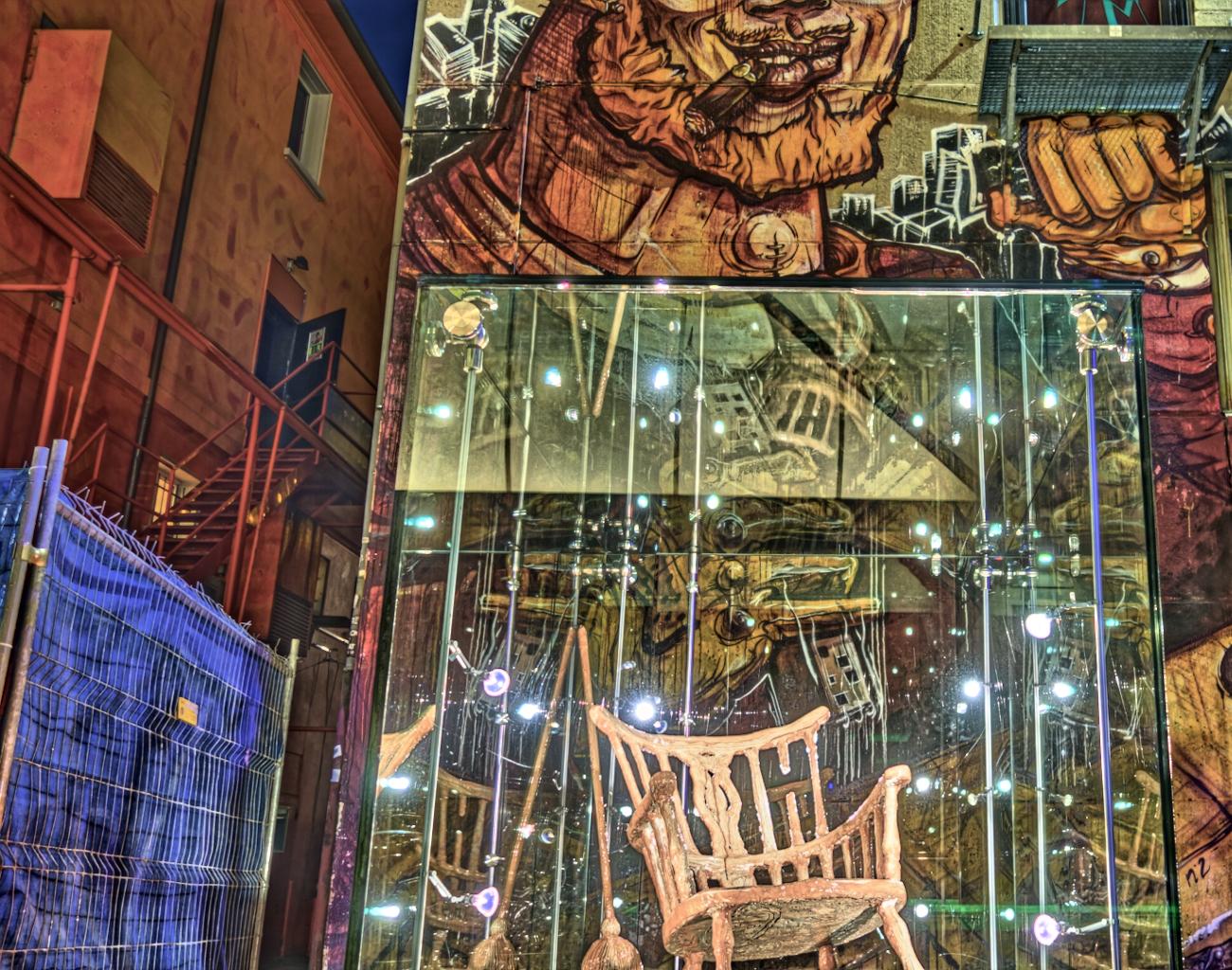}} \hskip0.3em
    \subfloat[$\rho=0.8$ ]{\includegraphics[width=0.32\linewidth]{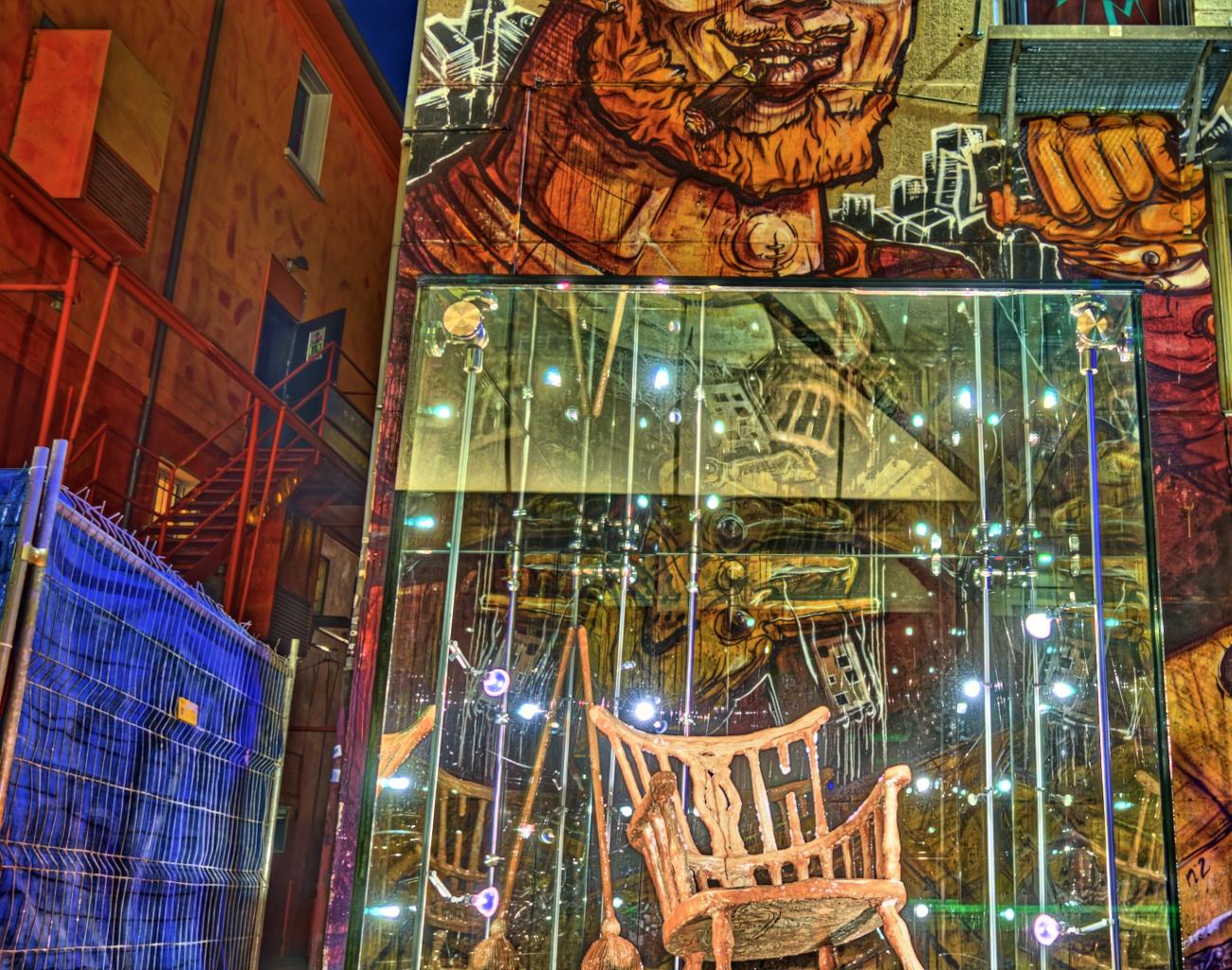}} \\ 
  \caption{Tone mapping results of \textbf{(a)}-\textbf{(c)} the ``Red Flowers'' and \textbf{(d)}-\textbf{(e)} the ``Show Window'' HDR scenes with different $\rho$ in Eq.~\eqref{eq:color1}.} 
  \label{fig:saturation}
  \vspace{-1em} 
\end{figure}

We last analyze the color saturation parameter $\rho$ in Eq.~\eqref{eq:color1} of PS-TMO, which can be adapted for different subjective preferences. We compare the tone mapping results with  $\rho\in\{0.4, 0.6, 0.8\}$ in Fig.~\ref{fig:saturation}, from which it is clear that a higher $\rho$ leads to a more color-saturated image. Empirically, we find that the default setting of $\rho=0.6$ works well across a variety of HDR scenes.

\begin{figure}[t]
  \centering
  \addtocounter{subfigure}{0}
    \subfloat[]{\includegraphics[width=0.48\linewidth]{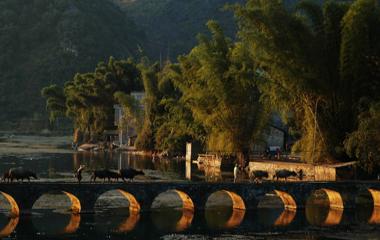}} \hskip0.3em
    \subfloat[]{\includegraphics[width=0.48\linewidth]{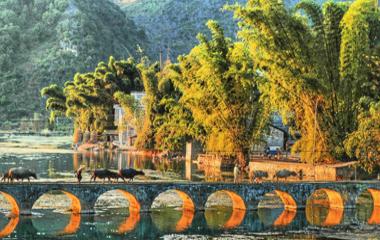}} \\ \vspace{-0.5em}
    \subfloat[]{\includegraphics[width=0.48\linewidth]{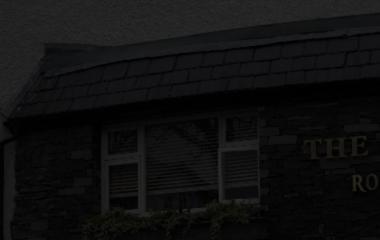}} \hskip0.3em
    \subfloat[]{\includegraphics[width=0.48\linewidth]{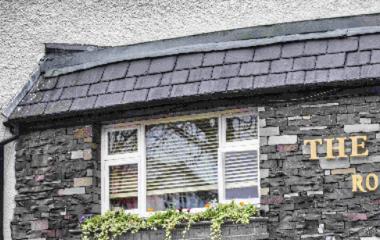}} \\ 
  \caption{Visual examples of low-light image enhancement. \textbf{(a)} and \textbf{(c)}: Low-light images. \textbf{(b)} and \textbf{(d)}: Enhanced images corresponding to (a) and (c), respectively, by PS-TMO.} 
  \label{fig:enhance}
  \vspace{-1em} 
\end{figure}

\section{Conclusion and Discussion}\label{sec:conclusion}
We have introduced a computational method for HDR image tone mapping, namely PS-TMO, based on lightweight tone mapping and fusion networks, optimized sequentially for two perceptual metrics, NLPD and MEF-SSIM. The tone mapping network is trained to generate the pseudo-multi-exposure image stack by varying the maximum luminance of the input HDR image. The fusion network is responsible for fusing the image stack into a final high-quality image that is high-contrast, well-exposed, and well-saturated. Without using the ground-truth LDR images for supervised training, PS-TMO matches and exceeds the state-of-the-art across a variety of HDR natural scenes. The perceptual advantages of PS-TMO are further verified by another perceptual quality metric - TMQI and in a form debiased subjective experiment.

The proposed PS-TMO is self-calibrated through MEF. In a similar spirit, we may artificially manipulate the light source in the scene (by linearly rescaling the maximum luminance $S_\mathrm{max}$) to endow PS-TMO (in particular the tone mapping network) with the capability of low-light and normal-light image enhancement (see Fig.~\ref{fig:enhance}), which is worthy of further investigation. Meanwhile, in our experiments, we assume a fixed display constraint with a minimum luminance of $I_\mathrm{min} = 5$ $\rm cd/m^{2}$ and a maximum luminance of $I_\mathrm{max}=300$ $\rm cd/m^{2}$, while the luminance ranges for displays on the market vary. Therefore, in the future, we will take steps to incorporate various display constraints into the proposed perceptual optimization framework.

{
\bibliographystyle{IEEEtran}
\bibliography{eg}
}

\vfill

\end{document}